\documentclass[a4paper,preprint, pre,showpacs, nofootinbib, floatfix]{revtex4-1}

\usepackage{graphicx}
\usepackage{amsfonts}
\usepackage{color}
\usepackage{amsmath}
\usepackage[maxfloats=100]{morefloats}

\usepackage{dsfont}

\newcommand{\argmax}{\mathrm{argmax}}
\newcommand{\argmin}{\mathrm{argmin}}

\newcommand{\x}{\ensuremath{\mathbf{x}}}
\newcommand{\y}{\ensuremath{\mathbf{y}}}
\newcommand{\z}{\ensuremath{\mathbf{z}}}
\newcommand{\m}{\ensuremath{\mathbf{m}}}

\newcommand{\Lmatrix}{\ensuremath{\mathbf{\Lambda}}}

\newcommand{\cmatrix}{\ensuremath{\mathbf{c}}}
\newcommand{\Cov}{\ensuremath{\mathbf{C}}}

\newcommand{\alpham}{\ensuremath{  \boldsymbol{\alpha}    }}

\newcommand{\1}{\mathbf{1}} 

\newcommand{\Ind}{\mathds{1}}

\newcommand{\thetav}{\ensuremath{\boldsymbol{\theta}}}

\newcommand{\thetam}{\ensuremath{\mathbf{\Theta}}}
\newcommand{\xmatrix}{\ensuremath{\mathbf{X}}}

\newcommand{\av}{  \ensuremath{\mathbf{a}}  }

\newcommand{\amatrix}{\ensuremath{\mathbf{A}}}
\newcommand{\nullv}{\ensuremath{\mathbf{0}}}
\newcommand{\rmd}{\mathrm{d}}
\newcommand{\rme}{\mathrm{e}}
\newcommand{\rmi}{\mathrm{i}}
\newcommand{\Tr}{\mathrm{Tr}}

\newcommand{\BMatrix}{\ensuremath{\mathbf{B}}}

\newcommand{\A}{{\ensuremath{\mathbf{D}}}}

\newcommand{\new}{}

\makeindex

\begin{document}

\title{Mean-field theory of Bayesian clustering} 

\author{Alexander Mozeika$^1$}
\author{Anthony CC Coolen$^{1,2}$}

\affiliation{ $^1$Institute for Mathematical and Molecular Biomedicine, King's College London, Hodgkin Building, London SE1 1UL, UK.
\\ $^2$Department of Mathematics,  King's College London, The Strand, London WC2R 2LS, UK.}

\date{\today}

\begin{abstract}
We show that model-based Bayesian clustering, the probabilistically most systematic approach to the partitioning of data, can be mapped into a statistical physics problem for a gas of particles, and as a result becomes amenable to a detailed quantitative analysis.  A central role in the resulting statistical physics framework is played by an entropy function.  We demonstrate that there is a relevant parameter regime where mean-field analysis of this function is exact, and that, under natural assumptions,  the  lowest entropy state of the hypothetical  gas corresponds  to the optimal clustering of data.  The byproduct of our analysis is a simple but effective clustering algorithm, which infers both the most plausible number of clusters in the data and the corresponding partitions. Describing  Bayesian clustering in statistical mechanical terms is found to be natural and surprisingly effective. 
\end{abstract}

\pacs{
02.50.Tt, 
05.10.-a, 
64.60.De 
}
\maketitle

\section{Introduction\label{section:intro}}
The need for clustering analysis  in scientific data exploration has grown significantly in recent years, due to the emergence of  large high-dimensional datasets  in areas such as high energy physics, astrophysics,  biology and post-genome medicine.   The aim of clustering analysis is to allocate similar data items, such as stars~\cite{Kuhn2014},  galaxies~\cite{Souza2017},  bacterial communities~\cite{Hanage2009}, or amino-acid sequences\new{~\cite{Martin2016}},  to the same category (or `cluster')  in an unsupervised way.   Inferring the true number of clusters reliably is crucial for the discovery of new data categories.  
Most current clustering methods, such as~\cite{Bishop2006,Frey2007,Rodriguez2014},  make no assumptions about  the data distribution, and are based on heuristic measures of similarity. Some allow for estimation of the number of clusters, but use empirical approaches to do so and ad-hoc evaluation criteria tested on benchmark datasets.

Model-based clustering assumes that each data point comes from one of a postulated number of populations with known distributions.  The archetypal example  is  the Gaussian Mixture Model (GMM)~\cite{Bishop2006}, which assumes Gaussian distributions. In such models Maximum likelihood (ML)  inference is typically used to find data partitions~\cite{Fraley2002}, but this is prone to overfitting~\cite{Bishop2006}.  The number of clusters $K$ is found upon adding a `penalty'  term to the log-likelihood function, such as AIC or BIC~\cite{Fraley2002}, sometimes with conflicting results~\cite{Souza2017}.  Bayesian inference of  GMM-generated data cures overfitting and  provides a systematic way to find $K$~\cite{Bishop2006}. However, computing  the posteriors  is  analytically intractable,  and one tends to resort to either variational mean-field approximation~\cite{Bishop2006} or computationally intensive  MCMC~\cite{Nobile2007}.  

A more general model-based Bayesian clustering protocol (SPD) was introduced in~\cite{Corander2009}.  Unlike GMM, it uses priors on the partitions to compute a maximum a posteriori probability (MAP) estimate of the data partitioning.  Both SPD and GMM Bayesian methods are usually  evaluated by  clustering synthetic and  benchmark  real-world data. This is not satisfactory;  one would prefer our knowledge and our confidence in clustering outcomes to be based on more than empirical tests. 

As a first step in this direction, in this \new{paper} we use  statistical physics to study model-based Bayesian clustering. This strategy was used in the past to study optimization problems,  see e.g.~\cite{Mezard2009},   and clustering \cite{Rose1990,Blatt1996}, but  not Bayesian clustering. Starting from the  SPD model,  we show that data partition  inference can be formulated in terms of a  quantity that can be seen as the entropy of a gas of  a particles (data-points), distributed over $K$  reservoirs (clusters).   In the regime of a large number of particles  we derive a mean-field  theory to describe this gas, and show that  its  lowest entropy state corresponds to the optimal MAP clustering  of data.   

\section{Model of data and Bayesian clustering\label{section:bayesian-clustering}}

Let us assume that we  observe the  sample   $\xmatrix=\{\x_1, \ldots, \x_N\}$, with $\x_i \in \mathbb{R}^d$ for all $i$, from the distribution 
\begin{eqnarray}
p\left(\xmatrix\vert\thetam,\Pi\right)=\prod_{\mu=1}^{\vert\Pi\vert}\prod_{i_\mu\in S_\mu} p(\x_{i_\mu}\vert\thetav_{\!\mu})\label{eq:prob-of-sample}.
\end{eqnarray}
This distribution is 
generated by the set (or `partition') $\Pi\!=\!\{S_1,\ldots,S_{\vert\Pi\vert}\}$,  with disjunct index sets (or `clusters') $S_\mu\!\neq\!\emptyset$, such that $S_\mu\!\cap\! S_\nu\!=\!\emptyset$ for $\mu\neq\nu$ and $\cup_{\mu=1}^{\vert\Pi\vert} S_\mu\!=\![N]$ with $[N]\!=\!\{1,\ldots,N\}$.  Any partition of data into $K$ clusters can be specified by  binary `cluster allocation'  variables $c_{i\mu}\!=\!\Ind\left[i\in S_\mu\right]$, where $i\!\in\![N]$  and $\mu\!\in\![K]$, forming an $N\!\times\! K$ partitioning matrix $\cmatrix$. This matrix satisfies by construction the following constraints: $\sum_{\mu=1}^Kc_{i\mu}\!=\!1$  for all  $i\!\in\![N]$,  and  $\sum_{i=1}^N c_{i\mu}\!\geq\!1$  for all  $\mu\!\in\![K]$.
Conversely, any $N\times K$ matrix $\cmatrix\!\in\!\{0,1\}^{NK}$ with binary entries that satisfies these constraints induces a partition  $\Pi(\cmatrix)\!=\!\{S_1(\cmatrix),\ldots,S_K(\cmatrix)\}$ of cardinality  $K$. If we also know the prior distributions of model parameters,  $p(\thetav_{\!\mu})$,   $p(\cmatrix\vert K)$ and $p(K)$, we can  use Bayes' theorem (see Appendix \ref{section:model} for details) to derive  the posterior distribution  
\begin{eqnarray}
p(\cmatrix,K\vert\xmatrix)=\frac{\rme^{-N\hat{F}_N\left(\cmatrix,\, \xmatrix\right)} p(\cmatrix\vert K) p(K) }{\sum_{\tilde{K}=1}^N\!p(\tilde{K})\!\sum_{  \tilde{\cmatrix}    }\rme^{-N\hat{F}_N\left(\tilde{\cmatrix},\, \xmatrix\right)} p( \tilde{\cmatrix} \vert \tilde{K})},
\label{eq:P(c|X)}
\end{eqnarray}
in which
\begin{eqnarray}
\hat{F}_N(\cmatrix,\, \xmatrix)&=& -\frac{1}{N}\log \left\langle\rme^{\sum_{\mu=1}^K \sum_{i=1}^N c_{i\mu}\log p\left(\x_{i}\vert\thetav_{\!\mu}\right)}\right\rangle_{\!\thetam} 
\label{def:F-hat}, 
\end{eqnarray}
with $\langle  f(\thetam)\rangle_{\thetam}\!=\!\int\! \big[\prod_{\mu=1}^K p(\thetav_{\!\mu})\,\rmd\thetav_{\!\mu}\big]f(\thetam)$.   
Expression (\ref{eq:P(c|X)}) can be used to infer the most probable partition $\Pi$ for each data sample. First, for each  $K\in[N]$ one computes 
\begin{eqnarray}
\hat{\cmatrix}\,\vert K&=&\argmax_{\cmatrix}\big\{
\rme^{-N\hat{F}_N(\cmatrix,\, \xmatrix)} p(\cmatrix\vert K)
\big\}.
 \label{eq:c|K}
\end{eqnarray}
Then one uses (\ref{eq:c|K}) to determine the estimate $\hat{\Pi}$ of $\Pi$:
\begin{eqnarray}
\hat{\Pi}&=&\argmax_{    \hat{\cmatrix}\, \vert K}\big\{ \rme^{-N\hat{F}_N(\hat{\cmatrix},\, \xmatrix)} p( \hat{\cmatrix} \vert K) p\left(K\right)  
\big\}.
 \label{eq:Pi-hat}
\end{eqnarray}
Clearly, a key role in our formulae is played by the function (\ref{def:F-hat}), 
which can be seen as an entropy of a gas  of $N$ `particles' (the data-points)  distributed over $K$  `reservoirs'  (clusters).  The particles can move from one  reservoir to another;  $c_{i\mu}$ tells us if  particle $i$  is in reservoir  $\mu$, and  the coordinates  $\x_i$ act as  a `quenched' disorder~\cite{Mezard1987}. We are then interested in the minimum entropy state $\argmin_{\cmatrix} \hat{F}_N(\cmatrix,\, \xmatrix)$. 

\section{Mean-field analysis of Bayesian clustering\label{section:mean-field-analysis}}
 
Let us first consider  the case where the cluster parameters are known.  \new{In this case the parameter prior} $p(\thetav_{\!\mu})$ is a delta function, and  (\ref{def:F-hat})  hence becomes
 \begin{eqnarray}
\hat{F}_N(\cmatrix,\xmatrix)&=&\!-\!\sum_{\mu=1}^K \frac{M_\mu(\cmatrix)}{N}\!\!\int\!\rmd \x~ \hat{Q}_\mu(\x\vert\cmatrix,\xmatrix)\log p(\x\vert \thetav_{\!\mu}),~~~~
\label{eq:F-hat-large-N}
\end{eqnarray}
which is now written in terms of \new{the number of particles in cluster $\mu$,} $M_\mu(\cmatrix)=\sum_{i=1}^N\!c_{i\mu}$, and the density   of particles  in cluster $\mu$, defined as
\begin{eqnarray}
\hat{Q}_\mu(\x\vert\cmatrix,\,\xmatrix)&=&\frac{1}{ M_\mu(\cmatrix)}\sum_{i=1}^N\!c_{i\mu} \delta(\x-\x_{i}).
 \label{def:Q-hat}
\end{eqnarray}
Suppose there are $L$ distributions $q_\nu(\x)$, such that for each $\nu$ we find $N_\nu$ particles with $\x_i$ sampled from $q_\nu(\x)$, with  $\sum_{\nu=1}^LN_\nu=N$ and $\lim_{N\rightarrow\infty}N_\nu/N=\gamma(\nu)$.  For large $N$ the density (\ref{def:Q-hat}) will then typically converge to 
\begin{eqnarray}
Q_\mu(\x)&=&\sum_{\nu=1}^L \alpha(\nu\vert\mu)\,  q_\nu(\x)\label{eq:Q-large-N}.
\end{eqnarray}
Here  $\alpha(\nu\vert\mu)\!=\!\alpha(\nu,\mu)/\alpha(\mu)$  is a  conditional probability, defined by  $\alpha(\mu)\!=\!\lim_{N\rightarrow\infty} M_\mu(\cmatrix)/N$ and $\alpha(\nu,\mu) \!=\!\lim_{N\rightarrow\infty}M_{\nu,\mu}(\cmatrix)/N$,  where   $M_\mu(\cmatrix)$ is the number of particles in cluster $\mu$ and  $M_{\nu,\mu}(\cmatrix)\!=\!\sum_{i_{\nu\in S_\mu(\cmatrix)}}  \Ind\left[ \x_{i_{\nu}} \!\sim \!q_\nu(\x)\right]$ is the number of those particles drawn from the distribution $q_\nu(\x)$ that are allocated by $\cmatrix$ to cluster $\mu$.  Clearly $\sum_{\mu\leq K}\alpha(\nu,\mu)\!=\!\gamma(\nu)$, $\sum_{\nu\leq L} \alpha(\nu,\mu)\!=\! \alpha(\mu)>0$ and  $\sum_{\nu\leq L}\sum_{\mu\leq K} \alpha(\nu,\mu)\!=\!1$.  If  (\ref{eq:Q-large-N}) holds for $N\to\infty$, then $\hat{F}_N(\cmatrix,\,\xmatrix)$ will for $N\to\infty$ converge to 
\begin{eqnarray}
F(\alpham)&=& \sum_{\mu=1}^K \sum_{\nu=1}^L \alpha(\nu, \mu) D(q_\nu \vert\vert  p_\mu)+   \sum_{\nu=1}^L \gamma(\nu)H(q_\nu).~~~
\label{eq:F-large-N}
\end{eqnarray}
Here $D(q_\nu \vert\vert  p_\mu)$ is the Kullback-Leibler distance between  $q_\nu(\x)$ and $p(\x\vert\thetav_{\!\mu})$, and $H(q_\nu)$  is a differential entropy~\cite{Cover2012}.  
The transparent and intuitive result (\ref{eq:F-large-N}) can be seen as a mean-field \new{(MF)} theory of $\hat{F}_N(\cmatrix,\, \xmatrix)$ (see Appendix \ref{section:P(F)}  for details). The $L\times K$ matrix $\alpham$, with entries $\alpha(\nu,\mu)$, acts as order parameter. More generally one would have  $P(F)=\int\!\rmd\alpham ~ P(\alpham)\, \delta(F \!- \! F(\alpham))$, where
\begin{widetext}
\begin{eqnarray}
P(\alpham)&=&\lim_{N\rightarrow\infty}\sum_{\cmatrix, \tilde{\cmatrix}}\,p(\cmatrix\vert K) \,q(\tilde{\cmatrix}\vert L)\prod_{\mu=1}^K\prod_{\nu=1}^L\delta\Big[\alpha(\nu, \mu)- \frac{1}{N}\sum_{i=1}^N \tilde{c}_{i\nu} c_{i\mu}  \Big]\!.
\label{def:P(alpha)}
\end{eqnarray}
\end{widetext}
Here $p(\cmatrix\vert K)$  and  $q(\tilde{\cmatrix}\vert L)$ are  the  assumed and the `true'  distributions of partitions, respectively.  We can limit ourselves to working with expression   (\ref{eq:F-large-N}), as opposed to the more involved (\ref{def:P(alpha)}), if $P(\alpham)$ is a delta function. 

We are interested in the state $\alpham$ for which the function $F(\alpham)$ is minimal. Firstly, from $D(q_\nu \vert\vert  p_\mu)\geq0$ it follows that $F(\alpham)\geq \sum_{\nu=1}^L \gamma(\nu)H(q_\nu)$.  The lower  bound is saturated when  $D(q_\nu \vert\vert  p_\mu)=0$, i.e. when $q_\nu(\x)=p(\x\vert \thetav_{\!\mu})$ for all $(\mu,\nu)$,  and the mapping between the sets $[L]$ and $[K]$ labelling  these distributions is bijective.  This can only happen when $L=K$ and $\alpha(\nu, \mu)=\gamma(\nu)\Ind\!\left[D(q_\nu \vert\vert  p_\mu)=0 \right]$, i.e. when the `true' partitioning of the data is recovered.

\begin{figure}[t]
\vspace*{-1mm} \hspace*{-2mm} \setlength{\unitlength}{0.278mm}
\begin{picture}(350,322)
\put(10,220){\includegraphics[height=100\unitlength,width=130\unitlength]{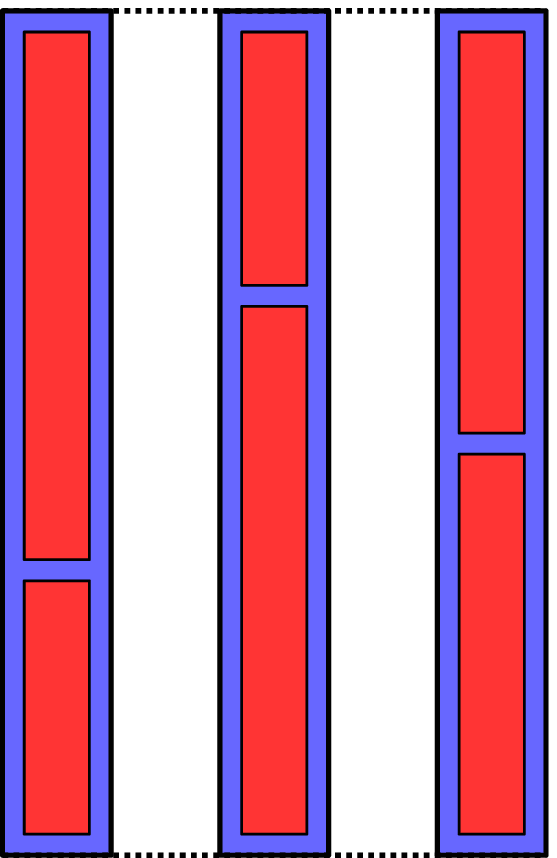}}
 \put(68 ,250){$q_\nu$}      \put(29 ,211){$1$} \put(73 ,211){$\mu$} \put(120 ,211){$K$}              
 
\put(167,220){\includegraphics[height=100\unitlength,width=130\unitlength]{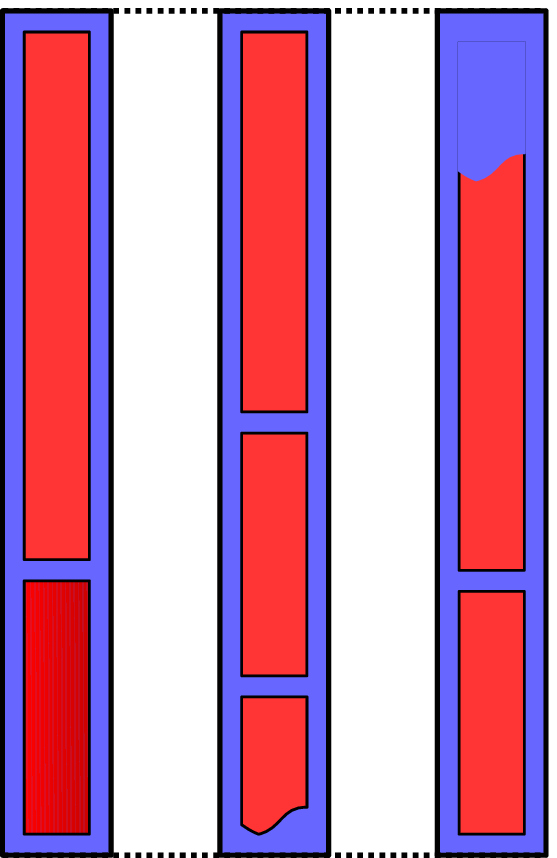}} 
\put(227 ,233){$q_\nu$}   \put(269 ,270){$q_\nu$}   \put(186 ,211){$1$} \put(233 ,211){$\mu$} \put(279 ,211){$K$}

\put(120,70){\includegraphics[height=107\unitlength,width=173\unitlength]{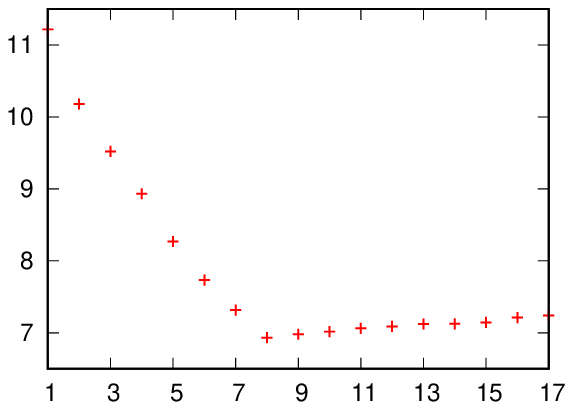}}
\put(0,0){\includegraphics[height=196\unitlength,width=317\unitlength]{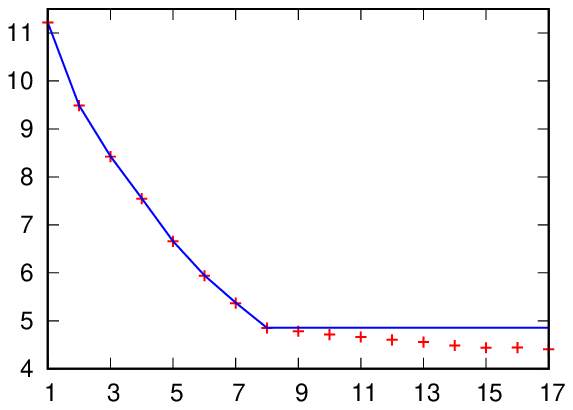}}
\put(8,110){$F$} \put(170,-7){$K$}
\put(111,91){
\rotatebox{90}{
\scriptsize{$\hat{F}_N+\log(K)$}
} 
}
\put(207,63){\scriptsize{$K$}}
\put(87,300){$(a)$} \put(245,300){$(b)$} \put(245,150){$(c)$}
\end{picture}
 \vspace*{-3mm}
\caption{(Color online) Bayesian clustering:   data (red rectangles)  from $L$ different distributions $q_{\nu}(\x)$ are allocated to $K$ clusters  (blue rectangles).  a)  For $K\leq L$, data from $q_{\nu}(\x)$ occupy  {\em at most} one cluster $\mu$.  b) For $K>L$, data from $q_{\nu}(\x)$ occupy {\em at least} one cluster. c) Minimum $F\equiv \min_{\alpham} F(\alpham)$ of the mean-field entropy  (blue line), shown as a function of $K$ and compared with the ground state entropy $\hat{F}_N\equiv\min_{\cmatrix}\hat{F}_N(\cmatrix ,\xmatrix)$ (red crosses),  computed for the data of Figure~\ref{figure:3d}.  The horizontal line corresponds to the lower bound $\sum_{\nu\leq L}\gamma(\nu)H(q_\nu)=4.853905$. Inset: the  sum of $\hat{F}_N$ and $\log(K)$  shown as a function of $K$.  The minimum of this sum  is obtained  when $K=L$.  \label{figure:phase} \vspace*{-5mm}
}
\end{figure}

 Secondly, from $D(q_\nu \vert\vert  p_\mu)\!\geq \! \min_{\tilde\mu}D(q_\nu \vert\vert  p_{\tilde\mu})$  we deduce 
 \begin{eqnarray}
 F(\alpham)\! \geq\!  \sum_{\nu=1}^L \gamma(\nu) \min_{\tilde\mu}D(q_\nu \vert\vert  p_{\tilde\mu})+\sum_{\nu=1}^L \gamma(\nu)H(q_\nu).
 \end{eqnarray}
 This lower bound is saturated when $\alpha(\nu, \mu)\!=\!\gamma(\nu)\Ind\!\left[\mu\!=\!\argmin_{\tilde\mu}D\left(q_\nu \vert\vert  p_{\tilde{\mu}}\right)\right]$ for all $(\mu,\nu)$. For  $K\!\leq\! L$, this state can be seen as the result of the following  `macroscopic' clustering  protocol: for all $\nu\in\{1,\ldots, L\}$, find the distribution  $p(\x\vert \thetav_{\!\mu})$ with the smallest distance $D(q_\nu \vert\vert  p_\mu)$ to $q_\nu(\x)$,  and assign all members of $\nu$ to cluster $\mu$. If $K\!<\!L$ this recipe will occasionally result in the data from more than one distribution being assigned to the same clusters, see Figure  \ref{figure:phase} (a),  but for  $K\!=\!L$, each cluster  would hold only one distribution.  Hence,  the protocol is able to recover the  true partitioning  even when the distributions  $q_\nu(\x)$ and $p(\x\vert\thetav_{\!\mu})$ are non-identical. 
%

The  inequality $D(q_\nu \vert\vert  p_\mu)\!\geq\min_{\tilde\nu}D(q_{\tilde{\nu}} \vert\vert  p_{\mu})$ gives  the lower bound $$F(\alpham)\!\geq\!\sum_{\mu=1}^K \alpha(\mu) \min_{\tilde\nu}D(q_{\tilde{\nu}} \vert\vert  p_{\mu})+   \sum_{\nu=1}^L \gamma(\nu)H(q_\nu),$$ which is saturated when  $\alpha(\nu, \mu)=\alpha(\mu)\Ind[\nu\!=\!\argmin_{\tilde\nu}D(q_{\tilde\nu} \vert\vert  p_{\mu})]$ for all $(\mu,\nu)$. This state would result from to the following protocol: for all $\nu\in\{1,\ldots,L\}$, find the distribution $p(\x\vert \thetav_{\!\mu})$  with the smallest
distance $D(q_\nu \vert\vert  p_\mu)$ to $q_\nu(\x)$,  and assign all members of $\mu$ to cluster $\nu$.   For $K\!>\!L$, this algorithm could allocate more than one distribution to the same cluster, see Figure~\ref{figure:phase}(b).   Furthermore, since $\sum_{\nu=1}^L \alpha(\mu)\Ind[\nu\!=\!\argmin_{\tilde\nu}D(q_{\tilde\nu} \vert\vert  p_{\mu})]=\alpha(\mu)$, the properties of $\alpha(\nu, \mu)$ imply validity of the set of $L$ linear equations  $\sum_{\mu=1}^K \alpha(\mu)\Ind[\nu\!=\!\argmin_{\tilde\nu}D(q_{\tilde\nu} \vert\vert  p_{\mu})]=\gamma(\nu)$, which is underdetermined and hence has either infinitely many solutions, or no solutions at all.  

We now consider  the case where the cluster parameters are unknown, and $p(\thetav_{\!\mu})>0$ for all $\{\thetav_{\!\mu}\}$. For $N\!\rightarrow\!\infty$,  the entropy  (\ref{def:F-hat})  is now strictly dominated via steepest descent by the following set of saddle point equations (see Appendix \ref{section:Laplace} for details):
\begin{eqnarray}
\frac{\partial}{\partial \theta_\mu(\ell)} \frac{1}{N} \sum_{i=1}^N\!c_{i\mu}  \log p\left(\x_{i}\vert\thetav_{\!\mu}\right)=0 \label{eq:dF0}.
\end{eqnarray}
Solving (\ref{eq:dF0})  for  Gaussian distributions $p(\x_{i}\vert\thetav_{\!\mu})\equiv \mathcal{N}\big(\x\vert\m_{\mu},\Lmatrix_{\mu}^{-1}\big)$, with mean 
$\m_{\mu}$ and inverse   covariance matrix $\Lmatrix_{\mu}$,  gives us (see Appendix \ref{section:Laplace}) : 
\begin{eqnarray}
\hat{F}_N(\cmatrix,\,\xmatrix)&=&\sum_{\mu=1}^K\frac{ M_\mu\left(\cmatrix\right)}{2N} \log\Big((2\pi\rme)^{d}\big\vert \Lmatrix_{\mu}^{-1}(\cmatrix,\,\xmatrix)\big\vert  \Big),  ~
 \label{eq:F-hat-Norm}
\end{eqnarray}
where  $\Lmatrix_{\mu}^{-1}(\cmatrix,\,\xmatrix)$ is the empirical covariance  matrix of the data in cluster $\mu$.  

\new{Since  $ \frac{1}{2}  \log \left((2\pi\rme)^{d}\left\vert \Lmatrix_{\mu}^{-1} \left(\cmatrix,\,\xmatrix\right)\right\vert  \right) $  is the differential entropy~\cite{Cover2012} of a Gaussian distribution with covariance matrix  $\Lmatrix_{\mu}^{-1}(\cmatrix,\,\xmatrix)$,} (\ref{eq:F-hat-Norm})  represents an average of $K$ entropies of Gaussian distributions, which for $N\!\to\!\infty$ will converge to the following mean-field entropy (see Appendix  \ref{section:P(F)}):
\begin{eqnarray}
F(\alpham)&=&  \sum_{\mu=1}^K\alpha(\mu) \frac{1}{2}\log \Big((2\pi\rme)^{d}\big\vert \Lmatrix_{\mu}^{-1}(\alpham)\big\vert  \Big),   \label{eq:F-Norm}
\end{eqnarray}
in which $\Lmatrix_{\mu}^{-1}(\alpham)$ denotes the covariance  matrix
\begin{eqnarray}
\Lmatrix_{\mu}^{-1}(\alpham)&=&\sum_{\nu=1}^L\! \alpha(\nu\vert\mu)\big\langle\!(\x\!-\!\m_{\mu}\!(\alpham))(\x\!-\!\m_{\mu}\!(\alpham))^{\!T}\big\rangle_\nu, 
~~~\label{eq:C-2}
\end{eqnarray}
with $\m_{\mu}(\alpham)=\sum_{\nu=1}^L \alpha(\nu\vert\mu)\langle\x\rangle_\nu$, and the short-hand  $\langle\{\cdots\}\rangle_\nu=\int\!\rmd\x~ q_\nu(\x) \{\cdots\}$. Note that  (\ref{eq:F-Norm})  also equals
\begin{eqnarray}
F(\alpham)&=&\!\sum_{\mu, \nu}\!\alpha(\nu, \mu) D(q_{\nu} \vert\vert \mathcal{N}_\mu (\alpham))+\sum_{\nu=1}^L\!\gamma(\nu)H(q_\nu),~~   \label{eq:F-Norm-2}
\end{eqnarray}
where $\mathcal{N}_\mu(\alpham)\equiv  \mathcal{N}\big(\x\vert\m_{\mu}(\alpham),\Lmatrix_{\mu}^{-1}(\alpham)\big)$.  Moreover, as shown in Appendix \ref{section:inform},
\begin{eqnarray}
F(\alpham) \geq  \sum_{\mu=1}^K\alpha(\mu)H(Q_\mu)\geq \sum_{\nu=1}^L\gamma(\nu)H(q_\nu).\label{eq:F-Norm-ineq}
\end{eqnarray}
%

The second  inequality in (\ref{eq:F-Norm-ineq}) has two consequences.  First,  if $K\leq L$ then for any state $\alpham$ that corresponds to either of the scenarios depicted  in Figures~\ref{figure:phase} (a,b),  we will have $F(\alpham) \geq \min_{K} \min_{\tilde{\alpham}} F(\tilde{\alpham})=\sum_{\nu\leq L}\gamma(\nu)H \left(q_\nu \right)$. The lower bound is satisfied when $L=K$  and $q_\nu(\x)$ is Gaussian. The `true'  parameters  $\alpham$ thus represent a locally stable state.  Second,  when $K>L$, the entropy $F(\alpham)$ can only increase with $L$. This follows from  (\ref{eq:F-Norm-2}) and $D(q_{\nu} \vert\vert \mathcal{N}_\mu(\alpham))\geq0$. If $q_\nu(\x)$ is not Gaussian, then $F(\alpham) \geq \sum_{\nu=1}^L\gamma(\nu)\frac{1}{2}  \log \left((2\pi\rme)^{d}\left\vert \Cov_\nu \right\vert  \right)$, where $\Cov_\nu$ is the covariance matrix of $ q_\nu(\x)$ (see Appendix \ref{section:inform}).  Equality corresponds to the state shown in  the Figure~\ref{figure:phase}(a) with  $L=K$, i.e. here the  `true'  data partitioning is recovered. 

The first inequality in (\ref{eq:F-Norm-ineq}) has an appealing geometric  interpretation.   The entropy $H (Q_\mu)$ of each cluster $\mu$ can for large $N$ be estimated by $(d/M_\mu(\cmatrix))\sum_{i=1}^N c_{i\mu}\log\rho_{i\mu}\left(\cmatrix\right) +\log\big(M_\mu(\cmatrix)\!-\!1\big)+\mbox{const.}$, where $\rho_{i\mu}(\cmatrix)=\min_{i\in S_\mu\left(\cmatrix\right)\setminus i}\vert\vert\x_i-\x_j\vert\vert$ (i.e. the Euclidean distance between particle $i$ and its nearest neighbour)~\cite{Kozachenko1987}.  The average entropy $\sum_{\mu=1}^K\alpha(\mu)H (Q_\mu)$ is hence estimated by $(d/N)\sum_{\mu=1}^K\sum_{i=1}^N c_{i\mu}\log\rho_{i\mu}(\cmatrix) +\sum_{\mu=1}^K\big( M_\mu(\cmatrix)/N\big) \log\big( M_\mu(\cmatrix)/N \big)  +\mbox{const.}$  This is minimized by any state  $\cmatrix$ which simultaneously maximises the entropy $-\sum_{\mu=1}^K \big(M_\mu(\cmatrix)/N\big) \log\big( M_\mu(\cmatrix)/N  \big)$, i.e.  `disperses' particles  maximally over  clusters,  and minimizes the nearest neighbour  distances $ \{\rho_{i\mu}(\cmatrix)\}$, i.e. favours  high particle `densities' in each cluster.  

The lower bound  $\sum_{\nu=1}^L \gamma(\nu)H\left(q_\nu\right)$  in  (\ref{eq:F-Norm-ineq})  is  saturated upon choosing any  bijective map  $\alpha:\nu\rightarrow\mu$, since this immediately gives us $F(\alpham)=\sum_{\nu=1}^L \gamma(\nu)H\left(q_\nu\right)$.  Such maps are special instances of the more general family 
\begin{eqnarray}
\alpha(\nu\vert\mu)&=&\frac{\Ind[\nu\in S_\mu]\gamma(\nu)}{\sum_{\tilde{\nu} \in S_\mu} \gamma(\tilde{\nu}) } \label{def:alpha}, 
\end{eqnarray}
where $\Pi=\{S_1,\ldots,S_K\}$  is any partitioning of $[L]$ into $K$ subsets.   Finding  $\min_{\alpham} F(\alpham)$ over all  possible matrices of the form (\ref{def:alpha})  by  enumeration of  all partitions of $[L]$ into $K$ subsets is feasible only for small  $L$, since the number of such partitions  is given by the Stirling number of the second kind $\mathcal{S}(L,K)$ which grows as $K^L$ for large $L$~\cite{Rennie1969}. 

One can also compute  $\min_{\alpham} F(\alpham)$ via the following  `greedy' algorithm.  Start with any partition $\Pi$ and compute $F(\alpham)$.  For all $x\in [L]$: consider all possible moves which do not create empty clusters, and execute the one which gives the largest decrease in $F(\alpham)$, then update $\Pi$.  Continue the last two steps until convergence of $F(\alpham)$ is observed. This macroscopic algorithm can also be implemented `microscopically'. At each  step: for all $i\in [N]$, consider all possible moves of  the particle  $i$ from its current cluster $S_\mu(\cmatrix)$ to a new cluster $S_\nu(\cmatrix)$ and select the one which reduces $\hat{F}_N(\cmatrix,\,\xmatrix)$ most.  To evolve from a non-ordered  state as in Figure~\ref{figure:phase}(b) to an `ordered' state as in  Figure~\ref{figure:phase}(a), this microscopic  algorithm has to move on average at least $N(K\!-\!1)/K$ particles  (see Appendix  \ref{section:algorithmic-cost}).  Each move was selected from among $N(K\!-\!1)$ possible moves,  so the  numerical complexity  is at least  of order $N^2(K\!-\!1)^2/K$.  

\section{Results of numerical experiments\label{section:numerical-experiments}}
\begin{figure}[t]
%
\hspace*{6mm} \setlength{\unitlength}{0.275mm}
\begin{picture}(350,155)
\put(0,0){\includegraphics[height=180\unitlength,width=280\unitlength]{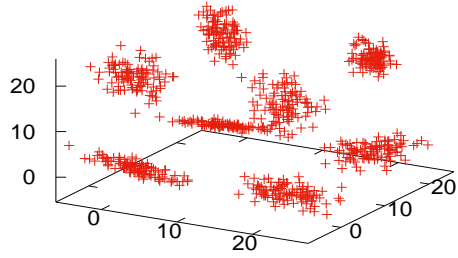}}
\end{picture}
 \vspace*{-10mm}
\caption{(Color online) Data used in our numerical experiments. We generated   $L=8$ clusters with $1000$ data-points each, of which $100$ are shown here. The data  in each cluster $(i,j,k)$ are generated from a distinct Gaussian distribution, with  mean $(\Delta i, \Delta j, \Delta k)$, where $i,j,k\in \{0,1\}$ ($\Delta=20$), and with covariance matrix sampled from the Wishart distribution with $4$ degrees of freedom and   precision matrix $\1$.   \label{figure:3d}\vspace*{-0mm}}
\end{figure}

Our mean-field theory for was derived under the assumption that $\hat{F}_N(\cmatrix,\xmatrix)$  is self-averaging for $N\to\infty$.  To investigate the correctness of   its predictions for finite sample sizes $N$, we studied low entropy states  of  (\ref{eq:F-hat-Norm}) as  obtained by the gradient descent algorithm on the  data of the Figure  \ref{figure:3d}.  For each $K\in[17]$ we ran the algorithm from $100$ different random initial states  $\cmatrix\,(0)$, and  computed  $\hat{F}_N(\cmatrix\,(\infty),\,\xmatrix)$ and the mean field  entropy  $F(\alpham)$   (\ref{eq:F-Norm}) for each.    

\begin{figure}[t]
\hspace*{11mm} \setlength{\unitlength}{0.28mm}
 \begin{picture}(350,155)
 \put(0,0){\includegraphics[height=150\unitlength]{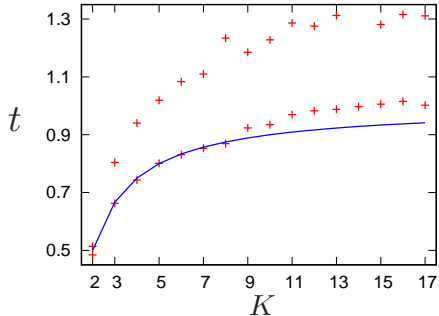}} 
 \put(-2,80){\large $t$}
  \put(111 ,-7){$K$}
\end{picture}
 \caption{\new{(Color online) Total (normalised) number of  `moves'  $t$ used by the gradient descent algorithm  to travel from a random unbiased partition to a final partition, i.e. the effective algorithmic runtime,  shown as a function of the assumed number of clusters $K$.  The minimum and maximum time (red crosses) obtained in $100$ runs  on the data of Figure \ref{figure:3d} are compared with the  average  lower bound $(K\!-\!1)/K$ (blue line).} }
 \label{figure:time} 
 \end{figure}

 For $K\!\leq \!L$, most final states $\cmatrix\,(\infty)$  allocate data from the same distribution correctly to the same cluster, see Figure~\ref{figure:phase} (a).  The values of  $\hat{F}_N(\cmatrix\,(\infty),\,\xmatrix)$ are those predicted by $F(\alpham)$, and indeed correspond to local minima and saddle points of $F(\alpham)$ (see Appendix \ref{section:numerics}).  Also, according to  Figure~\ref{figure:phase} (c), the value $\hat{F}_N\!=\!\min_{\cmatrix}\hat{F}_N(\cmatrix,\xmatrix)$ as estimated from $\cmatrix\,(\infty)$ is predicted accurately by $F\!=\!\min_{\alpham} F(\alpham)$.  Residual differences between  $\hat{F}_N$ and $F$ reflect finite size effects. These can be computed exactly when $K=L$, and when $\cmatrix\,(\infty)$ represents the true  partitioning of the data: the average and variance of $\hat{F}_N$ are in that case given  by $\sum_{\nu=1}^L\gamma(\nu)H(q_\nu)+Kd(d\!+\!1)/4N$ and $d/2N$, respectively (see Appendix \ref{section:log-det}).  Finally, we note that the number of particles  `moved'  by the algorithm in going from   $\cmatrix\,(0)$ to   $\cmatrix(\infty)$  is consistent  with the lower bound  $N(K\!-\!1)/K$, so the algorithmic complexity is quadratic in $N$,  see Figure~\ref{figure:time}.  

 If $K\!>\!L$,  the states  $\cmatrix\,(\infty)$  will allocate data from the same distribution to multiple clusters, see Figure~\ref{figure:phase} (b). Such states  are already present for small $K\!\leq\! L$, and proliferate as  $K$ is increased (see Appendix \ref{section:numerics}).   The  lower bound  $\sum_{\nu=1}^L\gamma(\nu)H(q_\nu)$ is now violated, and the gap between this bound and the value of $\hat{F}_N$ as obtained by gradient descent increases with $K$, see Figure~\ref{figure:phase} (c).  While some of the $\hat{F}_N(\cmatrix\,(\infty),\,\xmatrix)$ values are consistent with  $F(\alpham)$ (see Appendix \ref{section:numerics}), the mean-field theory fails to predict $\min_{\cmatrix}\hat{F}_N(\cmatrix ,\xmatrix)$ in this regime, due to the non-commutation of the $N\rightarrow\infty$  limit and  the $\min$ operator. 
 
\new{O}ur estimate of $\hat{F}_N=\min_{\cmatrix}\hat{F}_N(\cmatrix ,\xmatrix)$  can also be used  to infer the true number of clusters $L$.  Assuming uniform prior distributions of partitions $p(\cmatrix\vert K)=(K!  \mathcal{S}(L,K))^{-1}$ and cluster sizes $p(K)=N^{-1}\Ind[K\!\in\![N]]$ in the Bayesian formulae (\ref{eq:P(c|X)})-(\ref{eq:Pi-hat}), the total  entropy $\hat{F}_N +\frac{1}{N}\log(K! \mathcal{S}(L,K))\approx \hat{F}_N+\log(K)$ has its minimum at the correct value $K=L$, see inset in Figure~\ref{figure:phase} (c).

An  interesting and important question,  from a practical and a theoretical point view,  is how Bayesian clustering is affected by the `separation' between different clusters.  The simplest non-trivial case is to consider the clustering of $d$-dimensional data sampled from two isotropic Gaussian distributions  $\mathcal{N}(\m_1,\1)$ and  $\mathcal{N}(\m_2,\1)$.   
Here one can use the Euclidean distance $\vert\vert\m_1-\m_2\vert\vert=\Delta$, measured relative to the natural scale $\sqrt{d}$, as a measure of the degree of separation~\cite{Dasgupta1999} between the `clusters' centred at  $\m_1$ and $\m_2$. 
 For large $d$, most of the vectors $\x$  sampled from $\mathcal{N}\left(\m,\1\right)$ will be found  in the `sphere' of  radius $\sqrt{d}$  centred at  $\m$,  reflecting `concentration'  phenomena observed for large $d$. 
In particular if we  assume that $\x$ is sampled from $\mathcal{N}\left(\m,\Lmatrix\right)$, then  $\left\langle\vert\vert\x-\m\vert\vert^2\right\rangle=\Tr \,\Lmatrix$, and for $\lambda,\epsilon>0$:  
\begin{eqnarray}
&&{\rm Prob}\left(\vert\vert\x-\m\vert\vert^2\geq\Tr \,\Lmatrix +d\epsilon\right)\nonumber\\
&&~~~~~~~~~~={\rm Prob}\left(\rme^{\frac{\lambda}{2} \vert\vert\x-\m\vert\vert^2}\geq\rme^{\frac{\lambda}{2}(\Tr \,\Lmatrix+d\epsilon) }\right)\nonumber\\
&&~~~~~~~~~~~~~\leq    \left\langle\rme^{\frac{\lambda}{2} \vert\vert\x-\m\vert\vert^2}\right\rangle  \rme^{-\frac{\lambda}{2}(\Tr \,\Lmatrix+d\epsilon) } \nonumber\\
&&~~~~~~~~~~~~~~~~~~~~=  \rme^{-\frac{1}{2}\left(\log\vert\1-\lambda\Lmatrix\vert +\lambda(\Tr \,\Lmatrix+d\epsilon) \right)} \label{eq:concentr-ineq}.
\end{eqnarray}
 The upper bound in the above expression was obtained using Markov's inequality and properties of Gaussian integrals. For the choice $\Lmatrix =\1$, the above inequality, after  optimising  the upper bound with respect to $\lambda$,  gives us ${\rm Prob}\left(\vert\vert\x\!-\!\m\vert\vert^2\!\geq\!d(1+\epsilon)\right) \leq \rme^{-\frac{d}{2}\left(\log\frac{1}{1+\epsilon}-\epsilon\right)}$. 
 
\new{Let us now consider the MF  entropy  $\min_{\alpham} F(\alpham)$  for the distributions  $\mathcal{N}(\m_1,\1)$ and  $\mathcal{N}(\m_2,\1)$, with separation $\vert\vert\m_1-\m_2\vert\vert=\Delta$. For the assumed number of clusters $K=1$ this  entropy is given by 
\begin{eqnarray}
 F_1&=&\frac{d}{2}  \log \left(2\pi\rme\right)+\!\frac{1}{2}\!\log\Big| \1 \!+\!  \! \sum_{\nu=1}^2\!\gamma(\nu)\!\left( \m_\nu\!-\!\m \right)\!\left( \m_\nu\!-\!\m \right)^T\!  \Big|   \label{eq:F-MF-Norm-I-K1}
\end{eqnarray}
 where $\m=\sum_{\nu=1}^2\gamma(\nu)\, \m_\nu$, and $\gamma(\nu)$ is the fraction of data sampled from $\mathcal{N}(\m_\nu,\1)$. For $K=2$ we obtain 
\begin{eqnarray}
 F_2&=&   \frac{d}{2} \log \left(2\pi\rme\right),\label{eq:F-MF-Norm-I-K2}
\end{eqnarray}
which corresponds to the situation where the true clustering of data is recovered.  Furthermore, upon choosing $\m_1\!=\!\nullv$ and $\gamma(\nu)\!=\!\frac{1}{2}$ we obtain $F_1=\frac{d}{2} \log (2\pi\rme)+\frac{1}{2} \log[  1+(\frac{\Delta}{2})^2 ]$.  Thus in this case $F_1\geq F_2$, as required. However, if $\log(2)\geq \frac{1}{2} \log[ 1+(\frac{\Delta}{2})^2 ]$ then $F_2+\log(K)\geq F_1$ (note that we minimise $\min_{\alpham} F(\alpham) +\log(K)$ to infer true number of clusters), so that here we are unable to recover the correct number $K=2$  of clusters due to the cluster separation $\Delta$ being too small. This happens when  $\Delta\leq 2\sqrt{3}\approx 3.46$.  We expect that a similar analysis can be also performed for more general scenarios. } 

\begin{figure*}
 \setlength{\unitlength}{0.89mm}
 \begin{picture}(200,70)
\put(8,35){\includegraphics[width=190\unitlength,height=40\unitlength]{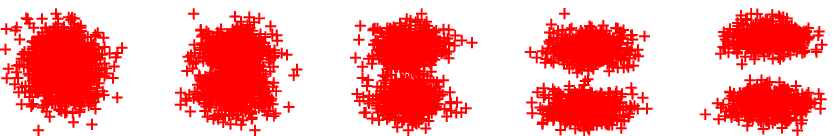}} 
 
 \put(0,0){\includegraphics[width=43.5\unitlength,height=35\unitlength]{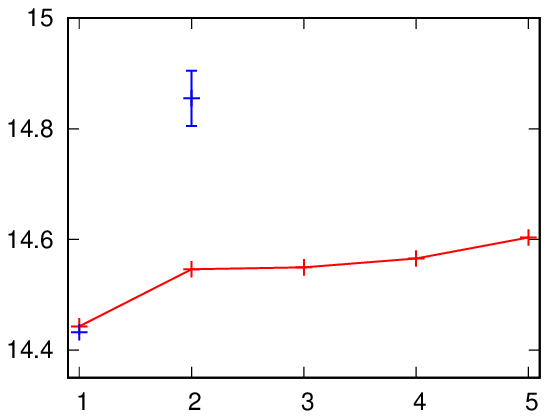}} 
 \put(40,0){\includegraphics[width=43.5\unitlength,height=35\unitlength]{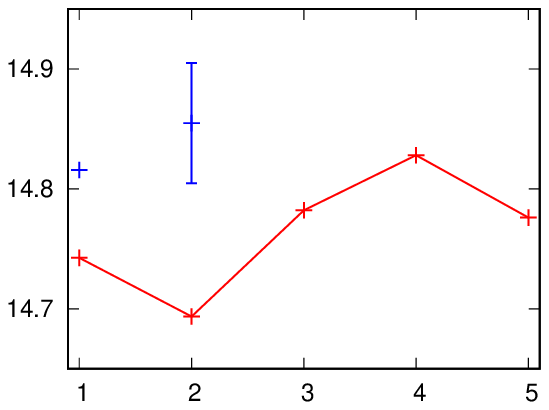}} 
 \put(80,0){\includegraphics[width=43.5\unitlength,height=35\unitlength]{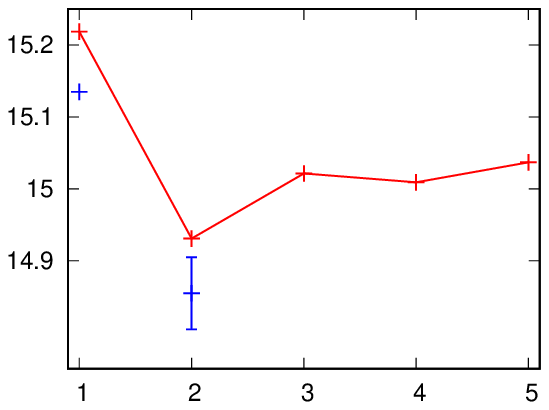}}
\put(120,0){\includegraphics[width=43.5\unitlength,height=35\unitlength]{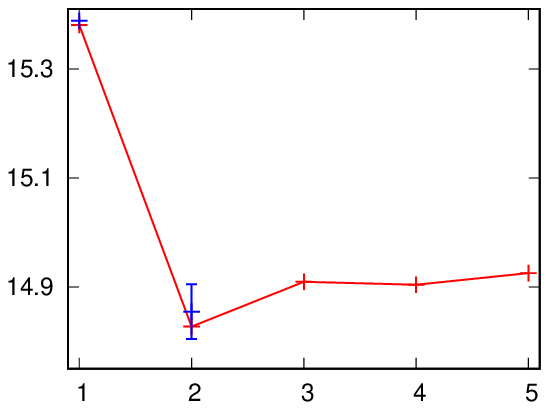}}
\put(160,0){\includegraphics[width=43.5\unitlength,height=35\unitlength]{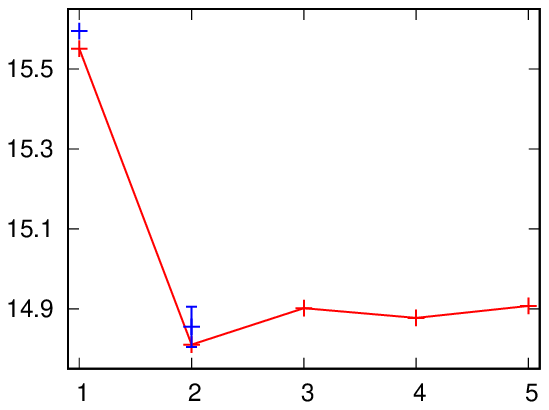}}
    
  \put(-2 ,9){\rotatebox{90}{
\small{$\hat{F}_N+\log K$}
} }  
      
      \put(22 ,-3){$K$} 
           \put(62 ,-3){$K$} 
                \put(102 ,-3){$K$} 
                     \put(142 ,-3){$K$} 
                          \put(182 ,-3){$K$} 
      
\end{picture}
\vspace*{2mm}
 \caption{\new{(Color online) Bayesian clustering of data  $\x\in \mathbb{R}^d$ generated  from the Gaussian distributions $\mathcal{N}(\m_1,\1)$ and  $\mathcal{N}(\m_2,\1)$, with separation  $\Delta=\vert\vert\m_1-\m_2\vert\vert$.  The data sample,  split equally between the constituent distributions,   is of size $N=2000$ and has $d=10$.  The data was generated for cluster separations $\Delta/\sqrt{d}\in\left\{  \frac{1}{2},1,\frac{3}{2},2,\frac{5}{2}   \right\}$. Top:  Data projected into two dimensions. The separation $\Delta$ of the clusters is increasing from the left to the right.  Bottom: the  sum $\hat{F}_N+\log K$ (red crosses connected by lines), where $\hat{F}_N\equiv\min_{\cmatrix}\hat{F}_N(\cmatrix ,\xmatrix)$, shown as a function of the assumed number of clusters $K$,  and compared  with the mean-field prediction $\min_\alpha F(\alpham)$ (blue crosses). For $K=2$, the mean-field prediction $\min_\alpha F(\alpham)=\frac{d}{2}\log(2\pi\rme)$ is plotted with the finite size corrections (error bars indicate one standard deviation). }}
 \label{figure:10dL2} 
 \end{figure*}

\new{Numerical experiments are in qualitative agreement with the predicted separation boundary $\Delta=2\sqrt{3}$, as can be seen in Figure \ref{figure:10dL2}. }
\new{In this Figure we also compare the  mean-field theory results  (\ref{eq:F-MF-Norm-I-K1}, \ref{eq:F-MF-Norm-I-K2}) with the results of numerical simulations. For  $K=1$ the discrepancy between theory and simulations is a finite size effect. In contrast,  for $K=2$ it is a combination of finite size effects  and the inability of the mean-field theory to account  for correlations between the data in clusters for small separations $\Delta$.  Such correlations are also responsible for a breakdown of the mean-field theory when $K>L$, see Figure \ref{figure:phase}. For larger separations $\Delta$ the theory is in good agreement with the simulations, see Figure \ref{figure:10dL2}, and discrepancies again reflect only finite size effects.  }

 \new{ 
The magnitude of the finite size effects can be estimated when  $K=L$ for any $d/N<1$, by the following argument.
For the empirical covariance matrix  $\hat{\Lmatrix}$ of a sample of $M$ $d$-dimensional data vectors generated from the Gaussian distribution  $\mathcal{N}(\m,\Lmatrix)$ the random quantity $\log\vert\hat{\Lmatrix}\vert$  will for large $M$ be described by the distribution $$\mathcal{N}(\log |\Lmatrix|+\tau(M,d), \sigma^2(M,d) ),$$ where  $\tau(M,d)=\sum_{\ell=1}^d \psi( \frac{M-\ell+1}{2})-d\log(\frac{M}{2})$ and~$ \sigma^2(M,d)=\sum_{\ell=1}^d\frac{2}{M-\ell+1}$~\cite{Cai2015}.  Assuming that  $K=L$ and that the clustering is perfect allows us to compute, by following steps similar to those followed in the Appendix \ref{section:log-det}, the average and variance    of  the entropy  (\ref{eq:F-hat-Norm}).  They are found to be given by $\min_\alpha F(\alpham)+\sum_{\nu=1}^L \gamma(\nu)\, \tau\left(\gamma(\nu)N,d\right)$ and $\frac{1}{4}\sum_{\nu=1}^L \gamma^2(\nu)\, \sigma^2\left(\gamma(\nu)N,d\right)$, respectively. }
 
 \new{When evaluated for real datasets,  the entropy function (\ref{eq:F-hat-Norm}) may also have value as an exploratory tool.  To show this, we consider the Wisconsin Diagnostic Breast Cancer (WDBC) dataset~\cite{Dua2017}, which describes  characteristics of cell nuclei in the images of cells  extracted from tumours~\cite{Street1993}, and contains  $N=569$ data-points of dimension $d=30$. This dataset has two  (linearly separable) classes, which we assume to be   the `true' clusters,  one is  `benign', represented by  $357$ data-points, and the other is `malignant', represented by  $212$ data-points~\cite{Street1993}. }
  \new{A first simple unsupervised method which one might apply to this dataset is hierarchical clustering, which uses pairwise distances between the data-points to build a  hierarchy of clusters, see e.g.~\cite{Friedman2001}.  The agglomerative version of this algorithm, with Euclidean distances,  separates this data  into clusters of sizes $549$  and  $20$ at the $K=2$ clusters level of  hierarchy, into clusters of sizes  $549$,  $19$ and   $1$ at the $K=3$  clusters level of  hierarchy, into clusters of sizes $438$,  $111$,  $19$ and   $1$ at the $K=4$  clusters level of  hierarchy, etc.  Hence, upon  assuming (correctly) that $K=2$,  one cannot recover the true clusters  of the WDBC data with this algorithm.}   
\new{Alternatively, the $K$-Means clustering algorithm, see e.g.~\cite{Bishop2006}, which minimises the squared Euclidean distance between the points in a cluster,  `finds' in the WDBC dataset  (again upon assuming  $K=2$) clusters of sizes $438$ and  $131$.  Upon comparing these  with the true clusters, we observe  that $K$-Means `misclassifies'  $83$ data-points in total.   It is interesting that the clusters found by  $K$-Means were  also present in the four  clusters generated via  hierarchical clustering.} 

\begin{figure}[t]
%
 \setlength{\unitlength}{0.262mm}
\hspace*{5mm}\begin{picture}(350,170)
\put(0,0){\includegraphics[height=180\unitlength,width=280\unitlength]{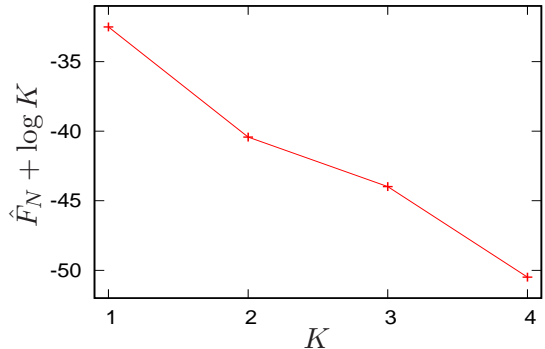}}
  \put(-7 ,51){\rotatebox{90}{
\small{$\hat{F}_N+\log K$}
} }  
      \put(141 ,-5){$K$} 
\end{picture}
%
\caption{\new{(Color online)  The  sum $\hat{F}_N+\log K$, where $\hat{F}_N\equiv\min_{\cmatrix}\hat{F}_N(\cmatrix ,\xmatrix)$, as computed for the Wisconsin Diagnostic Breast Cancer data \cite{Dua2017} (red crosses connected by lines), shown as a function of the assumed number of clusters $K$. These results suggest that the true number of clusters in this dataset is at least $K=4$. }\label{figure:Wisconsin}}
\end{figure}

\new{Using instead the gradient descent minimisation of (\ref{eq:F-hat-Norm}) as a clustering protocol suggests that there are more than $K=4$ clusters\footnote{For $K>4$,  this approach favours small clusters, i.e. we are in non-asymptotic  regime,  which  suggests that  a full  Bayesian framework is more appropriate for this data.}  in the WDBC dataset (see Figure \ref{figure:Wisconsin}).  For $K=2$ the algorithm outputs clusters of sizes $328$ and $241$, which is, compared with the hierarchical and $K$-Means results, much closer to the true sizes $357$ and  $212$ of the  WDBC dataset.  Now $57$ data-points were misclassified, which can be explained by the non-sphericity of clusters in this dataset. }
\new{In particular, for  any data covariance matrix $\hat{\Sigma}$   the ratio  $\mathcal{S}(\hat{\Sigma})=\Tr^2(\hat{\Sigma})/d\Tr(\hat{\Sigma}^2)$ can be used as a measure of `sphericity' of data, see e.g.~\cite{Jung2009}.  We note that $1/d\leq     \mathcal{S}(\hat{\Sigma})         \leq1$,   and that the lower bound $1/d$ is saturated  only  when a few eigenvalues dominate all others  for large $d$, i.e. when only a few `directions' in $\mathbb{R}^d$  contribute to the variability  in the data.  The upper bound is saturated when all  eigenvalues are equal, i.e. all directions in $\mathbb{R}^d$ contribute equally to the variability. The sphericity values of the `benign' and `malignant' clusters in the WDBC dataset are given by $0.034$ and $0.036$,  respectively, so the data in these clusters is highly non-spherical.  This indeed suggests that the entropy function (\ref{eq:F-hat-Norm}), derived upon assuming arbitrary multivariate Gaussian distributions of a data in the clusters, is better equipped to deal with this scenario than hierarchical or $K$-Means clustering.}

\section{Summary\label{section:summary}}

In conclusion, in this paper we have demonstrated that mapping Bayesian clustering of data to a statistical mechanical problem is not only possible, but in fact also quite intuitive and fruitful. It enables us to identify objectively the most plausible number of clusters in a dataset, and to obtain transparent interpretations and explanations of why and how conventional clustering methods (which are quite often based on ad-hoc definitions) may or may not fail to detect clusters correctly, dependent on the quantitative features of the data. 

One possible extension of this work,  currently in progress, is a more general analytical treatment of this Bayesian clustering problem, in which the distribution  $P(F)=\int\!\rmd\alpham ~ P(\alpham)\, \delta(F \!- \! F(\alpham))$ is no longer assumed to converge to a delta distribution for large $N$. This will allow us allow us to tackle also the nontrivial regime where  $N,d\rightarrow\infty$ with $N/d$ finite, and to  correct  the present mean-field theory in the $K>L$ regime.

\begin{acknowledgments} 
\noindent
This work was supported by the Medical Research Council of the  United Kingdom (grant MR/L01257X/1). 
\end{acknowledgments}

\appendix

 \section{Model of data and Bayesian clustering\label{section:model}}
Let us assume that we  observe the  sample   $\xmatrix=\{\x_1, \ldots, \x_N\}$, where $\x_i \in \mathbb{R}^d$ for all $i$, drawn from the distribution 
\begin{eqnarray}
p\left(\xmatrix\vert\thetam,\Pi\right)=\prod_{\mu=1}^{\vert\Pi\vert}\prod_{i_\mu\in S_\mu} p(\x_{i_\mu}\vert\thetav_{\!\mu})\label{eq:prob-of-sample-app},
\end{eqnarray}
generated by the partition $\Pi=\left\{S_1, S_2,\ldots,S_{\vert\Pi\vert}\right\}$,  where the index sets $S_\mu\neq\emptyset$ obey $S_\mu\cap S_\nu=\emptyset$ for $\mu\neq\nu$, and $\cup_{\mu=1}^{\vert\Pi\vert} S_\mu=[N]$, with the short-hand $[N]=\{1,\ldots,N\}$.  Furthermore, we assume that each  parameter $\thetav_{\!\mu}$ is sampled randomly and independently from the distribution $p(\thetav_{\!\mu})$, and that  we are also given the prior distribution of $\Pi$, $P(\Pi)$.  This allows us to write down  the joint distribution 
\begin{eqnarray}
p\left(\xmatrix,\thetam,\Pi\right)&=&p\left(\xmatrix\vert\thetam,\Pi\right)p\left(\Pi\right)\prod_{\mu=1}^{\vert\Pi\vert} p(\thetav_{\!\mu}),
\label{eq:joint-prob-using-S}
\end{eqnarray}
where $\thetam=\{\thetav_1,\ldots,\thetav_{\vert\Pi\vert}\}$.
Upon integrating out the parameters $\thetav_{\!\mu}$  in the above we obtain the distribution 
\begin{eqnarray}
p\left(\xmatrix, \Pi\right)&=&\left\langle p\left(\xmatrix\vert\thetam,\Pi\right)     \right\rangle_{\thetam\vert\Pi} p\left(\Pi\right),\label{eq:P(X,S)}
\end{eqnarray}
where $\left\langle  f\left(\thetam\right) \right\rangle_{\thetam\vert\Pi}=\int  f\left(\thetam\right) \!\big\{\prod_{\mu=1}^{\vert\Pi\vert} p(\thetav_{\!\mu})\,\rmd\thetav_{\!\mu}\big\}$.  From this follows the conditional distribution
\begin{eqnarray}
p\left( \Pi\vert\xmatrix\right)&=&\frac{ p\left(\xmatrix\vert\Pi\right) p\left(\Pi\right)}{\sum_{\tilde\Pi} p(\xmatrix\vert\tilde\Pi) p(\tilde\Pi)}\label{eq:P(S|X)}
\end{eqnarray}
with 
\begin{eqnarray}
p\left(\xmatrix\vert \Pi\right)&=&\left\langle p\left(\xmatrix\vert\thetam,\Pi\right)     \right\rangle_{\thetam\vert\Pi} \label{eq:P(X|S)}.
\end{eqnarray}

Let us next consider the  `partition function'
\begin{eqnarray}
&&\sum_{\Pi} p\left(\xmatrix\vert\Pi\right) p\left(\Pi\right)\nonumber\\[-1mm]
&&~~~~~=\sum_{K=1}^N\sum_{\Pi} p\left(\xmatrix\vert\Pi\right) p\left(\Pi\right)\Ind\left[\vert\Pi\vert=K\right]\\
&&~~~~~=\sum_{K=1}^N\sum_{\Pi} p\left(\xmatrix\vert\Pi\right) p\left(\Pi\vert K\right)p(K)\nonumber,
\end{eqnarray}

where we have defined the two distributions 
\begin{eqnarray}
p\left(\Pi\vert K\right)&=&\frac{p\left(\Pi\right)\Ind\left[\vert\Pi\vert=K\right]}
{\sum_{\tilde\Pi}p(\tilde\Pi)\Ind\big[\vert\tilde\Pi\vert=K\big]}\label{eq:P(S|k)}
\\[1mm]
p(K)&=&\sum_{\Pi}p\left(\Pi\right)\Ind\left[\vert\Pi\vert=K\right]\nonumber.
\end{eqnarray}
Furthermore, if we define $\Pi_K$ to be a partition $\Pi$ with $\vert\Pi\vert=K$, i.e.  $\Pi_K=\{S_1,\ldots,S_K\}$, then 
\begin{eqnarray}
&&\sum_{\Pi} p\left(\xmatrix\vert\Pi\right) p\left(\Pi\right)\nonumber\\[-1mm]
&&~~~~~=\sum_{K=1}^N p(K)\sum_{\Pi_K} p\left(\xmatrix\vert\Pi_K\right) p\left(\Pi_K\vert K\right)\label{eq:P(X)}
\end{eqnarray}
%

and the distribution of $\Pi_K$ is given by 
\begin{eqnarray}
p\left(\Pi_K\vert (\xmatrix\right)&=&\frac{p\left(\xmatrix\vert\Pi_K\right) p\left(\Pi_K\vert K\right) p(K)}{\sum_{\tilde{K}=1}^N p(\tilde{K})\sum_{\tilde{\Pi}_{\tilde{K}}} p(\xmatrix\vert\tilde{\Pi}_{\tilde{K}}) p(\tilde{\Pi}_{\tilde{K}}\vert\tilde{K})}.\label{eq:P(Sk|X)}
\end{eqnarray}
The mode of this distribution is located at 
\begin{eqnarray}
\hat{\Pi}_K&=&\argmax_{\Pi_K} \Big\{p\left(\xmatrix\vert\Pi_K\right) p\left(\Pi_K\vert K\right)\Big\}. \label{eq:Sk)}
\end{eqnarray}
from which, in turn, it follows that the mode of the distribution  (\ref{eq:P(S|X)}) is located at 
\begin{eqnarray}
\hat{\Pi}&=&\argmax_{\hat{\Pi}_K} \Big\{p(\xmatrix\vert\hat{\Pi}_K) p(\hat{\Pi}_K\vert K) p(K)
\Big\}.\label{eq:S)}%
\end{eqnarray}
To see this one considers 
%
%
\begin{eqnarray}
\hat{\Pi}&=& \argmax_{\Pi} \left\{p\left(\xmatrix\vert\Pi\right) p\left(\Pi\right)\right\}\nonumber\\
%
%
&=&\argmax_{\Pi}\Big\{ \left\{p\left(\xmatrix\vert\Pi_1\right) p\left(\Pi_1\right)\right\},\ldots\nonumber\\
&&~~~~~~~~~\ldots,\left\{p\left(\xmatrix\vert\Pi_K\right) p\left(\Pi_K\right)\right\},\ldots\nonumber\\
&&~~~~~~~~~~~~~~~~\ldots,\left\{p\left(\xmatrix\vert\Pi_N\right) p\left(\Pi_N\right)\right\}\Big\}\nonumber,
\end{eqnarray}
%
%
where $\left\{p\left(\xmatrix\vert\Pi_K\right) p\left(\Pi_K\right)\right\}$ is a set generated by  $\{\Pi_K\}$.  
Clearly, $\max_{\Pi_K}\left\{p\left(\xmatrix\vert\Pi_K\right) p\left(\Pi_K\right)\right\}=p(\xmatrix\vert\hat{\Pi}_K) p(\hat{\Pi}_K)$, in which $\hat{\Pi}_K=\argmax_{\Pi_K}\left\{p(\xmatrix\vert\Pi_K) p\left(\Pi_K\right)\right\}$,  from which follows that 
\begin{eqnarray}
\hat\Pi&=&\argmax_{\hat{\Pi}_K}\left\{p(\xmatrix\vert\hat{\Pi}_K) p(\hat{\Pi}_K)\right\}
\nonumber
\\
&=&\argmax_{\hat{\Pi}_K} \Big\{p(\xmatrix\vert\hat{\Pi}_K) p(\hat{\Pi}_K\vert K) p(K)\Big\}.
\end{eqnarray}

Any partition $\Pi_K$ of the data into $K$ clusters can be specified by the  binary  `allocation'  variables $c_{i\mu}=\Ind\left[i\in S_\mu\right]$, where $i\in[N]$  and $\mu\in[K]$, forming the matrix $\cmatrix$   with $\left[\cmatrix\right]_{i\mu}=c_{i\mu}$.   Hence $\Pi_K\equiv\Pi_K\left(\cmatrix\right)=\{S_1\left(\cmatrix\right),\ldots,S_K\left(\cmatrix\right)\}$.  Conversely, an $N\!\times\! K$ matrix $\cmatrix$ with binary entries is a partition only if it satisfies the constraints $\sum_{\mu=1}^Kc_{i\mu}=1\mbox{ for all } i\in[N]$ and $\sum_{i=1}^N c_{i\mu}\geq1 \mbox{ for all }  \mu\in[K]$. 
The simplest distribution  implementing these constraints  is  the uniform distribution  
\begin{widetext}
\begin{eqnarray}
p\left(\cmatrix\vert K\right)&=&\frac{\left\{\prod_{i=1}^N \Ind\left[\sum_{\nu=1}^Kc_{i\nu}=1\right]\right\}\!\!\left\{\prod_{\mu=1}^K\Ind\left[\sum_{j=1}^N c_{j\mu}\geq1\right]\right\}}{\sum_{ \tilde{\cmatrix}}\left\{\prod_{i=1}^N \Ind\left[\sum_{\nu=1}^K\tilde{c}_{i\nu}=1\right]\right\}\!\!\left\{\prod_{\mu=1}^K\Ind\left[\sum_{j=1}^N \tilde{c}_{j\mu}\geq1\right]\right\}}\label{eq:P(c|k)-uniform}.
\end{eqnarray}
\end{widetext}
The denominator in this expression gives the total number of partitions of the set $[N]$ into $K$  subsets  $\mathcal{S}(N,K)$, i.e. it equals the Stirling number of the second kind times the number $K!$ of subset permutations. Thus the probability of each individual partition $\cmatrix$  is  given by $1/K!\,\mathcal{S}(N,K)$.  We note that for $N\rightarrow\infty$ and $K\in O(N^0)$ we have $N^{-1}\log(K!\mathcal{S}(N,K))\to \log(K)$~\cite{Rennie1969}.

Using this new notation allows us to write the distribution $p(\xmatrix\vert\Pi_K)$  as
\begin{eqnarray}
p(\xmatrix\vert\Pi_K)&\equiv&p\left(\xmatrix\vert \cmatrix,K\right)\label{eq:P(X|c)}\\[1mm]
%
%
&=&\left\langle\rme^{\sum_{\mu=1}^K \sum_{i=1}^N c_{i\mu}\log p\left(\x_{i}\vert\thetav_{\!\mu}\right)}\right\rangle_{\thetam}\nonumber\\
&=&\rme^{-N\hat{F}_N\left(\cmatrix,\, \xmatrix\right)}\nonumber,
\end{eqnarray}
where  $\left\langle  f\left(\thetam\right) \right\rangle_{\thetam}=\int  f\left(\thetam\right) \!\big\{\prod_{\mu=1}^K p(\thetav_{\!\mu})\,\rmd\thetav_{\!\mu}\big\}$,  and we defined the log-likelihood
\begin{eqnarray}
\hat{F}_N(\cmatrix,\, \xmatrix)&=& -\frac{1}{N}\log \left\langle\rme^{\sum_{\mu=1}^K \sum_{i=1}^N c_{i\mu}\log p\left(\x_{i}\vert\thetav_{\!\mu}\right)}\right\rangle_{\!\thetam}
\nonumber
\\&&\label{def:F-hat-app}.
\end{eqnarray}
 Furthermore, combining  $p\left(\cmatrix,K\right)=p\left(\cmatrix\vert K\right)p(K)$ with  (\ref{eq:P(X|c)}) gives us  the joint distribution 
\begin{eqnarray}
p\left(\xmatrix, \cmatrix, K\right)&=&\rme^{-N\hat{F}_N\left(\cmatrix,\, \xmatrix\right)} p\left(\cmatrix,K\right)\label{eq:P(X,c)}
\end{eqnarray}
from which we can derive the conditional distribution 
\begin{eqnarray}
&&p\left(\cmatrix,K\vert\xmatrix\right)\nonumber\\
&&~~~~=\frac{\rme^{-N\hat{F}_N(\cmatrix,\, \xmatrix)} p(\cmatrix\vert K) p(K) }{\sum_{\tilde{K}=1}^N\!p(\tilde{K})\!\sum_{  \tilde{\cmatrix}    }\rme^{-N\hat{F}_N\left(\tilde{\cmatrix},\, \xmatrix\right)} p( \tilde{\cmatrix} \vert \tilde{K})}\label{eq:P(c|X)-app}.
\end{eqnarray}
For $K\in[N]$ the mode of this distribution is located at 
\begin{eqnarray}
\hat{\cmatrix}\,\vert K&=&\argmax_{\cmatrix}\, p\left(\cmatrix,K\vert \xmatrix\right)\nonumber\\
&=&\argmax_{\cmatrix}\big\{\rme^{-N\hat{F}_N\left(\cmatrix,\, \xmatrix\right)} p\left(\cmatrix\vert K\right)\big\} \label{eq:c|k}
\end{eqnarray}
and hence the mode of (\ref{eq:P(S|X)}) is given by 
\begin{eqnarray}
\hat{\Pi}&=&\argmax_{    \hat{\cmatrix}\, \vert K}\big\{\rme^{-N\hat{F}_N\left(\cmatrix,\, \xmatrix\right)} p\left( \hat{\cmatrix} \vert K\right) p(K) 
\big\}  \label{eq:Pi-hat-app}
\end{eqnarray}
which is our MAP estimator of the partition of data $\Pi$.
\vspace*{3mm}

\section{Laplace approximation\label{section:Laplace}}
Let us consider the log-likelihood  density  (\ref{def:F-hat}). We note that $\hat{F}_N\left(\cmatrix,\, \xmatrix\right)=\sum_{\mu=1}^K \hat{F}_\mu^N\left(\cmatrix,\, \xmatrix\right)$, where 
\begin{eqnarray}
\hat{F}_\mu^N\left(\cmatrix,\, \xmatrix\right)&=&-\frac{1}{N}\!\log\!\! \int\!\!  \rme^{-N\Phi_\mu(\thetav_{\!\mu}\vert\cmatrix,\,\xmatrix)} p(\thetav_{\!\mu})  \rmd \thetav_{\!\mu}        \label{def:F-hat-mu}\\
\Phi_\mu(\thetav_{\!\mu}\vert\cmatrix,\,\xmatrix)&=&-\frac{1}{N} \sum_{i=1}^N\!c_{i\mu}\!\log p\left(\x_{i}\vert\thetav_{\!\mu}\right)\nonumber   
\end{eqnarray}
$\hat{F}_\mu^N(\cmatrix,\, \xmatrix)$ 
is a log-likelihood density of cluster $\mu$. For large $N$ it can be evaluated by the Laplace method~\cite{DeBruijn1970}:
\begin{widetext}
\begin{eqnarray}
\hat{F}_\mu^N\left(\cmatrix,\, \xmatrix\right)&=&-\frac{1}{N}\log \Bigg(\frac{\int  \rme^{-N\Phi_\mu(\thetav_{\!\mu}\vert\cmatrix,\,\xmatrix)} p(\thetav_{\!\mu})  \rmd \thetav_{\!\mu}}{\int  \rme^{-N\Phi_\mu( \tilde{\thetav}_\mu\vert\cmatrix,\,\xmatrix)}   \rmd  \tilde{\thetav}_\mu} \!\int \! \rme^{-N\Phi_\mu( \tilde{\thetav}_\mu\vert\cmatrix,\,\xmatrix)}   \rmd \tilde{\thetav}_\mu\! \Bigg) \nonumber    \\
&&=\Phi_\mu(\thetav_{\!\mu}^*\vert\cmatrix)\label{eq:F-hat-mu-large-N},
\end{eqnarray}
%
\end{widetext}
where 
\begin{eqnarray}
\thetav_{\!\mu}^*&=&\argmin_{\thetav} \Phi_\mu(\thetav\vert\cmatrix,\,\xmatrix).
\end{eqnarray}
The stationarity condition $\frac{\partial}{\partial \theta_\mu(\ell)}\Phi_\mu(\thetav\vert\cmatrix,\,\xmatrix)=0$ for all $\ell$, from which to solve $\thetav_{\!\mu}^*$,  gives us the equations 
\begin{eqnarray}
\frac{\partial}{\partial \theta_\mu(\ell)} \frac{1}{N} \sum_{i=1}^N\!c_{i\mu}  \log p\left(\x_{i}\vert\thetav_{\!\mu}\right)=0 \label{eq:dF0-app}.
\end{eqnarray}
Let us now evaluate (\ref{eq:dF0-app}) for the multivariate Gaussian  distributions
 \begin{eqnarray}
\mathcal{N}(\x\vert\m_{\mu},\Lmatrix_{\mu}^{-1})&=&\frac{\rme^{-\frac{1}{2} (\x-\m_{\mu})^T\Lmatrix_\mu(\x-\m_{\mu})}}{|2\pi\Lmatrix_{\mu}^{-1}|^{\frac{1}{2}}}         \label{def:Gauss},
\end{eqnarray}
 with the means $\m_{\mu}$  and the inverse covariance matrices $\Lmatrix_{\mu}$. Upon assuming that $ p\left(\x_{i}\vert\thetav_{\!\mu}\right)\equiv \mathcal{N}(\x\vert\m_{\mu},\Lmatrix_{\mu}^{-1})$,  the desired log-likelihood  density becomes
\begin{eqnarray}
&&-\frac{1}{N} \sum_{i=1}^N\!c_{i\mu}  \log\mathcal{N}\left(\x_i\vert\m_{\mu},\Lmatrix_{\mu}^{-1}\right)= 
\nonumber\\[-1mm]
&&~~~~~~~~~~~~~\frac{1}{2N} \sum_{i=1}^N\!c_{i\mu}(\x_i-\m_{\mu})^T\Lmatrix_\mu(\x_i-\m_{\mu})
\nonumber\\
&&~~~~~~~~~~~~~
-\frac{  M_\mu\left(\cmatrix\right)   }{2N}\log \left((2\pi)^{-d}\left\vert \Lmatrix_\mu\right\vert  \right)  
\label{eq:F-hat-mu-Normal},
\end{eqnarray}
Here $M_\mu(\cmatrix)\!=\!\sum_{i=1}^N c_{i\mu}\!=\!\vert S_\mu\left(\cmatrix\right)\vert$ denotes the number of data points in cluster $\mu$.  Solving the equations  $\frac{\partial}{\partial m_{\mu\ell}}  \sum_{i=1}^N\!c_{i\mu}  \log\mathcal{N}(\x_i\vert\m_{\mu},\Lmatrix_{\mu}^{-1})=0$ and  
$$\frac{\partial}{\partial \left[\Lmatrix_\mu\right]_{s\ell} } \sum_{i=1}^N\!c_{i\mu}  \log\mathcal{N}(\x_i\vert\m_{\mu},\Lmatrix_{\mu}^{-1})=0$$ gives us  
\begin{eqnarray}
\m_{\mu}  &=&\frac{1}{  M_\mu\left(\cmatrix\right) } \sum_{i=1}^N\!c_{i\mu}\x_i\label{eq:theta-Norm-large-N}
\\
\Lmatrix_{\mu}^{-1}&=&\frac{1}{   M_\mu\left(\cmatrix\right) } \sum_{i=1}^N\!c_{i\mu}\left(\x_i-\m_{\mu}\right)\left(\x_i-\m_{\mu}\right)^T\!,~
\label{eq:Cov-Normal-large-N}
\end{eqnarray}
i.e. the empirical mean and covariance of the data in cluster $\mu$. Using the above results in equation (\ref{eq:F-hat-mu-Normal}) we then obtain the log-likelihood  density  \new{(\ref{eq:F-hat-Norm}).}

\section{Distribution  of log-likelihood --  A  `field  theory' approach\label{section:P(F)} }
Let us assume that the data $\xmatrix=\{\x_1,\ldots,\x_N\}$ are sampled from the distribution 
\begin{eqnarray}
p(\xmatrix\vert L)&=&\sum_{\tilde{\cmatrix}}q(\tilde{\cmatrix}\vert L) \left\{\prod_{\nu=1}^L \prod_{i_\nu\in S_\nu(\tilde{\cmatrix})} q_{\nu}(\x_{i_\nu})\right\}   \label{eq:P(X)-1},
\end{eqnarray}
where $q(\tilde{\cmatrix}\vert L)$ is the  `true' distribution of the  partitions $\tilde{\cmatrix}$ of  size $L$. We are interested in computing the distribution of  log-likelihoods
\begin{eqnarray}
P_N(F)&=&\sum_{\cmatrix}p(\cmatrix\vert K) \! \int\! \rmd \xmatrix ~p(\xmatrix\vert L)
\delta\big(F\!-\!\hat{F}_N(\cmatrix,\xmatrix)\big)
\nonumber
\\[-1mm]&&
\label{def:P(F)} 
\\
\hat{F}_N\left(\cmatrix,\,\xmatrix\right)&=&-\frac{1}{N}\sum_{\mu=1}^K\sum_{i=1}^N\!c_{i\mu}\!\log p\left(\x_{i}\vert\thetav_{\!\mu}\right).  \label{def:F-hat-large-N}
\end{eqnarray}
Here $p(\cmatrix\vert K)$ is our `assumed' distribution of the partition $\cmatrix$ of  size $K$.  Let us now evaluate $P_N(F)$ further:
\begin{widetext}
\begin{eqnarray}
P_N(F)&=&\sum_{\cmatrix, \tilde{\cmatrix}}p(\cmatrix\vert K) q(\tilde{\cmatrix}\vert L) \int  \left  \{\prod_{\nu=1}^L \prod_{i_\nu\in S_\nu(\tilde{\cmatrix})} q_{\nu}(\x_{i_\nu})\right\} \delta\left(F-\hat{F}_N(\cmatrix,\xmatrix)\right) \rmd \xmatrix\label{eq:P(F)-1}
\end{eqnarray}
We note that the sum over $\tilde{\cmatrix}$  inside the function $\hat{F}_N(\cmatrix,\xmatrix)$ can be written in the following form 
\begin{eqnarray}
%
-\hat{F}_N\left(\cmatrix,\xmatrix\right)&=&\sum_{\mu=1}^K\frac{\vert  S_\mu(\cmatrix) \vert}{N}\int \frac{1}{\vert  S_\mu(\cmatrix) \vert} \sum_{i_\mu\in S_\mu(\cmatrix)}  \delta\left(\x-\x_{i_\mu}\right)\log p\left(\x\vert\thetav_{\!\mu}\right)\rmd\x\nonumber\\
&=&\sum_{\mu=1}^K\!\frac{\vert  S_\mu(\cmatrix) \vert}{N}\!\!\!\int\! \!\!\frac{1}{\vert  S_\mu(\cmatrix) \vert}\sum_{\nu=1}^L\sum_{i_{\mu\nu}\in S_\mu(\cmatrix) \cap  S_\nu(\tilde{\cmatrix})} \delta\!\left(\x-\x_{i_{\mu\nu}}\right)\!\log p\!\left(\x\vert\thetav_{\!\mu}\right)\!\rmd\x\nonumber\\
&=&\sum_{\mu=1}^K \frac{\vert  S_\mu(\cmatrix) \vert}{N} \int Q_\mu(\x\vert\cmatrix,\tilde{\cmatrix},\xmatrix)  \log p\left(\x\vert\thetav_{\!\mu}\right)\rmd\x,
\end{eqnarray}
where we have defined the density 
\begin{eqnarray}
Q_\mu(\x\vert\cmatrix,\tilde{\cmatrix},\xmatrix)&=&\frac{1}{\vert  S_\mu(\cmatrix) \vert}\!\sum_{\nu=1}^L  \sum_{i_{\mu\nu}\in S_\mu(\cmatrix) \cap  S_\nu(\tilde{\cmatrix})}\!\! \!\!\!\delta\!\left(\x-\x_{i_{\mu\nu}}\right)\label{def:Q(x)}
\end{eqnarray}
 Using the above form in (\ref{eq:P(F)-1}) we obtain 
\begin{eqnarray}
P_N(F)
&=&\sum_{\cmatrix, \tilde{\cmatrix}}p(\cmatrix\vert K) \,q(\tilde{\cmatrix}\vert L)\int\!\rmd \xmatrix \left\{\prod_{\nu=1}^L \prod_{i_\nu\in S_\nu(\tilde{\cmatrix})} q_{\nu}(\x_{i_\nu})\right\}  \nonumber\\
&&~~~~~~~~~~~~~~~~~~~~~\times\delta\left(F+    \sum_{\mu=1}^K \frac{\vert  S_\mu(\cmatrix) \vert}{N} \int\! Q_\mu(\x\vert\cmatrix,\tilde{\cmatrix},\xmatrix) \log p\left(\x\vert\thetav_{\!\mu}\right)\rmd\x\right) \nonumber\\
&=&\sum_{\cmatrix, \tilde{\cmatrix}}p(\cmatrix\vert K) \,q(\tilde{\cmatrix}\vert L) \left\{\prod_{\mu=1}^K\prod_{\x}\int \!\rmd Q_\mu(\x) \right\}  P_N\left[\{Q_\mu(\x)\}\vert\cmatrix, \tilde{\cmatrix}\,\right]\nonumber\\
  &&~~~~~~~~~~~~~~~~~~~~~\times\delta\left(\!F+  \!  \sum_{\mu=1}^K\! \frac{\vert  S_\mu(\cmatrix) \vert}{N}\int\! \!Q_\mu(\x)\! \log p\left(\x\vert\thetav_{\!\mu}\right)\!\rmd\x\right),
 \label{eq:P(F)-2}
\end{eqnarray}
were we have defined the (functional) distribution  
\begin{eqnarray}
P_N\left[\{Q_\mu(\x)\}\vert\cmatrix, \tilde{\cmatrix}\,\right]&=&\int  \left\{\prod_{\nu=1}^L \prod_{i_\nu\in S_\nu(\tilde{\cmatrix})} q_{\nu}(\x_{i_\nu})\right\}   \nonumber\\
 &&~~~~~~~~~~\times\left\{\prod_{\mu=1}^K\prod_{\x}\delta\left[ Q_\mu(\x)-Q_\mu(\x\vert\cmatrix,\tilde{\cmatrix},\xmatrix)\right]\right\} \rmd \xmatrix\label{def:P}.
\end{eqnarray}
Let us next consider 
\begin{eqnarray}
&&P_N\left[\{Q_\mu(\x)\}\vert\cmatrix, \tilde{\cmatrix}\,\right]\nonumber\\
&&~~~~~~~~~=\int \!\rmd \xmatrix  \left\{\prod_{\nu=1}^L \prod_{i_\nu\in S_\nu(\tilde{\cmatrix})} q_{\nu}(\x_{i_\nu})\right\}  \left  \{\prod_{\mu=1}^K\prod_{\x} \int\frac{\rmd \hat{Q}_\mu(\x)}{2\pi/N}\right   \}\nonumber\\
&&~~~~~~~~~~~~~~~~~~\times\rme^{\rmi N\sum_{\mu=1}^K\int\hat{Q}_\mu(\x)\left[ Q_\mu(\x)-Q_\mu(\x\vert\cmatrix,\tilde{\cmatrix},\xmatrix)\right]\rmd\x}\nonumber\\
&&~~~~~~~~~=\left  \{\prod_{\mu=1}^K\prod_{\x}\int \frac{\rmd \hat{Q}_\mu(\x)}{2\pi/N}\right   \} \rme^{\rmi N\sum_{\mu=1}^K\int\hat{Q}_\mu(\x)Q_\mu(\x)\rmd\x}   \nonumber\\
&&~~~~~~~~~~~~~~~~~~\times   \int \!\rmd \xmatrix \left  \{\prod_{\nu=1}^L \prod_{i_\nu\in S_\nu(\tilde{\cmatrix})} q_{\nu}(\x_{i_\nu})\right \}  \rme^{\sum_{\nu=1}^L \sum_{\mu=1}^K \frac{N}{\vert  S_\mu(\cmatrix) \vert} \sum_{i_{\mu\nu}\in S_\mu(\cmatrix) \cap  S_\nu(\tilde{\cmatrix})} -\rmi \hat{Q}_\mu\left(\x_{i_{\mu\nu}}\right) }\nonumber\\
&&~~~~~~~~~=\left\{\prod_{\mu=1}^K\prod_{\x}\int \frac{\rmd \hat{Q}_\mu(\x)}{2\pi/N}\right\} \rme^{\rmi N\sum_{\mu=1}^K\int\hat{Q}_\mu(\x)Q_\mu(\x)\rmd\x}   \nonumber\\
&&~~~~~~~~~~~~~~~~ \times     \prod_{\nu=1}^L \prod_{\mu=1}^K \prod_{i_{\mu\nu}\in S_\mu(\cmatrix) \cap  S_\nu(\tilde{\cmatrix})}  \int\! q_{\nu}\left(\x_{i_{\mu\nu}}\right)  \rme^{-\rmi\frac{ N}{\vert  S_\mu(\cmatrix) \vert}  \hat{Q}_\mu\left(\x_{i_{\mu\nu}}\right) }     \rmd \x_{i_{\mu\nu}}    \nonumber\\
&&~~~~~~~~~=\left\{\prod_{\mu=1}^K\prod_{\x}\int \!\!\frac{\rmd \hat{Q}_\mu(\x)}{2\pi/N}\right\} \nonumber\\
&&~~~~~~~~~~~~~~~~ \times \rme^{\rmi N\sum_{\mu=1}^K\!\int\!\hat{Q}_\mu(\x)Q_\mu(\x)\rmd\x+N\sum_{\mu=1}^K \sum_{\nu=1}^L    \frac{ \vert S_\mu(\cmatrix) \cap  S_\nu(\tilde{\cmatrix})\vert}{N}  \log\!\int\!  \rmd \x~q_{\nu}\left(\x\right)  \rme^{-\rmi \frac{N  \hat{Q}_\mu(\x)}{\vert S_\mu(\cmatrix) \vert}     }}\label{eq:P-1}.
\end{eqnarray}

\end{widetext}
Thus for $P_N\left[Q\vert \alpham(\cmatrix, \tilde{\cmatrix})\right]\equiv P_N\left[\{Q_\mu(\x)\}\vert\cmatrix, \tilde{\cmatrix}\,\right]$ we have 
\begin{eqnarray}
P_N\left[Q\vert  \alpham(\cmatrix, \tilde{\cmatrix})\right]&=&    \int\mathcal{D} \hat{Q}\,    \rme^{N\Psi\left[Q, \hat{Q}\vert\alpham(\cmatrix, \tilde{\cmatrix})\,\right]}\label{eq:P-2},
\end{eqnarray}
where
\begin{eqnarray}
\Psi \big[Q, \hat{Q}\vert\alpham(\cmatrix, \tilde{\cmatrix})\,\big]&=& \rmi \sum_{\mu=1}^K \int\hat{Q}_\mu(\x)Q_\mu(\x)\rmd\x\nonumber\\
&&~~~~~~+  \sum_{\mu=1}^K\sum_{\nu=1}^L   \alpha(\nu, \mu\vert\cmatrix, \tilde{\cmatrix})\log\!\!\int\!\! q_{\nu}\left(\x\right)  \,\rme^{\frac{ -\rmi}{ \alpha(\mu\vert\cmatrix)   }  \hat{Q}_\mu\left(\x\right) }     \rmd \x,
\label{def:Psi}
\end{eqnarray}
with the usual short-hand for the path integral measure, $\int\mathcal{D} \hat{Q}\equiv\left\{\prod_{\mu=1}^K\prod_{\x}\int [\rmd \hat{Q}_\mu(\x)/(2\pi/N)]\right\} $. In the above formula we have also introduced the matrix $\alpham(\cmatrix, \tilde{\cmatrix})$, with entries $[\alpham(\cmatrix, \tilde{\cmatrix})]_{\nu\mu}= \alpha(\nu, \mu\vert\cmatrix, \tilde{\cmatrix})$,  where in turn  $\alpha(\nu, \mu\vert\cmatrix, \tilde{\cmatrix})=N^{-1}\vert S_\mu(\cmatrix) \!\cap\!  S_\nu(\tilde{\cmatrix})\vert$.  We note that  $\cup_{\mu=1}^K \left(S_\mu(\cmatrix) \cap  S_\nu(\tilde{\cmatrix})\right)=S_\nu(\tilde{\cmatrix})$ and that $\cup_{\nu=1}^L \left(S_\mu(\cmatrix) \cap  S_\nu(\tilde{\cmatrix})\right)=S_\mu(\cmatrix)$.  From these properties it follows that the entries $\alpha(\nu, \mu\vert\cmatrix, \tilde{\cmatrix})\geq0$ can be interpreted as representing  a joint distribution, i.e. $\sum_{\mu=1}^K\sum_{\nu=1}^L \alpha(\nu, \mu\vert\cmatrix, \tilde{\cmatrix})=1$, with the marginals $\sum_{\nu=1}^L \alpha(\nu, \mu\vert\cmatrix, \tilde{\cmatrix})=\alpha(\mu\vert\cmatrix)=\vert S_\mu(\cmatrix)\vert/N$ and  $\sum_{\mu=1}^K\alpha(\nu, \mu\vert\cmatrix, \tilde{\cmatrix})=\alpha(\nu\vert\tilde{\cmatrix})=\vert S_\nu(\tilde{\cmatrix})\vert/N$.
Using all these ingredients in equation (\ref{eq:P(F)-3}) then leads us to 
\begin{widetext}
\begin{eqnarray}
P_N(F)&=&\sum_{\cmatrix, \tilde{\cmatrix}}p(\cmatrix\vert K) \,q(\tilde{\cmatrix}\vert L)  \int\mathcal{D} Q\,  P_N\left[Q\vert\alpham(\cmatrix, \tilde{\cmatrix})\right] \delta\left(F+    \sum_{\mu=1}^K \frac{\vert  S_\mu(\cmatrix) \vert}{N} \int Q_\mu(\x) \log p\left(\x\vert\thetav_{\!\mu}\right)\rmd\x\right)\nonumber\\
&=&\int\rmd\alpham\, P_N(\alpham) \int\mathcal{D} Q\,  P_N\left[Q\vert\alpham\right]  \delta\!\left(\!F\!+\!    \sum_{\mu=1}^K\!\! \alpha(\mu)\! \!\int\! Q_\mu(\x) \log p\left(\x\vert\thetav_{\!\mu}\right)\rmd\x\!\right)\label{eq:P(F)-3},
\end{eqnarray}
\end{widetext}
where we have defined the integral measure $\int\mathcal{D} Q\equiv \big\{\prod_{\mu=1}^K\prod_{\x}\int \rmd Q_\mu(\x) \big\}$ as well as the short-hand $\int\!\rmd\alpham\equiv \prod_{\mu=1}^K\prod_{\nu=1}^L\int\rmd\alpha(\nu,\mu)$.  The distribution of $\alpham$ is given by 
\begin{eqnarray}
P_N(\alpham)&=&\sum_{\cmatrix, \tilde{\cmatrix}}\,p(\cmatrix\vert K) \,q(\tilde{\cmatrix}\vert L)  \prod_{\mu=1}^K\!\prod_{\nu=1}^L\!\delta\!\left[\alpha(\nu, \mu)\!-\!\alpha(\nu, \mu\vert\cmatrix,\! \tilde{\cmatrix})\right].
\nonumber
\\[-2mm]&&
\label{def:P(alpha)-app}
\end{eqnarray}
Now for any smooth function $g$ we can consider the following average:
\begin{widetext}
\begin{eqnarray}
\int P_N(F)\, g(F)\,\rmd F&=&\int\rmd\alpham\, P_N(\alpham) \int\mathcal{D} Q\,  P_N\left[Q\vert\alpham\right]  ~g\Big(\!-\sum_{\mu=1}^K \alpha(\mu) \int Q_\mu(\x) \log p\left(\x\vert\thetav_{\!\mu}\right)\rmd\x\Big)\nonumber\\
&=&\int\rmd\alpham\, P_N(\alpham) \frac{\int\mathcal{D} Q\,  P_N\left[Q\vert\alpham\right]}{\int\mathcal{D} \tilde{Q}\,  P_N\big[\tilde{Q}\vert\alpham\big]}  ~g\Big(\!-\sum_{\mu=1}^K \alpha(\mu) \int Q_\mu(\x) \log p\left(\x\vert\thetav_{\!\mu}\right)\rmd\x\Big)\nonumber\\
&=&\int\rmd\alpham\, P_N(\alpham) \frac{    \int\mathcal{D} Q\,   \int\mathcal{D} \hat{Q}\,    \rme^{N\Psi\left[Q, \hat{Q}\vert\alpham\right]}   }{\int\mathcal{D} \tilde{Q}\,   \int\mathcal{D} \hat{Q}\,    \rme^{N\Psi\left[\tilde{Q}, \hat{Q}\vert\alpham\right]}} \nonumber\\
&&~~~~~~~~~~~~~~~~~~~~~~~\times g\Big(\!-\sum_{\mu=1}^K\! \alpha(\mu)\!\! \int \!\!Q_\mu(\x) \log p\left(\x\vert\thetav_{\!\mu}\right)\rmd\x\Big)\label{eq:<g>}.
\end{eqnarray}
\end{widetext}
Let us assume that $P_N(\alpham)\rightarrow P(\alpham)$ as $N\rightarrow\infty$.  Furthermore we expect that in this limit  the functional integral in the above equation is dominated by the extremum of the functional $\Psi$ and hence for the distribution $ P(F)=\lim_{N\rightarrow\infty} P_N(F)$ we obtain 
\begin{widetext}
\begin{eqnarray}
&&\int P(F)\, g(F)\,\rmd F=\int P(\alpham) ~g\Big(\!-\sum_{\mu=1}^K \alpha(\mu)\int\! \!Q_\mu(\x\vert\alpham)\log p\left(\x\vert\thetav_{\!\mu}\right)\rmd\x\Big)\rmd\alpham\label{eq:<g>-large-N},
\end{eqnarray}
\end{widetext}
where $Q_\mu(\x\vert\alpham)$ is a solution of the saddle-point equations $\delta\Psi[Q, \hat{Q}\vert\alpham]/\delta  Q_\mu(\x)=0$ and  $\delta\Psi[Q, \hat{Q}\vert\alpham]/\delta  \hat{Q}_\mu(\x)=0$. Solving the latter gives us the following two equations:  
\begin{eqnarray}
 \rmi  \hat{Q}_\mu(\x)&=&0\label{eq:SP1}
 \\
Q_\mu(\x)&=& \sum_{\nu=1}^L   \frac{\alpha(\nu, \mu)}{\alpha(\mu)}   \frac{q_{\nu}\left(\x\right)   \rme^{\frac{ -\rmi}{ \alpha(\mu)   }  \hat{Q}_\mu\left(\x\right) }}{\int q_{\nu}\left(\x^\prime\right)  \,\rme^{\frac{ -\rmi}{ \alpha(\mu)   }  \hat{Q}_\mu\left(\x^\prime\right) }     \rmd \x^\prime}
~~~\label{eq:SP2}
\end{eqnarray}
from which follows the equation 
\begin{eqnarray}
Q_\mu(\x\vert\alpham)&=& \sum_{\nu=1}^L   \alpha(\nu\vert \mu) q_{\nu}\left(\x\right)\label{eq:SP-Q}, 
\end{eqnarray}
where $ \alpha(\nu\vert\mu)=\alpha(\nu, \mu)/\alpha(\mu)$ is a conditional distribution.  From the above we conclude that 
\begin{eqnarray}
P(F)&=&\int\!\rmd\alpham~ P(\alpham)
\label{eq:P(F)-large-N}
\\[-1mm]
&&\times~\delta\Big(F+\sum_{\mu=1}^K\!\sum_{\nu=1}^L \alpha(\nu,\mu)\! \! \int\!\! q_{\nu}\!\left(\x\right) \log p\left(\x\vert\thetav_{\!\mu}\right) \rmd\x\Big).
\nonumber
\end{eqnarray}
If we assume that $P(\alpham)$ is a delta function,  this gives us the mean-field (MF) log-likelihood 
\begin{eqnarray}
F(\alpham)&=&-\!\sum_{\mu=1}^K\!\sum_{\nu=1}^L\! \alpha(\nu,\mu)\!\!  \int\!\! q_{\nu}\!\left(\x\right) \log p\left(\x\vert\thetav_{\!\mu}\right)\! \rmd\x 
~~~~~\label{eq:F-MF}
\end{eqnarray}
\new{which is seen to be equivalent to (\ref{eq:F-large-N}).}
Let us next consider the distribution (\ref{def:P(F)})  of the log-likelihood density  (\ref{eq:F-hat-Norm}):
\begin{eqnarray}
P_N(F)&=&\sum_{\cmatrix}p(\cmatrix\vert K)  \int\!\rmd \xmatrix~p(\xmatrix\vert L)
\label{def:P(F)-Norm-1}
\\[-1mm]
&&\times
\delta\Big(F-      \sum_{\mu=1}^K\frac{ \vert S_\mu\left(\cmatrix\right)\vert}{2N}  \log\!\left(\!(2\pi\rme)^{d}\!\left\vert \Lmatrix_{\mu}^{-1}(\cmatrix, \xmatrix)\right\vert  \right)\! \!  \Big) 
\nonumber
%
\end{eqnarray}
 where  $\Lmatrix_{\mu}^{-1}(\cmatrix, \xmatrix)$ is the covariance matrix of the data in cluster $\mu$, which can be written in the form  
\begin{eqnarray}
\Lmatrix_{\mu}^{-1}(\cmatrix, \xmatrix)&=&\frac{1}{ \vert S_\mu\left(\cmatrix\right)\vert}\!  \sum_{i_\mu\in S_\mu\left(\cmatrix\right)}\!\! \!\!\left(\x_{i_\mu}\!\!-\!\m_{\mu}\left(\cmatrix\right)\!\right) \left(\x_{i_\mu}\!\!-\!\m_{\mu}\left(\cmatrix\right)\!\right)^T\label{eq:Cov-1},
\end{eqnarray}
 where $\m_{\mu}(\cmatrix) =\frac{1}{  \vert S_\mu\left(\cmatrix\right)\vert    }  \sum_{i_\mu\in S_\mu\left(\cmatrix\right)}  \x_{i_\mu}$. 
 Further manipulation of $P_N(F)$ gives
 \begin{widetext}
\begin{eqnarray}
P_N(F)&=&\sum_{\cmatrix, \tilde{\cmatrix}}p(\cmatrix\vert K)\, q(\tilde{\cmatrix}\vert L)  \int    ~
\left\{\prod_{\nu=1}^L \prod_{i_\nu\in S_\nu(\tilde{\cmatrix})} q_{\nu}(\x_{i_\nu})\right\}\nonumber\\
&&~~~~~~~~~~~~~~\times\delta\Big(F- \sum_{\mu=1}^K\frac{ \vert S_\mu\left(\cmatrix\right)\vert}{2N} \log \left((2\pi\rme)^{d}\left\vert \Lmatrix_{\mu}^{-1}(\cmatrix, \xmatrix)\right\vert  \right)   \Big)\rmd \xmatrix~~~       
  \label{eq:P(F)-Norm-2}
\end{eqnarray}
and the covariance matrix can be written in the form 
\begin{eqnarray}
\Lmatrix_{\mu}^{-1}\left(\cmatrix, \xmatrix\right)&=&\frac{1}{ \vert S_\mu\left(\cmatrix\right)\vert}  \sum_{\nu=1}^L\sum_{i_{\nu\mu}\in    S_\mu(\cmatrix) \cap  S_\nu(\tilde{\cmatrix}) }  
\left(\x_{i_{\nu\mu}}-\m_{\mu}\left(\cmatrix\right)\right) \!\left(\x_{i_{\nu\mu}}-\m_{\mu}\left(\cmatrix\right)\right)^T\nonumber\\
&=&\int\!\rmd\x ~ Q_\mu(\x\vert\cmatrix,\tilde{\cmatrix},\xmatrix) \left(\x\!-\!     \int\! Q_\mu(\y\vert\cmatrix,\tilde{\cmatrix},\xmatrix)\,\y\,\rmd\y    \right) \nonumber\\
&&~~~~~~~~~~~~~~~~~~~~~~~\times\left(\x\!-\!   \int \!Q_\mu(\z\vert\cmatrix,\tilde{\cmatrix},\xmatrix)\,\z\,\rmd\z        \right)^T\!\!\!\!.  \label{eq:Cov-2} 
\end{eqnarray}
\end{widetext}
 From the above it is clear  that  $\hat{F}_N$ is a functional of the density $Q_\mu(\x\vert\cmatrix,\tilde{\cmatrix},\xmatrix) $, defined in (\ref{def:Q(x)}),  and the matrix $\alpham(\nu,\mu\vert\cmatrix\tilde{\cmatrix})$. Following the same steps as in deriving equations (\ref{eq:P(F)-2})-(\ref{eq:P(F)-3}) gives us   
\begin{eqnarray}
P_N(F)&=&\int\rmd\alpham\, P_N(\alpham) \int\mathcal{D} Q\,  P_N\left[Q\vert\alpham\right]
\label{eq:P(F)-Norm-3}
\\&&\times~ \delta\Big(F-    \sum_{\mu=1}^K \alpha(\mu) \frac{1}{2}  \log \left((2\pi\rme)^{d}\left\vert \Lmatrix_{\mu}^{-1} \left[Q\right]   \right\vert  \right)     \Big),
\nonumber
\end{eqnarray}
where 
\begin{eqnarray}
\Lmatrix_{\mu}^{-1}\left[Q\right]&=&\int\!\rmd\x~ Q_\mu(\x) \Big(\x\!- \!    \int\! Q_\mu(\y)\,\y\,\rmd\y  \Big)
\nonumber
\\&&~~~~~~~~~~~~~~~~~~~~~~~\times \Big(\x\!- \!  \int\! Q_\mu(\z)\,\z\,\rmd\z        \Big)^T   \label{eq:Cov-3} .
\end{eqnarray}
Furthermore, for $N\rightarrow\infty$, using a similar argument as outlined in equations (\ref{eq:<g>})-(\ref{eq:P(F)-large-N}), we obtain 
\begin{eqnarray}
P(F)&=&\int\! \rmd\alpham  ~ P(\alpham)
\label{eq:P(F)-Norm-large-N}
\\[-1mm]
&&\times~\delta\Big(F-    \sum_{\mu=1}^K \alpha(\mu)\frac{1}{2}  \log \left((2\pi\rme)^{d}\left\vert \Lmatrix_{\mu}^{-1} \left(\alpham\right)   \right\vert  \right)     \Big)
\nonumber
\end{eqnarray}
where the covariance matrix $ \Lmatrix_{\mu}^{-1}(\alpham) $ is defined by 
\begin{eqnarray}
\Lmatrix_{\mu}^{-1}(\alpham)&=&\sum_{\nu=1}^L\! \alpha(\nu\vert\mu)\Big\langle\!\left (\x-\m_{\mu}\!\left(\alpham\right)\right) \left(\x-\m_{\mu}\!\left(\alpham\right)\right)^T\Big\rangle_\nu,\label{eq:C-2-app}
\end{eqnarray}
where $\m_{\mu}\left(\alpham\right)=\sum_{\nu=1}^L \alpha(\nu\vert\mu)\langle\x\rangle_\nu$ is the mean, and we used  the short-hand $\langle\{\cdots\}\rangle_\nu=\int q_\nu(\x) \{\cdots\}\rmd\x$.  Assuming that $P(\alpham)$ is a delta function subsequently gives us the MF log-likelihood expression \new{(\ref{eq:F-Norm}).}
\vspace*{3mm}

\section{Proofs of information-theoretic inequalities\label{section:inform}}

In this section we compute lower bounds for the  MF entropy  (\ref{eq:F-Norm}). First, we show that $F(\alpham) $ satisfies the inequalities
\begin{eqnarray}
F(\alpham) \geq  \sum_{\mu=1}^K\alpha(\mu)H \left(Q_\mu \right)\geq \sum_{\nu=1}^L\gamma(\nu)H \left(q_\nu \right).\label{eq:F-Norm-ineq-app}
\end{eqnarray}
Let us  consider the  Kullback-Leibler distance~\cite{Cover2012}  $D\left(Q_\mu\vert\vert\mathcal{N}_\mu\right)$  between the mixture  $Q_\mu(\x)=\sum_{\nu=1}^L \alpha(\nu\vert\mu) q_\nu(\x)$ and the Gaussian distribution  $ \mathcal{N}\left(\x\vert\m_{\mu},\Lmatrix_{\mu}^{-1}\right)$:
\begin{widetext}
\begin{eqnarray}
D\left(Q_\mu\vert\vert\mathcal{N}_\mu\right)&=&\int\!\rmd \x \sum_{\nu=1}^L \alpha(\nu\vert\mu)\,  q_\nu(\x) 
\log \left(  \frac{ \sum_{\nu=1}^L \alpha(\nu\vert\mu)\,  q_\nu(\x)}{\mathcal{N}\left(\x\vert\m_{\mu},\Lmatrix_{\mu}^{-1}\right)}\right) 
\nonumber
\\
&=& -H \left(Q_\mu \right)-\sum_{\nu=1}^L \alpha(\nu\vert\mu)  
\int\!\rmd \x~ q_\nu(\x) \log  \mathcal{N}\left(\x\vert\m_{\mu},\Lmatrix_{\mu}^{-1}\right) 
\nonumber
\\
&=& -H \left(Q_\mu \right) -\sum_{\nu=1}^L \alpha(\nu\vert\mu)
\int\! \rmd \x ~ q_\nu(\x) \log \left( \frac{\rme^{-\frac{1}{2} (\x-\m_{\mu})^T\Lmatrix_\mu(\x-\m_{\mu})}}{\left\vert2\pi\Lmatrix_{\mu}^{-1}\right\vert^{\frac{1}{2}}}
\right)
\nonumber
\\
&=&-H \left(Q_\mu \right) + \frac{1}{2}\log \left( (2\pi)^d \left\vert\Lmatrix_{\mu}^{-1}\right\vert\right)\nonumber\\
&&~~~~~~~~~~~+\frac{1}{2}\sum_{\nu=1}^L \alpha(\nu\vert\mu)\int\! \rmd \x~q_\nu(\x)  
(\x-\m_{\mu})^T\Lmatrix_\mu(\x-\m_{\mu})
\nonumber
\\
&=&-H \left(Q_\mu \right) + \frac{1}{2}\log \left( (2\pi)^d \left\vert\Lmatrix_{\mu}^{-1}\right\vert\right)\nonumber\\
 &&~~~~~~~~~~~~~~~~~~~~~~~+\frac{1}{2}\Tr\Bigg\{\Lmatrix_\mu\sum_{\nu=1}^L \alpha(\nu\vert\mu) \int  q_\nu(\x)(\x-\m_{\mu}) (\x-\m_{\mu})^T\rmd \x \Bigg\} \nonumber
\end{eqnarray}
\end{widetext}
Let us define the mean and covariance of the distribution $Q_\mu(\x)\!=\!\sum_{\nu=1}^L \alpha(\nu\vert\mu) q_\nu(\x)$ as  $\m_{\mu}\!=\!\int Q_\mu(\x)\,\x\, \rmd \x$ and $\Lmatrix_{\mu}^{-1}\!=\!\int Q_\mu(\x)(\x\!-\!\m_{\mu}) (\x\!-\!\m_{\mu})^T\rmd \x$. Then  $D\left(Q_\mu\vert\vert\mathcal{N}_\mu\right)=-H \left(Q_\mu \right) +\frac{1}{2}\log \left( (2\pi\rme)^d \left\vert\Lmatrix_{\mu}^{-1}\right\vert\right)$ and from the simple property $D\left(Q_\mu\vert\vert\mathcal{N}_\mu\right)\geq0$ we immediately deduce that
\begin{eqnarray}
F(\alpham)&\geq&  \sum_{\mu=1}^K\alpha(\mu)H \left(Q_\mu \right)\label{eq:F-Norm-lb}.
\end{eqnarray}
Furthermore,  for the average entropy we find the following inequality
\begin{widetext}
\begin{eqnarray}
\sum_{\mu=1}^K\alpha(\mu)H \left(Q_\mu \right)&=&\sum_{\mu=1}^K\alpha(\mu)  \sum_{\nu_1=1}^L\alpha(\nu_1\vert\mu)\int q_{\nu_1}(\x) \log\Big(1/\sum_{\nu_2=1}^L\alpha(\nu_2\vert\mu) \,q_{\nu_2}(\x)\Big)\rmd\x\nonumber\\
&=&\sum_{\mu=1}^K\sum_{\nu_1=1}^L\alpha(\nu_1, \mu)\nonumber\\
&&~~\times\int q_{\nu_1}(\x) \Big\{\log q_{\nu_1}(\x)-\log q_{\nu_1}(\x)+\log\Big(1/\sum_{\nu_2=1}^L\alpha(\nu_2\vert\mu) \,q_{\nu_2}(\x)\Big) \Big\}\rmd\x\nonumber\\
&=&\sum_{\mu=1}^K\sum_{\nu=1}^L\alpha(\nu, \mu) D\left(q_{\nu} \vert\vert Q_\mu\right) + \sum_{\nu=1}^L\gamma(\nu)H \left(q_\nu \right)
\nonumber
\\
&\geq & \sum_{\nu=1}^L\gamma(\nu)H \left(q_\nu \right)\label{eq:aver-mixture-lb}.
\end{eqnarray}
\end{widetext}

Secondly, for the average  entropy
\begin{eqnarray}
F_0&=&\sum_{\nu=1}^L\gamma(\nu)\frac{1}{2}  \log \left((2\pi\rme)^{d}\left\vert \Cov_\nu \right\vert  \right) \label{eq:F-0},
\end{eqnarray}
where $\Cov_\nu= \left\langle \x\,\x^T\right \rangle_\nu- \left\langle \x\right \rangle_\nu\, \left\langle\x\right \rangle_\nu^T$ is the covariance matrix of  $ q_\nu(\x)$, we can show that  the following holds:
\begin{eqnarray}
F(\alpham)\geq F_0\label{eq:F-0-lb}
\end{eqnarray}
for all $\alpham$.  
The above equality follows from properties of the covariance matrix 
\begin{eqnarray}
\Lmatrix_{\mu}^{-1}(\alpham)&=& 
\sum_{\nu=1}^L \alpha(\nu\vert\mu)\, \Cov_\nu    \label{eq:C-3}  \\
&& +\sum_{\nu=1}^L \alpha(\nu\vert\mu)\left (\langle \x \rangle_\nu\!-\!\m_{\mu}\left(\alpham\right)\right)\left(\langle \x \rangle_\nu\!-\!\m_{\mu}\left(\alpham\right)\right)^T. \nonumber
\end{eqnarray}
To prove (\ref{eq:F-0-lb}) we first derive the inequality 
\begin{eqnarray}
 \log\left\vert\sum_{\nu=1}^L\alpha(\nu) \A_\nu \right\vert&\geq&\sum_{\nu=1}^L\alpha(\nu) \log\left\vert \A_\nu \right\vert \label{eq:log-det-ineq}
\end{eqnarray}
for symmetric positive definite matrices $\A_\nu$ and  $\sum_{\nu=1}^L \alpha(\nu)=1$, where $\alpha(\nu)\geq0$.
This inequality  can be derived by repeated  application of Minkowski's inequality for determinants, viz. $\vert\A +\BMatrix \vert^{\frac{1}{d}}\geq  \vert\A  \vert^{\frac{1}{d}} +\vert\BMatrix    \vert^{\frac{1}{d}}$ for symmetric positive definite matrices $\A$ and $\BMatrix$:
\begin{widetext}
\begin{eqnarray}
 \left\vert\sum_{\nu=1}^L\alpha(\nu) \A_\nu \right\vert^{\frac{1}{d}}   &=&   \left\vert \alpha(1) \A_1+ \sum_{\nu=2}^L\alpha(\nu) \A_\nu \right\vert^{\frac{1}{d}} \nonumber\\
 &&~~~~~~~~~~~~~~ \geq   \alpha(1) \left\vert \A_1\right\vert^{\frac{1}{d}}+ \left\vert\sum_{\nu=2}^L\alpha(\nu) \A_\nu \right\vert^{\frac{1}{d}}  \geq\sum_{\nu=1}^L\alpha(\nu)  \left\vert\A_\nu \right\vert^{\frac{1}{d}}  \label{eq:Minkowski-ineq}
\end{eqnarray}
from which follows  the result 
\begin{eqnarray}
 \log\left\vert\sum_{\nu=1}^L\alpha(\nu) \A_\nu \right\vert&\geq&  d\log\left(\sum_{\nu=1}^L\alpha(\nu)  \left\vert\A_\nu \right\vert^{\frac{1}{d}} \right)\geq   \sum_{\nu=1}^L\alpha(\nu) \log\left\vert \A_\nu \right\vert \label{eq:Jensen-ineq}.
\end{eqnarray}
The last step in this argument relied on Jensen's inequality~\cite{Cover2012}.  Let us now  apply (\ref{eq:log-det-ineq})  to the difference  of entropies 
\begin{eqnarray}
2\left(F(\alpham)-F_0\right)&=&-\sum_{\nu=1}^L\gamma(\nu)  \log \left\vert \Cov_\nu \right\vert +\sum_{\mu=1}^K\alpha(\mu) \log \left\vert \Lmatrix_{\mu}^{-1}\left(\alpham\right) \right\vert\label{eq:F-F0}\\
&=&-\sum_{\nu=1}^L\gamma(\nu)  \log \left\vert \Cov_\nu \right\vert 
+\sum_{\mu=1}^K\alpha(\mu)\nonumber\\
&&~~~~~~~~\times \log \Bigg     \vert 
\sum_{\nu=1}^L \alpha(\nu\vert\mu)  \Big(\Cov_\nu+\left (\langle \x \rangle_\nu-\m_{\mu}\left(\alpham\right)\right)\!\left(\langle \x \rangle_\nu-\m_{\mu}\left(\alpham\right)\right)^T\Big)
 \Bigg  \vert \nonumber\\
 &&~~\geq -\sum_{\nu=1}^L\gamma(\nu)  \log \left\vert \Cov_\nu \right\vert 
+\sum_{\mu=1}^K\alpha(\mu) \sum_{\nu=1}^L \alpha(\nu\vert\mu)\nonumber\\
&&~~~~~~~~~~~~\times\log \left   \vert \Cov_\nu+\left (\langle \x \rangle_\nu-\m_{\mu}\left(\alpham\right)\right)\!\left(\langle \x \rangle_\nu-\m_{\mu}\left(\alpham\right)\right)^T
 \right  \vert \nonumber\\
 &&~~\geq -\sum_{\nu=1}^L\gamma(\nu)  \log \left\vert \Cov_\nu \right\vert 
+d\sum_{\mu=1}^K\alpha(\mu) \sum_{\nu=1}^L \alpha(\nu\vert\mu)\nonumber\\
&&~~~~~~~~~~~~~~\times\log \! \left(\! \left     \vert \Cov_\nu\right\vert^{\frac{1}{d}}\!+\!\left\vert\!\left (\!\langle \x \rangle_\nu\!-\!\m_{\mu}\left(\alpham\right)\!\right)\!\left(\!\langle \x \rangle_\nu   \!-\!   \m_{\mu}\left(\alpham\right)\!\right)^T
 \right  \vert^{\frac{1}{d}} \!\right).\nonumber
 %
%
\end{eqnarray}
\end{widetext}
The last line in the above, obtained by  Minkowski's inequality, is equal to  zero, and hence  $F(\alpham)\geq F_0$ for all $\alpham$.   
\vspace*{3mm}


\section{Algorithmic cost of  ordering  random unbiased partitions \label{section:algorithmic-cost}  } 
Let us assume that we have $N$  `particles'  of $L$ different  `colours'  which are distributed into $K$ different  reservoirs.   The probability that a particle  has colour $\nu\in[L]$ is $\gamma(\nu)$ and that it is in the reservoir  $\mu$ is $1/K$.  Assuming that colour and reservoir allocation are independent events, the probability of   `configuration'  $\amatrix=(\av_1,\ldots,\av_N)$, where $\av_i=(a_i(1) , a_i(2))$ with the colour $a_i(1)\in [L]$ and reservoir number $a_i(2)\in [K]$  of the particle  $i$, is given by
\begin{eqnarray}
P(\amatrix)&=&\prod_{i=1}^N P(\av_i)\mbox{,}
\\
P(\av_i)&\equiv& P(a_i(1)=\nu , a_i(2)=\mu)=\frac{\gamma(\nu)}{K}\label{def:prob-rand-clust}.
\end{eqnarray}
The total number of particles in reservoir $\mu$ is given by $N_{\mu}(\amatrix)=\sum_{i=1}^N\delta_{\mu; a_i(2)}$. 
Let us now consider the joint distribution of particle numbers in reservoirs  
\begin{eqnarray}
P(N_1,\ldots,N_K)&=&\sum_{\amatrix}P(\amatrix)\prod_{\mu=1}^K\delta_{N_{\mu};  N_{\mu}(\amatrix) }\\
~~~&=&     K^{-N} \sum_{a_1(2),\ldots,a_N(2)}  \prod_{\mu=1}^K\delta_{N_{\mu};  \sum_{i=1}^N\delta_{\mu; a_i(2)}    }\nonumber\\
 &=&     K^{-N} \frac{N!}{ \prod_{\mu=1}^K N_{\mu}!}\nonumber,
\end{eqnarray}
where $\sum_{\mu=1}^K N_{\mu}=N$. The probability of observing the event that at least one reservoir is empty is given by  
\begin{eqnarray}
&&\hspace*{-10mm}
1-P(N_1>0,\ldots,N_K>0)
\nonumber
\\
~~~&=&1-\sum_{N_1>0,\ldots,N_K>0}  K^{-N} \frac{N!}{ \prod_{\mu=1}^K N_{\mu}!}
\nonumber
\\
&=& K^{-N} \Big( \sum_{N_1\geq0,\ldots,N_K\geq0}  \frac{N!}{ \prod_{\mu=1}^K N_{\mu}!}
\nonumber
\\&&\hspace*{30mm}
-\sum_{N_1>0,\ldots,N_K>0}  \frac{N!}{ \prod_{\mu=1}^K N_{\mu}!}  \Big) \nonumber\\
 &=& \sum_{\ell=1}^{K-1} {{K}\choose{\ell}}  \Big(1-\frac{\ell}{K}\Big)^N.               
\end{eqnarray}
Thus the probability of this event decays exponentially with increasing $N$ and, as  $N\rightarrow\infty$, the sequence $a_1(2),\ldots, a_N(2)$, sampled from the distribution (\ref{def:prob-rand-clust}) is,  with high probability,  a \emph{partition} of  the set $[N]$   into  $K$ subsets (or clusters).  Furthermore,  the entropy density $N^{-1}\log(K^N)=\log K$ of such sequences approaches the entropy density $N^{-1}\log \left(K!\,\mathcal{S}(N,K)\right)$ of the random  partitions sampled uniformly  from  (\ref{eq:P(c|k)-uniform}).

 Let us assume that $K\leq L$.  The total number of particles of colour $\nu$, and the number of particles of colour $\nu$ in reservoir $\mu$ are given, respectively, by  $N_{\nu}(\amatrix)=\sum_{i=1}^N\delta_{\nu; a_i(1)}$ and $N_{\nu\mu}(\amatrix)=\sum_{i=1}^N\delta_{\nu; a_i(1)}\delta_{\mu; a_i(2)}$. The number of particles of colour $\nu$ which are \emph{not} in reservoir $\mu$ is the difference $N_{\nu}(\amatrix)-N_{\nu\mu}(\amatrix)$. Suppose that  each reservoir  has a preference  for particles  of  a particular colour (or colours), i.e. there is an onto mapping $\nu\rightarrow\mu(\nu)$ between colours and reservoirs, then the total number of particles which are not in  `their' reservoirs, i.e. the number of particles which are to be  `moved'  in order for  all particles  to be  in  reservoirs to which they  belong,  is given by the difference $\sum_{\nu=1}^L \left(N_{\nu}(\amatrix)-N_{\nu\mu(\nu)}(\amatrix)\right)=N-\sum_{\nu=1}^LN_{\nu\mu(\nu)}(\amatrix)$.

We are interested in the average and variance of $N-\sum_{\nu=1}^LN_{\nu\mu(\nu)}(\amatrix)$.   The average is given by 
\begin{eqnarray}
\left\langle N\!-\!\sum_{\nu=1}^LN_{\nu\mu(\nu)}(\amatrix)\right\rangle_{\amatrix}
%
%
&=&N\!-\!   \sum_{\nu=1}^L    \sum_{i=1}^N   \left\langle  \delta_{\nu; a_i(1)}\delta_{\mu(\nu); a_i(2)}   \right\rangle_{\!\amatrix} \nonumber\\
&=&N-   \sum_{\nu=1}^L    \sum_{i=1}^N     \frac{\gamma(\nu)}{K}
\nonumber
\\&=& N\frac{K-1}{K}
%
\end{eqnarray}
and the variance is given by 
\begin{eqnarray}
Var\Big\{N\!-\!\sum_{\nu=1}^LN_{\nu\mu(\nu)}(\amatrix) \Big\}&=& Var\Big\{\sum_{\nu=1}^LN_{\nu\mu(\nu)}(\amatrix) \Big\}\nonumber\\
%
%
&=&  \left\langle  \Big( \sum_{\nu=1}^LN_{\nu\mu(\nu)}(\amatrix) \!-\!\frac{N}{K}\Big)^2            \right\rangle_{\!\amatrix}\nonumber\\
%
%
&=&\frac{N}{K}  \left(1- \frac{1}{K}  \right).
\end{eqnarray}
The average in the penultimate line of the above was computed as follows
\begin{eqnarray}
&&\hspace*{-7mm}
 \left\langle  \Big( \sum_{\nu=1}^LN_{\nu\mu(\nu)}(\amatrix) \Big)^2   \right\rangle_{\!\amatrix}  
 \nonumber
 \\
 ~~
&=& \sum_{\nu} \sum_{i_1, i_2}\!\left\langle    \delta_{\nu; a_{i_1}\!(1)} \delta_{\mu(\nu); a_{i_1}\!(2)}    \delta_{\nu; a_{i_2}\!(1)} \delta_{\mu(\nu); a_{i_2}\!(2)}  \right\rangle_{\!\amatrix}     +  \nonumber\\
&& \hspace*{-2mm}\sum_{\nu_1\neq \nu_2} \sum_{i_1, i_2=1}\! \!\Big  \langle  \!  \delta_{\nu_1; a_{i_1}(1)} \delta_{\mu(\nu_1); a_{i_1}(2)}    \delta_{\nu_2; a_{i_2}(1)} \delta_{\mu(\nu_2); a_{i_2}(2)} \! \Big\rangle_{\!\amatrix}        \nonumber\\
&=&  \frac{ N }{K} \!+\! \frac{N(N\!-\!1)}{K^2}    \sum_{\nu=1}^L \gamma^2(\nu) \!+\! \frac{N(N\!-\!1)}{K^2} \sum_{\nu_1\neq \nu_2} \gamma(\nu_1)\gamma(\nu_2) \nonumber\\
&=&  \frac{ N }{K} + \frac{N(N-1)}{K^2}. 
\end{eqnarray}
From the above derivations it follows that  for a random unbiased partition to be ordered, i.e. for particles of the same colour to occupy at most one reservoir, a fraction of  particles has to be moved that is on average $\langle 1-\frac{1}{N}\sum_{\nu=1}^LN_{\nu\mu(\nu)}(\amatrix)\rangle_{\amatrix}=(K\!-\!1)/K$,  with variance $Var\{1-\frac{1}{N}\sum_{\nu=1}^LN_{\nu\mu(\nu)}(\amatrix) \}=(1\!-\! K^{-1})/NK$.

\section{Details of numerical experiments\label{section:numerics}}  

\begin{figure}[t]
\hspace*{-5mm}
 \setlength{\unitlength}{0.65mm}
 \begin{picture}(151,128)
  \put(0,64){\includegraphics[height=70\unitlength]{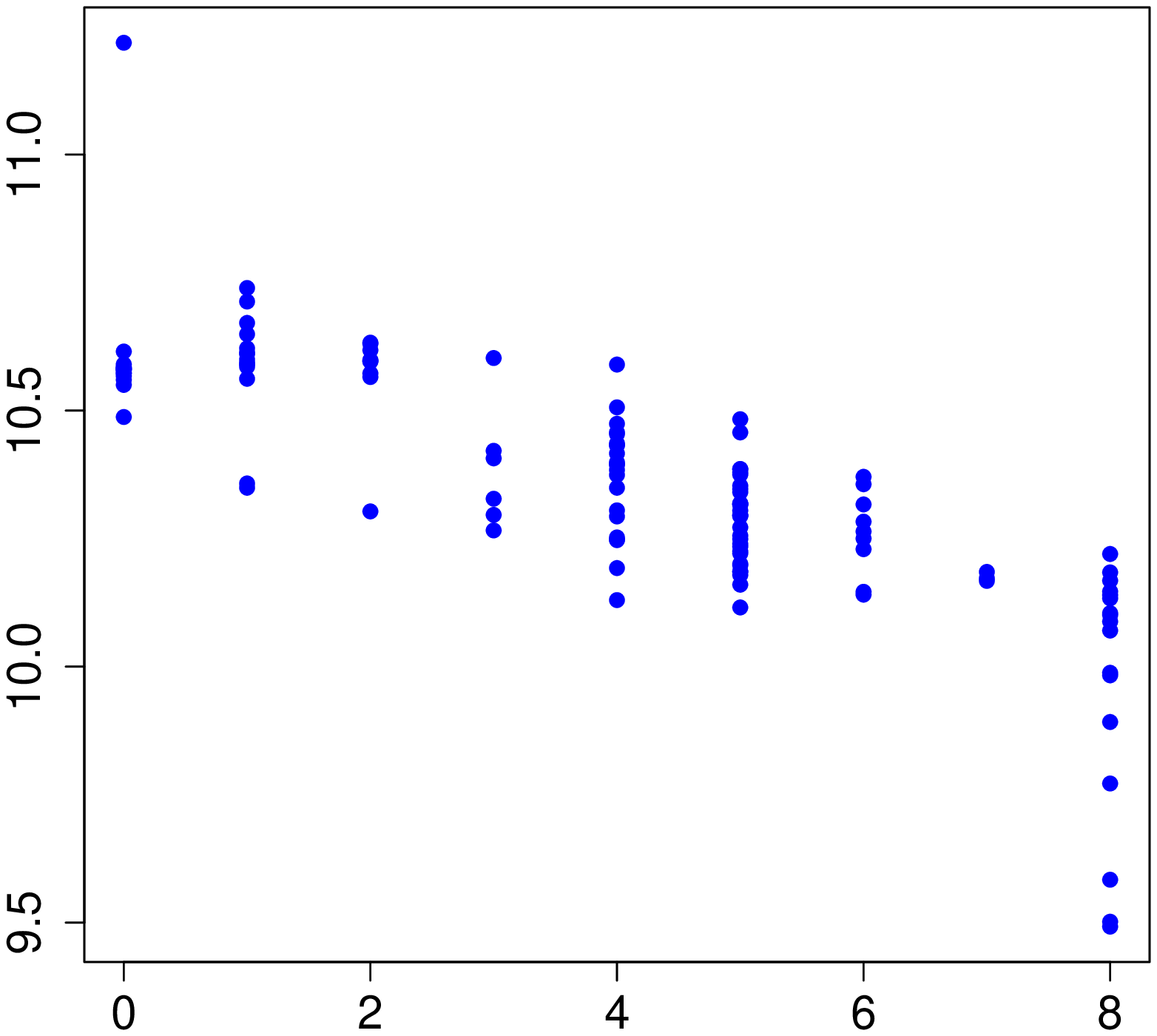}}	 \put(15 ,119){\small $F(\alpham)$} \put(31 ,65){$\mathcal{N}_{-}(\alpham)$}
   \put(71,64){\includegraphics[height=70\unitlength]{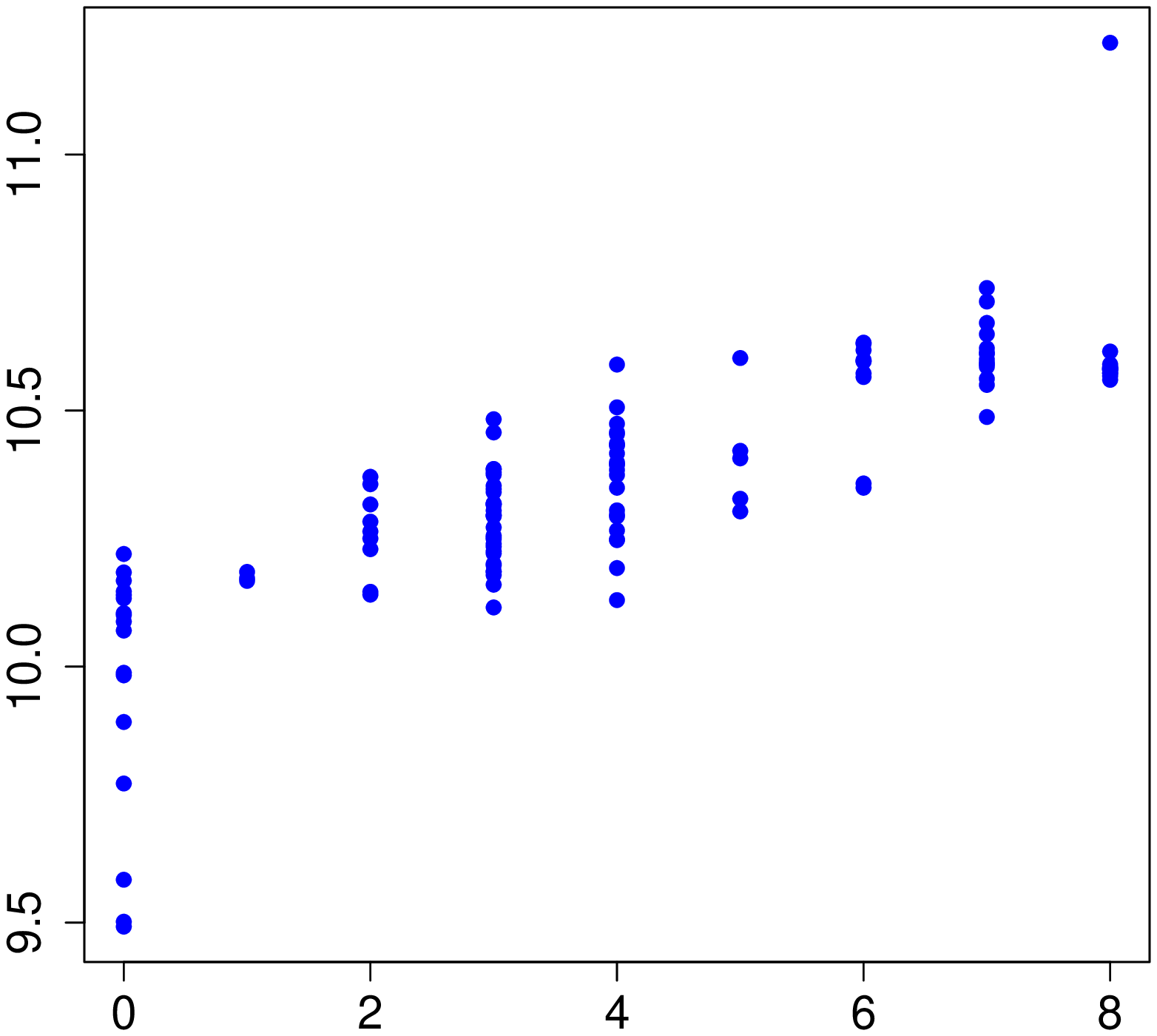}}  \put(86,119){\small $F(\alpham)$}   \put(102 ,65){\small $\mathcal{N}_{+}(\alpham)$}

 \put(0,0){\includegraphics[height=70\unitlength]{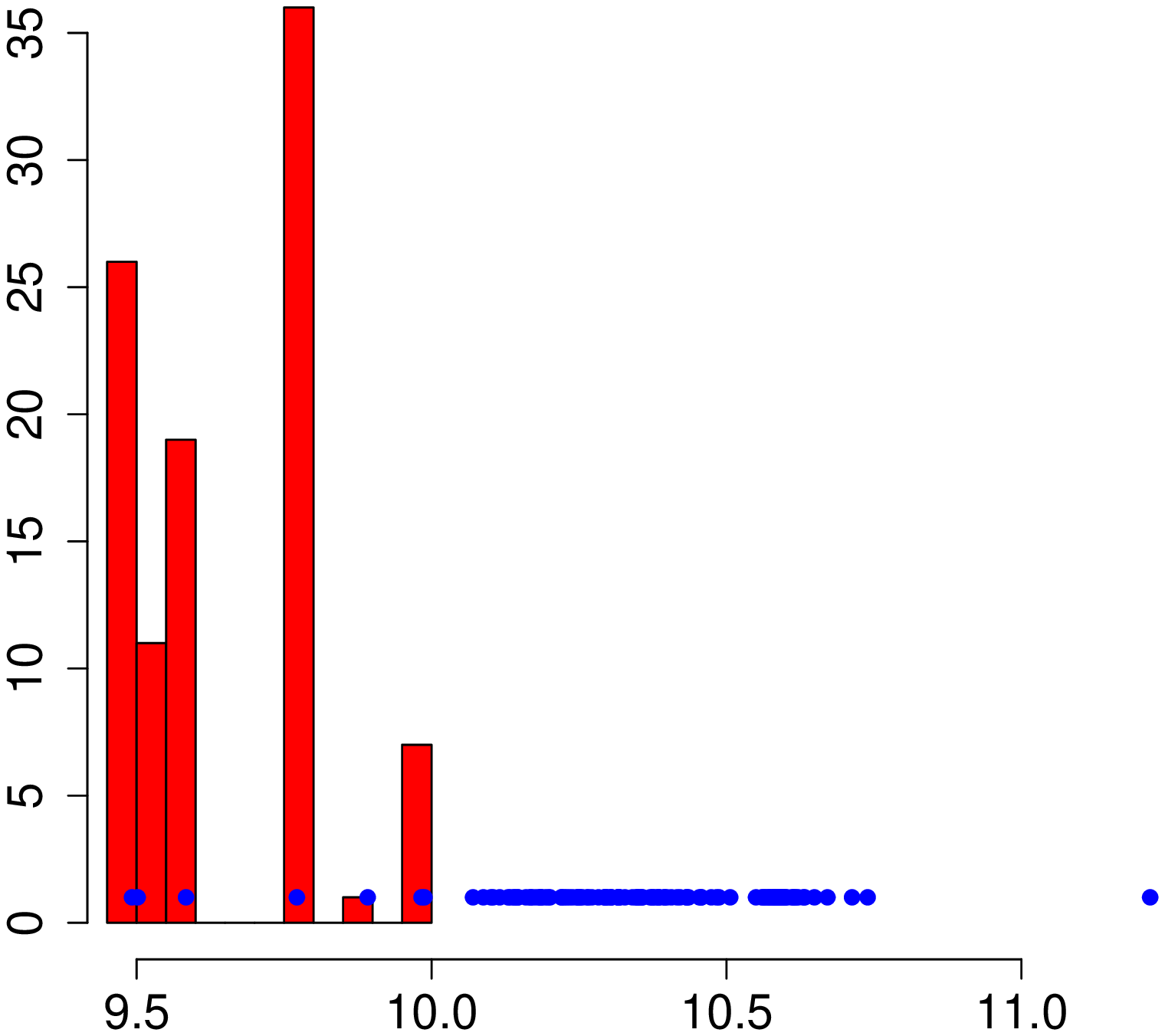}}
 \put(71,0){\includegraphics[height=70\unitlength]{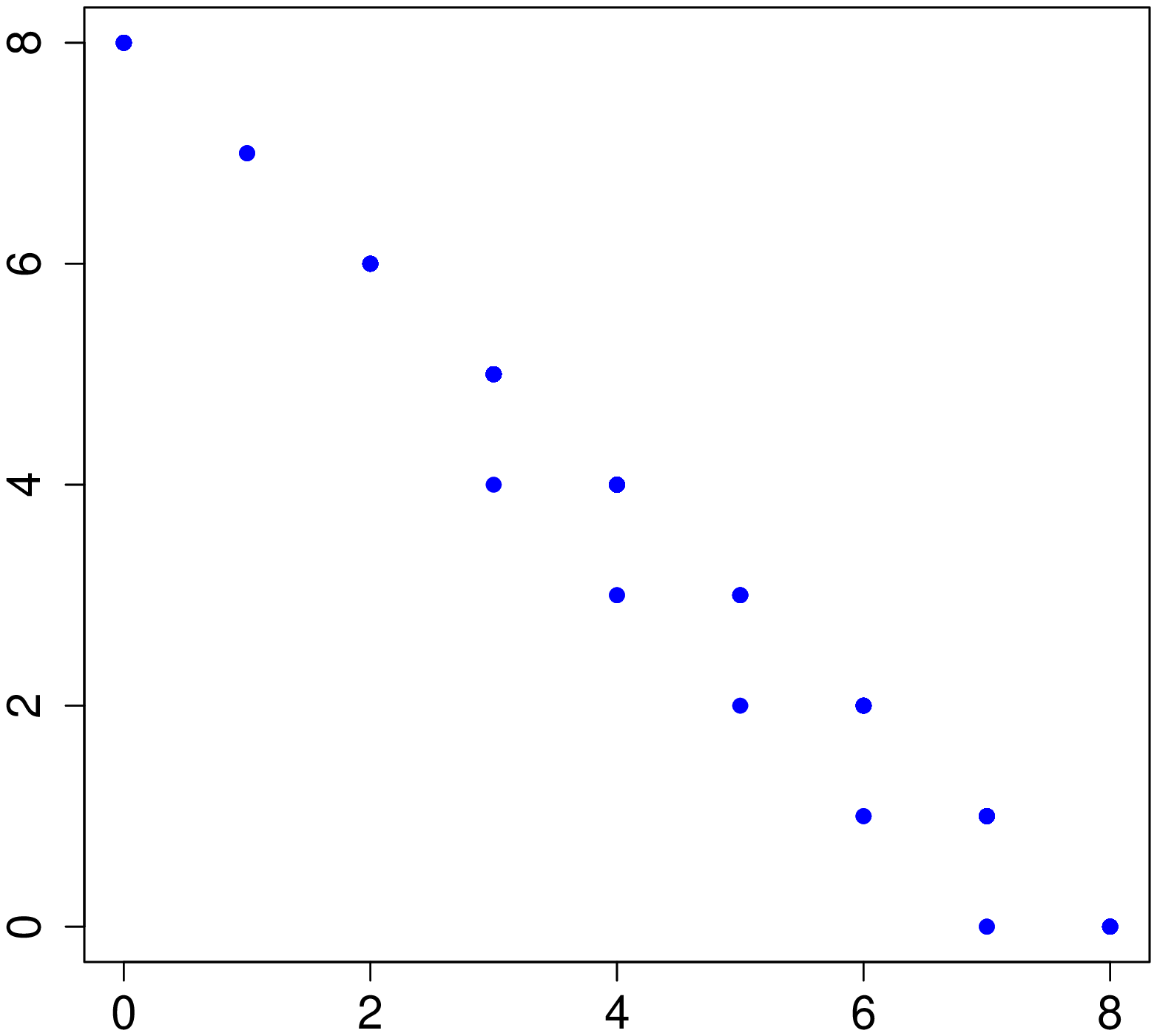}}
       \put(101 ,1){\small $\mathcal{N}_{+}(\alpham)$}  
     \put(60 ,40){\small $\mathcal{N}_{-}(\alpham)$} 
  \put(20,1){\small $\hat{F}_N(\cmatrix\,(\infty), \xmatrix)$}   
      
\end{picture}
 \caption{(Color online) Top left: $F(\alpham)$ as a function of the number of  $F$-increasing  directions $\mathcal{N}_{-}(\alpham)$.  Top: right: $F(\alpham)$ as a function of the number of  $F$-decreasing  directions $\mathcal{N}_{+}(\alpham)$.
 Bottom left:  histogram of log-likelihood values $\hat{F}_N(\cmatrix\,(\infty), \xmatrix)$, obtained  by running gradient descent  from a $100$ different random unbiased partitions, with  the assumed number $K=2$ of clusters.  Blue filled circles correspond to the MF log-likelihood, $F(\alpham)$, computed for all possible values of  $\alpha(\nu, \mu)= \Ind[\nu\!\in\! S_\mu]\gamma(\nu)$. Bottom right:  $\mathcal{N}_{-}(\alpham)$ as a function of $\mathcal{N}_{+}(\alpham).$}
 \label{figure:FhistK2} 
 \end{figure}

 \begin{figure}[t]
 \setlength{\unitlength}{0.67mm}
 \hspace*{-4mm}
 \begin{picture}(151,53)
 \put(0,0){\includegraphics[height=50\unitlength]{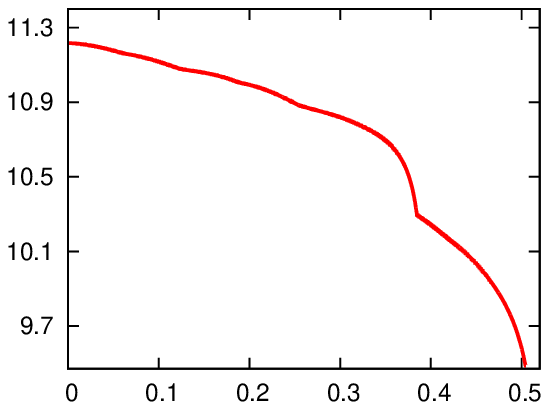}}
 \put(67,0){\includegraphics[height=50\unitlength]{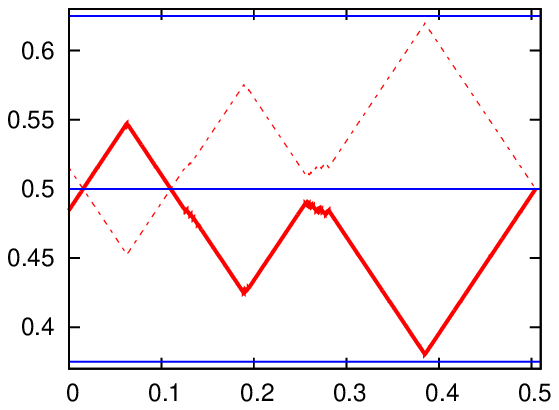}}
 \put(49,40){$\hat{F}_N(t)$}  
 \put(37 ,-3){$t$}   \put(105 ,-3){$t$}   
 \put(84 ,40){$\alpha(\mu\vert t)$}
 \end{picture}
\vspace*{-2mm}
 \caption{(Color online) Evolution of the log-likelihood, $\hat{F}_N(t)\equiv\hat{F}_N(\cmatrix(t),\xmatrix)$,  and the fraction of data in cluster $\mu$, $\alpha(\mu\vert t)\equiv\alpha(\mu\vert\cmatrix(t))$, where $\mu=\{1,2\}$, shown as functions of time (normalised number of   `moves') in the gradient descent algorithm evolving from a random unbiased initial partition. The  assumed number of clusters is $K=2$. Blue horizontal lines correspond to the  levels $3/8$, $4/8$ and $5/8$.  } 
 \label{figure:dynK2} 
 \end{figure}
 
  In this section we study  the performance of the simplest algorithm that minimises the log-likelihood function  (\ref{eq:F-hat-Norm}) via gradient descent, for the data described in Figure  \ref{figure:3d}. 
  The algorithm is implemented as follows: 
 \begin{enumerate}
\item Start with any initial  partition $\Pi(\cmatrix\,(0))=\left\{ S_1(\cmatrix\,(0)),\ldots,S_K(\cmatrix\,(0)) \right\}$, and compute the log-likelihood $\hat{F}_N(\cmatrix\,(0),\xmatrix)$.
\item For all $i\in [N]$, consider all possible moves of $i$ from  its current cluster $S_\mu(\cmatrix)$ to a new cluster $S_\nu(\cmatrix)$ and  compute the new value $\hat{F}_N(\cmatrix,\xmatrix)$ for each. 
\item  Select and execute a move which gives the largest decrease in $\hat{F}_N(\cmatrix,\xmatrix)$, and update $\Pi(\cmatrix)$.
\item Continue the last two steps while the value of $\hat{F}_N(\cmatrix, \xmatrix)$ continues to change. 
\item Output the partition $\Pi(\cmatrix\,(\infty))$ and the value of $\hat{F}_N\left(\cmatrix\,(\infty),\,\xmatrix\right)$.
\end{enumerate}

Using as initial states random  partitions of data $\cmatrix\,(0)$, where each  $i\in [N]$ has a probability  $1/K$ of being allocated to one of the $K$ clusters\footnote{In section \ref{section:algorithmic-cost}  we proved  that  for $N\rightarrow\infty$ the matrix $\cmatrix$ constructed in this way  is,  with high probability,  a \emph{partition}  of the set  $[N]$  into the $K$  subsets.},  we run the above algorithm for each value of $K\in[17]$ for $100$ different initalisations  $\cmatrix\,(0)$ and select the final partition, $\cmatrix\,(\infty)$,  with the smallest value of  $\hat{F}_N\equiv \hat{F}_N(\cmatrix\,(\infty),\xmatrix)$. The latter  is our estimate of  $\min_{\cmatrix}\hat{F}_N(\cmatrix ,\xmatrix)$.   We also compute, with the same parameters used to generate  our data, the mean-field log-likelihood  $F(\alpham)$ via equation (\ref{eq:F-Norm}).    

\begin{figure}[t]
\hspace*{-5mm}
 \setlength{\unitlength}{0.65mm}
 \begin{picture}(151,128)
  \put(0,64){\includegraphics[height=70\unitlength]{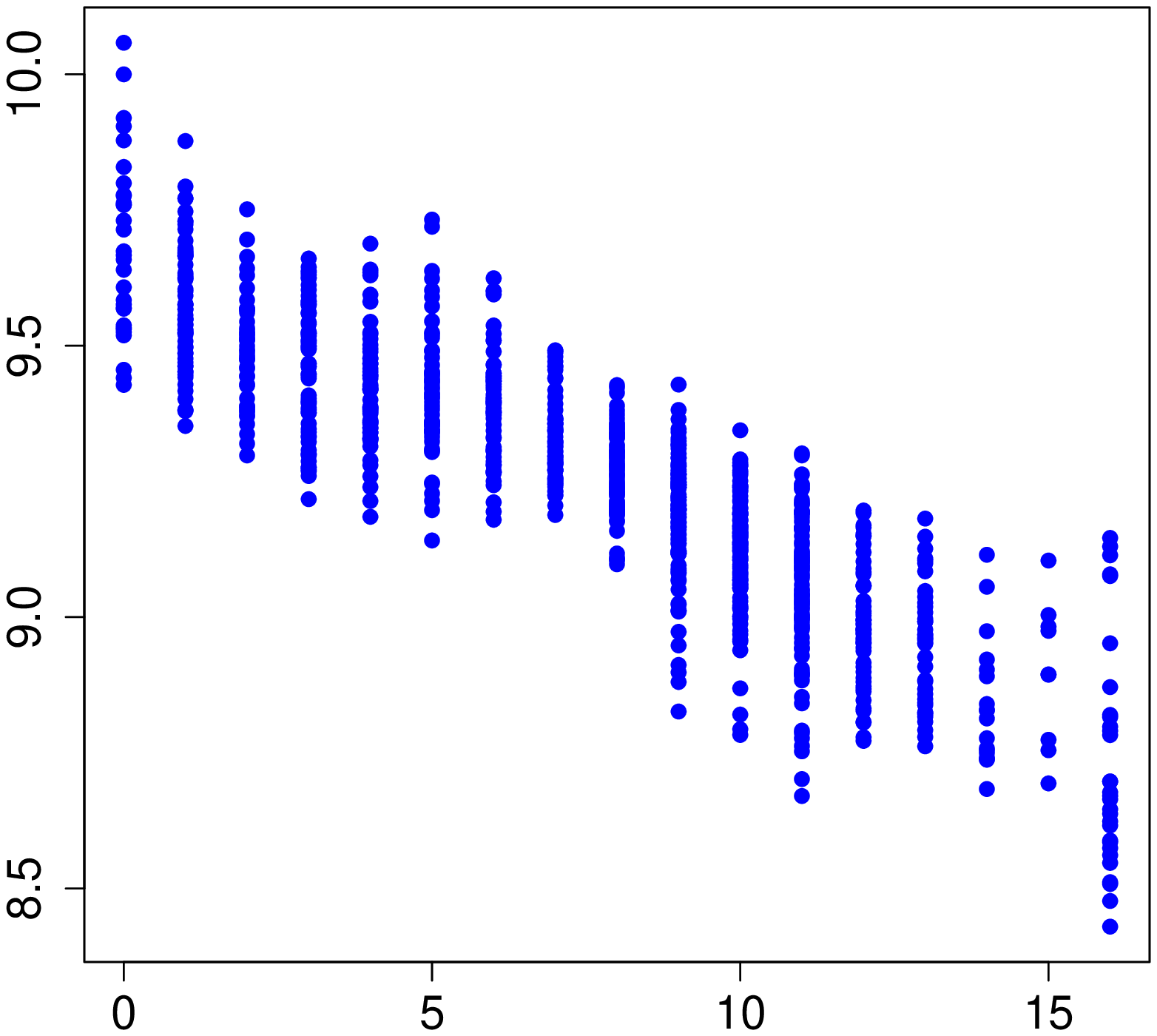}}	 \put(15 ,119){\small $F(\alpham)$} \put(31 ,65){$\mathcal{N}_{-}(\alpham)$}
   \put(71,64){\includegraphics[height=70\unitlength]{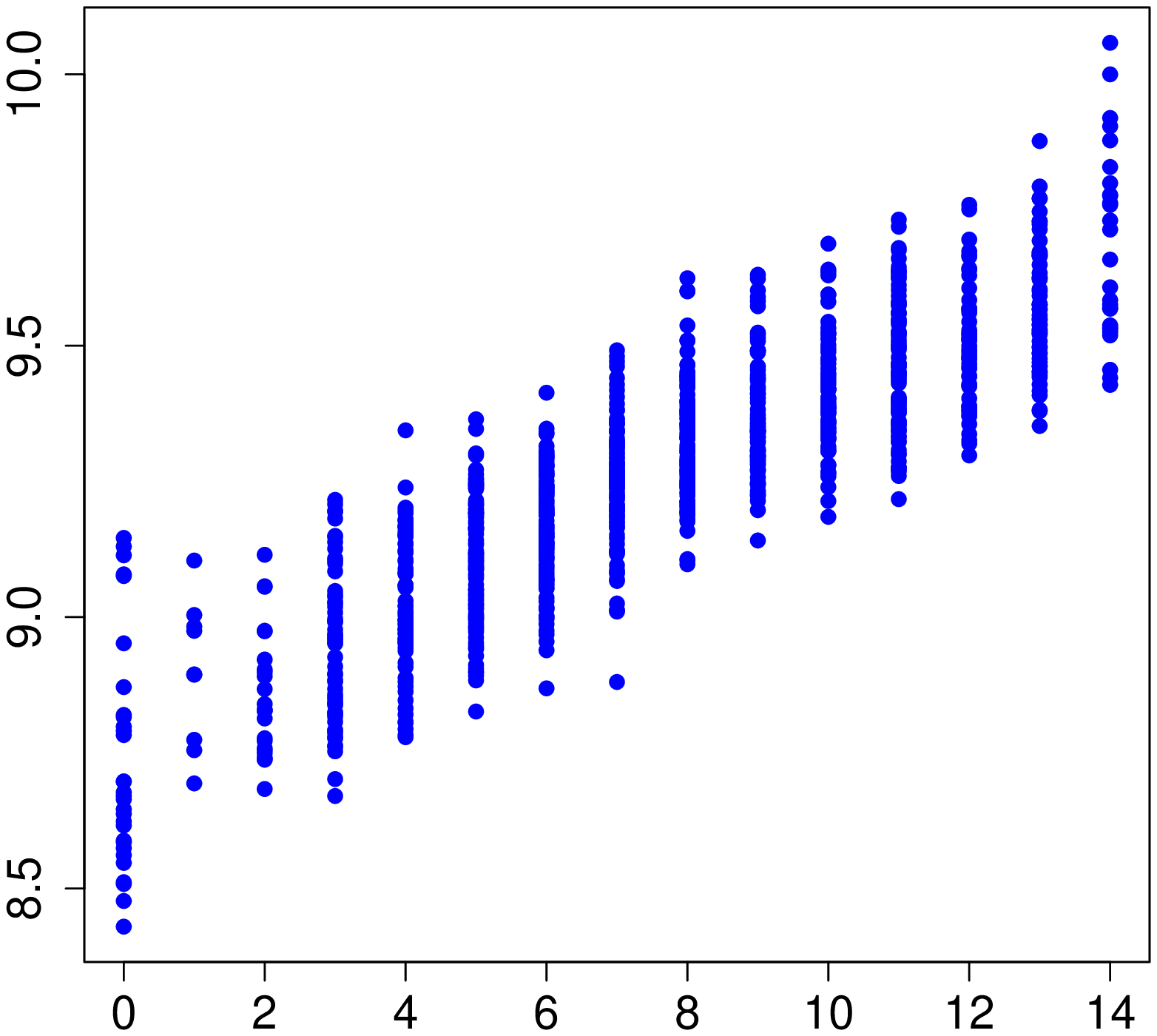}}  \put(86,119){\small $F(\alpham)$}   \put(102 ,65){\small $\mathcal{N}_{+}(\alpham)$}

 \put(0,0){\includegraphics[height=70\unitlength]{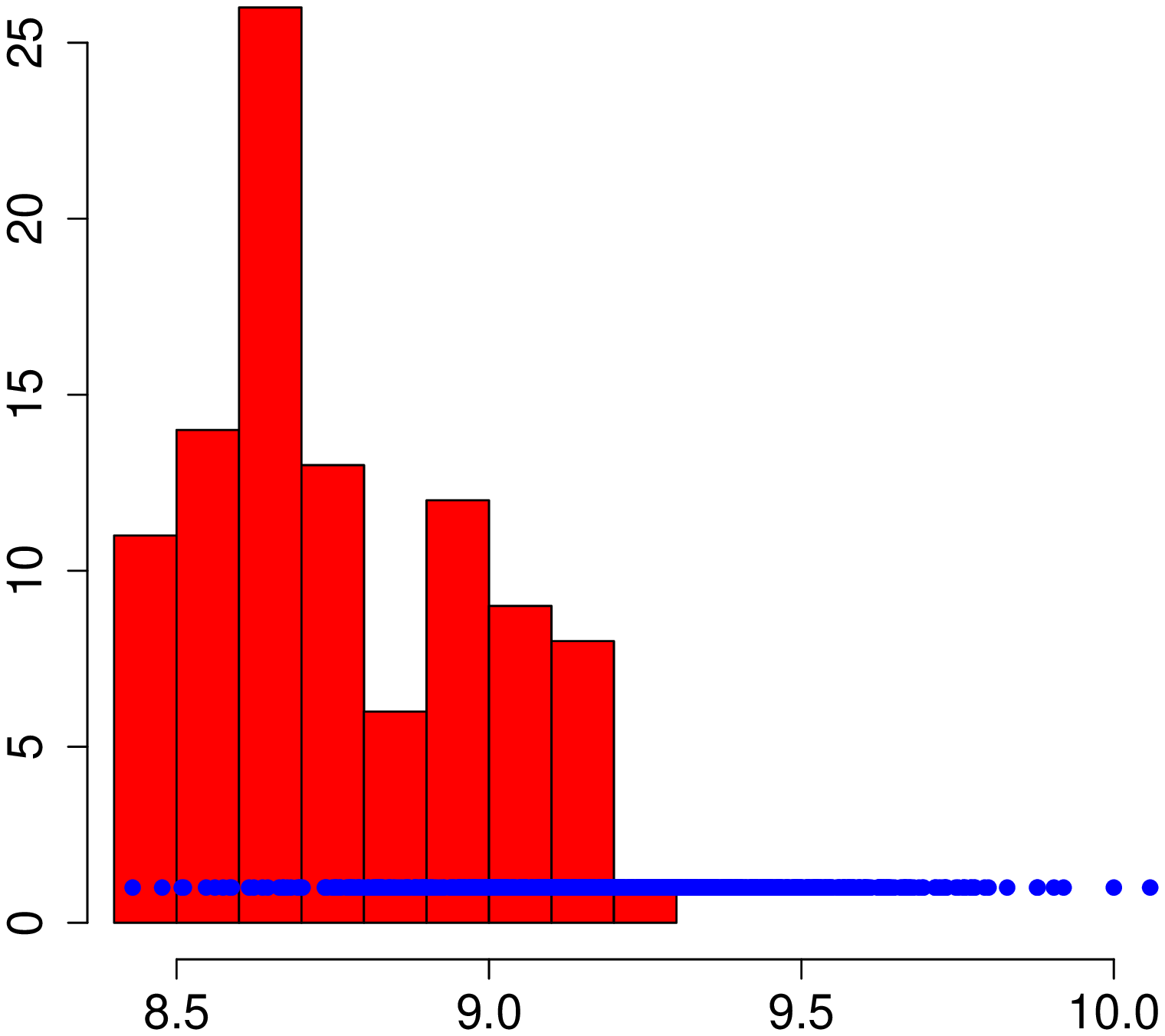}}
 \put(71,0){\includegraphics[height=70\unitlength]{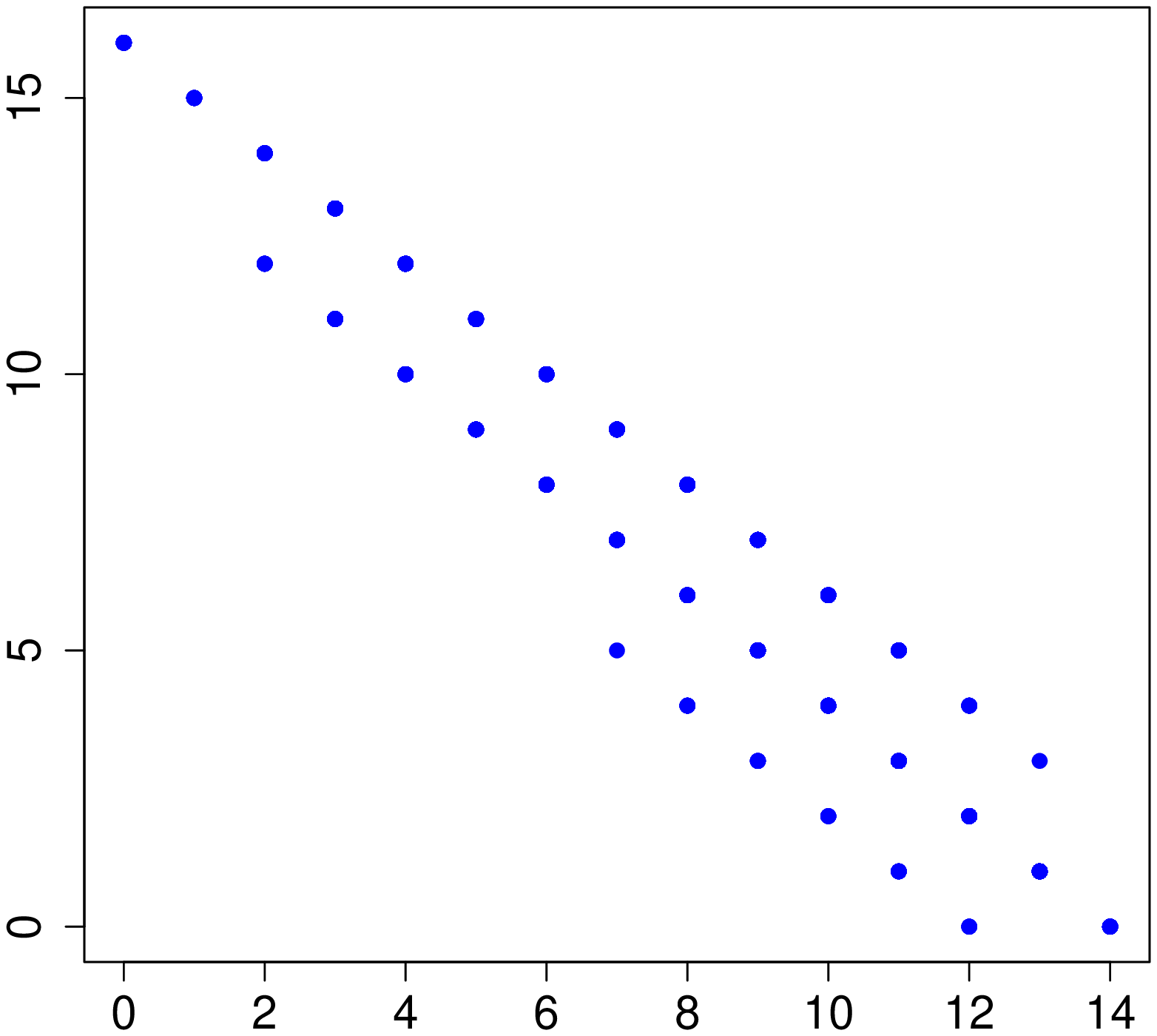}}
       \put(101 ,1){\small $\mathcal{N}_{+}(\alpham)$}  
     \put(60 ,40){\small $\mathcal{N}_{-}(\alpham)$} 
  \put(20,1){\small $\hat{F}_N(\cmatrix\,(\infty), \xmatrix)$}   
      
\end{picture}
 \caption{(Color online) Top left: $F(\alpham)$ as a function of the number of  $F$-increasing  directions $\mathcal{N}_{-}(\alpham)$.  Top: right: $F(\alpham)$ as a function of the number of  $F$-decreasing  directions $\mathcal{N}_{+}(\alpham)$.
 Bottom left:  histogram of log-likelihood values $\hat{F}_N(\cmatrix\,(\infty), \xmatrix)$, obtained  by running gradient descent  from a $100$ different random unbiased partitions, with  the assumed number $K=3$ of clusters.  Blue filled circles correspond to the MF log-likelihood, $F(\alpham)$, computed for all possible values of  $\alpha(\nu, \mu)= \Ind[\nu\!\in\! S_\mu]\gamma(\nu)$. Bottom right:  $\mathcal{N}_{-}(\alpham)$ as a function of $\mathcal{N}_{+}(\alpham).$}
 \label{figure:FhistK3} 
 \end{figure}

\begin{figure}[t]
\hspace*{-5mm}
 \setlength{\unitlength}{0.65mm}
 \begin{picture}(151,128)
  \put(0,64){\includegraphics[height=70\unitlength]{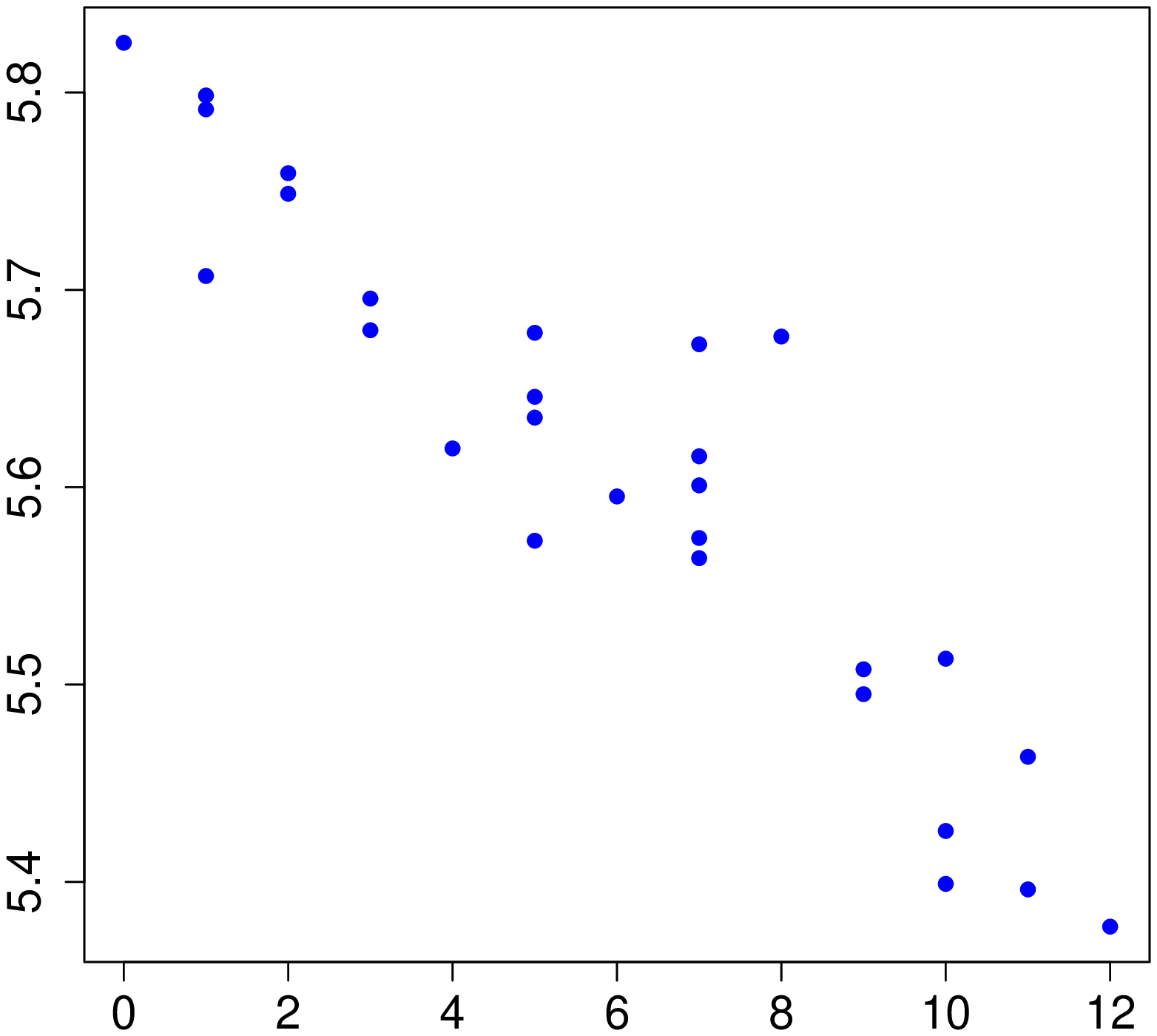}}	 \put(15 ,119){\small $F(\alpham)$} \put(31 ,65){$\mathcal{N}_{-}(\alpham)$}
   \put(71,64){\includegraphics[height=70\unitlength]{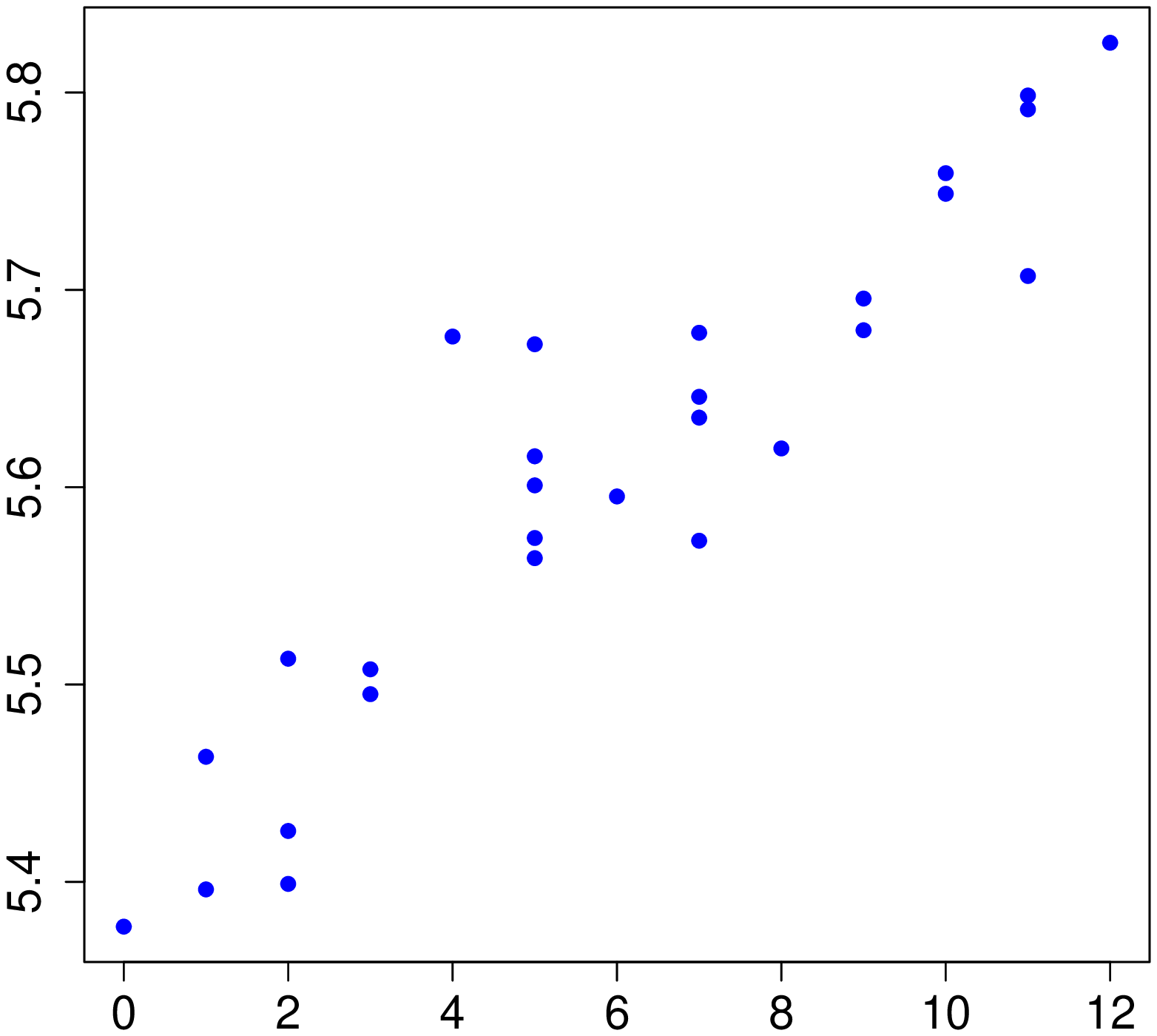}}  \put(86,119){\small $F(\alpham)$}   \put(102 ,65){\small $\mathcal{N}_{+}(\alpham)$}

 \put(0,0){\includegraphics[height=70\unitlength]{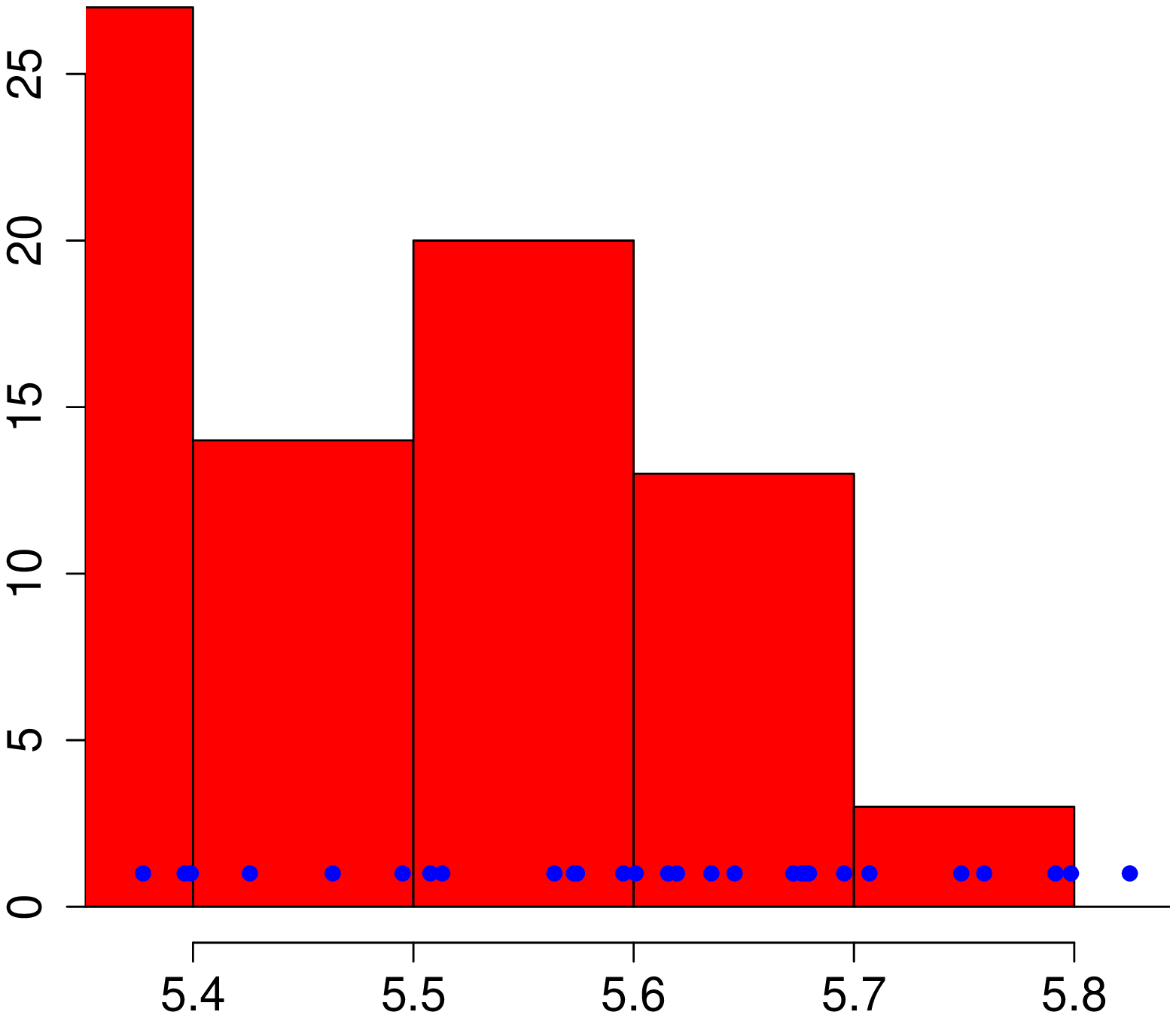}}
 \put(71,0){\includegraphics[height=70\unitlength]{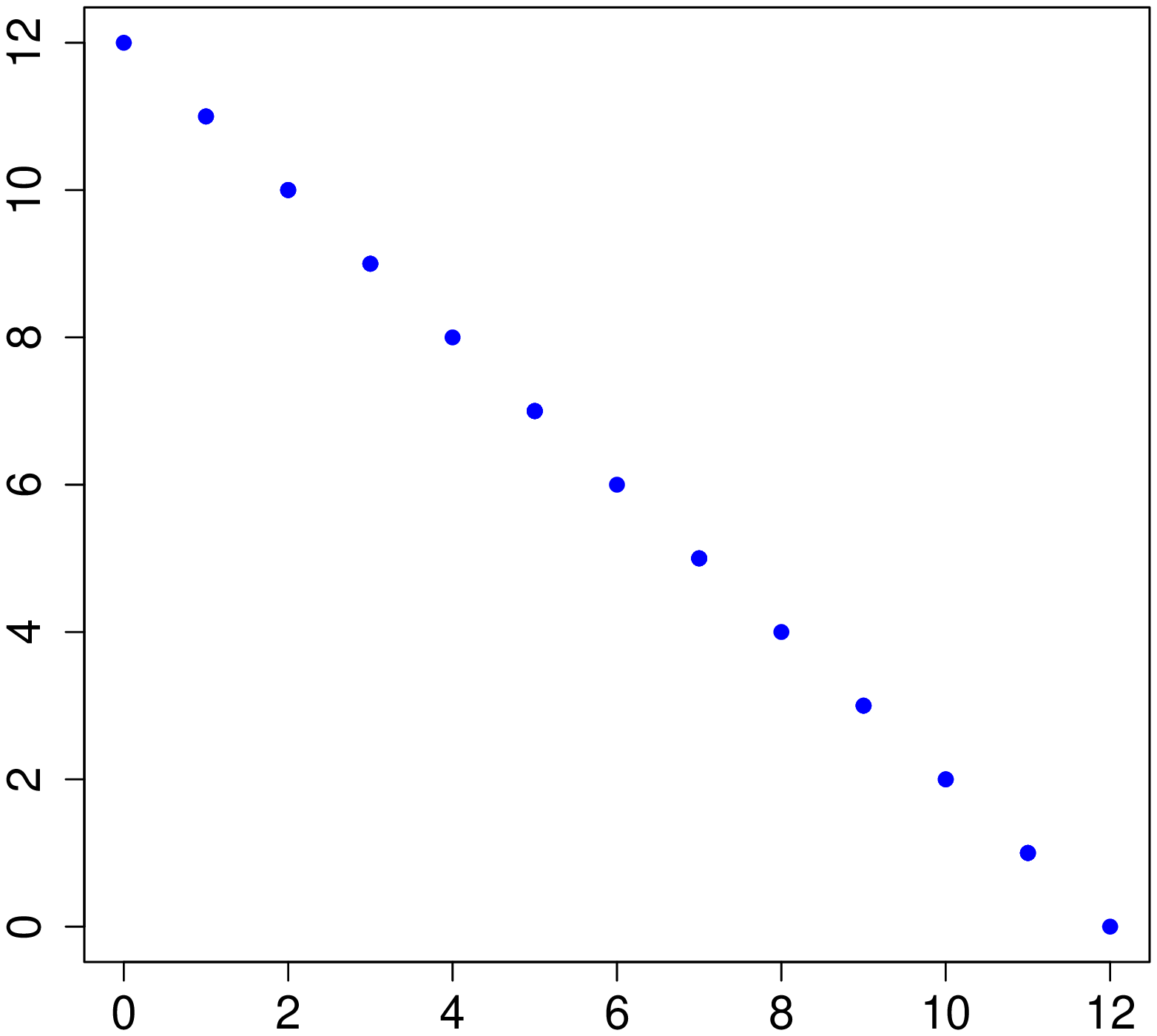}}
       \put(101 ,1){\small $\mathcal{N}_{+}(\alpham)$}  
     \put(60 ,40){\small $\mathcal{N}_{-}(\alpham)$} 
  \put(20,1){\small $\hat{F}_N(\cmatrix\,(\infty), \xmatrix)$}   
      
\end{picture}
 \caption{(Color online) Top left: $F(\alpham)$ as a function of the number of  $F$-increasing  directions $\mathcal{N}_{-}(\alpham)$.  Top: right: $F(\alpham)$ as a function of the number of  $F$-decreasing  directions $\mathcal{N}_{+}(\alpham)$.
 Bottom left:  histogram of log-likelihood values $\hat{F}_N(\cmatrix\,(\infty), \xmatrix)$, obtained  by running gradient descent  from a $100$ different random unbiased partitions, with  the assumed number $K=7$ of clusters.  Blue filled circles correspond to the MF log-likelihood, $F(\alpham)$, computed for all possible values of  $\alpha(\nu, \mu)= \Ind[\nu\!\in\! S_\mu]\gamma(\nu)$. Bottom right:  $\mathcal{N}_{-}(\alpham)$ as a function of $\mathcal{N}_{+}(\alpham).$}
 \label{figure:FhistK7} 
 \end{figure}

\begin{figure}[t]
 \setlength{\unitlength}{0.67mm}
 \begin{picture}(100,75)
 \put(0,0){\includegraphics[height=73\unitlength]{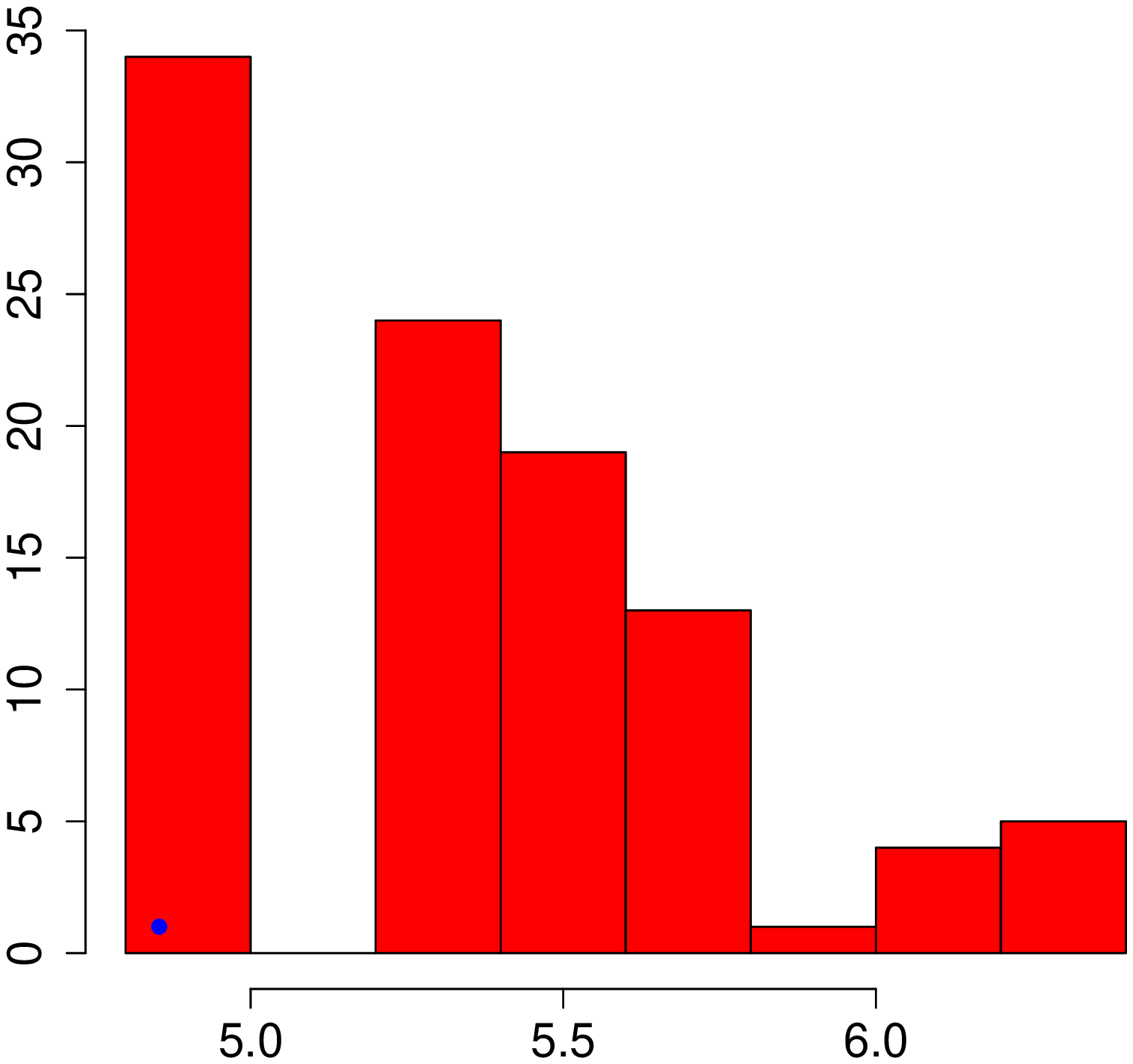}}
   \put(22 ,1){$\hat{F}_N(\cmatrix\,(\infty), \xmatrix)$} 
\end{picture}
 \caption{(Color online) Histogram of  the log-likelihood values obtained  by running the gradient descent algorithm  from a $100$ different random unbiased partitions,   with  the assumed number $K=8$ of clusters.  The blue filled circle corresponds to the MF lower bound $\sum_{\nu=1}^L\gamma(\nu)H(q_\nu)=4.853905$. }
 \label{figure:FhistK8} 
 \end{figure}

 When $K\leq L$, the log-likelihood $\hat{F}_N(\cmatrix\,(\infty),\xmatrix)$ is dominated by  partitions $\cmatrix\,(\infty)$  corresponding  to local minima and  saddlepoints of  $F(\alpham)$. The matrix $\alpham$ is defined by the entries $\left[\alpham\right]_{\nu\mu}\!=\!\Ind[\nu\!\in\! S_\mu]\gamma(\nu)$, generated by partitions $\Pi=\{S_1,\ldots,S_K\}$ of the set $[L]$ into $K$ subsets.  The total number of partitions is given by  $\mathcal{S}(L,K)$.  To enumerate all partitions we use the algorithm of~\cite{Djokic1989}.   We classify turning points  of $F(\alpham)$ as follows. For a given $\Pi$ and its associated matrix $\alpham$ we count the number $\mathcal{N}_{+}(\alpham)$ of  elementary `moves' into  the new partition $\tilde{\Pi}$ and $\tilde{\alpham}$ (in a single elementary  `move', a member of the set $S_\mu$, with $\vert S_\mu\vert>1$, is moved into  the set $S_\nu$) for which $F(\alpham)>F(\tilde{\alpham})$,  and the number $\mathcal{N}_{-}(\alpham)$ of moves for which $F(\alpham)<F(\tilde{\alpham})$.  If  $\mathcal{N}_{+}(\alpham)=0$ the state $\alpham$ is a (possibly local) minimum, and if $\mathcal{N}_{-}(\alpham)=0$ the state $\alpham$ is a (possibly local)  maximum. All other cases  are saddle points.  In Figures \ref{figure:FhistK2}, \ref{figure:FhistK3},  \ref{figure:FhistK7} and \ref{figure:FhistK8}  we compare $\hat{F}_N\left(\cmatrix\,(\infty),\,\xmatrix\right)$ with  $F(\alpham)$. 

  \begin{figure}[t]
 \setlength{\unitlength}{0.67mm}
 \hspace*{-4mm}
 \begin{picture}(151,53)
 \put(0,0){\includegraphics[height=50\unitlength]{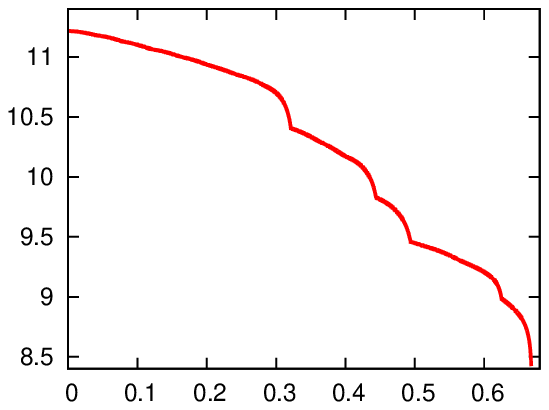}}
 \put(67,0){\includegraphics[height=50\unitlength]{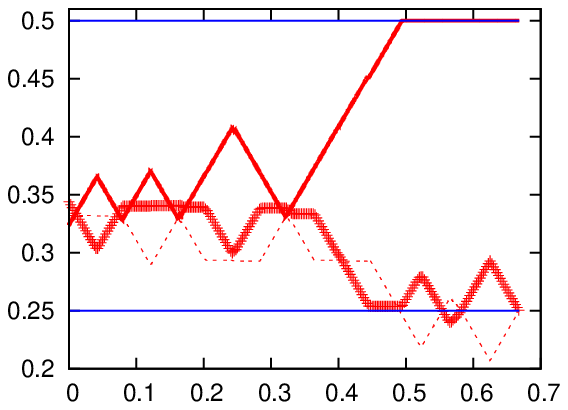}}
 \put(49,40){$\hat{F}_N(t)$}  
 \put(37 ,-3){$t$}   \put(105 ,-3){$t$}   
 \put(84 ,40){$\alpha(\mu\vert t)$}
 \end{picture}
\vspace*{-2mm}
 \caption{(Color online) Evolution of the log-likelihood, $\hat{F}_N(t)\equiv\hat{F}_N(\cmatrix(t),\xmatrix)$,  and the fraction of data in cluster $\mu$, $\alpha(\mu\vert t)\equiv\alpha(\mu\vert\cmatrix(t))$, where $\mu=\{1,2,3\}$, shown as functions of time (normalised number of   `moves') in the gradient descent algorithm evolving from a random unbiased initial partition. The  assumed number of clusters is $K=3$. Blue horizontal lines correspond to the  levels $3/8$, $4/8$ and $5/8$.  } 
 \label{figure:dynK3} 
 \end{figure}

\begin{figure}[t]
 \setlength{\unitlength}{0.67mm}
 \hspace*{-4mm}
 \begin{picture}(151,53)
 \put(0,0){\includegraphics[height=50\unitlength]{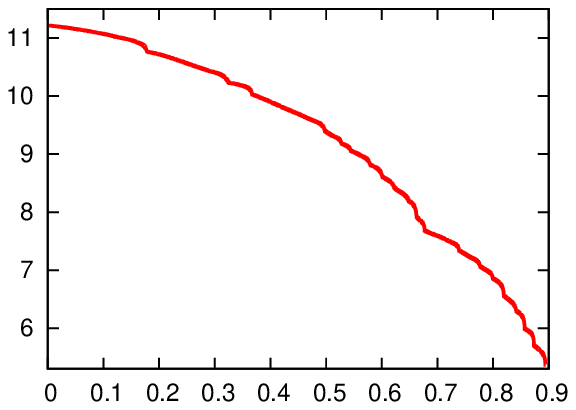}}
 \put(67,0){\includegraphics[height=50\unitlength]{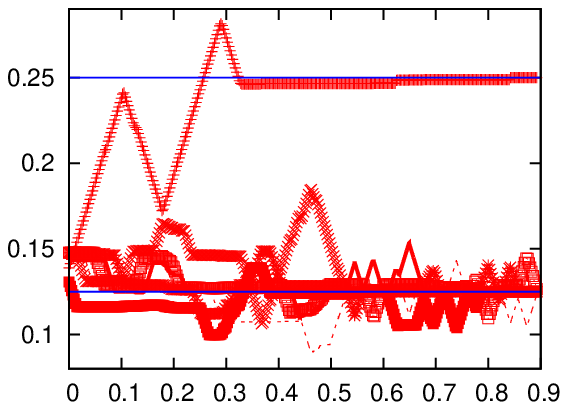}}
 \put(49,40){$\hat{F}_N(t)$}  
 \put(37 ,-3){$t$}   \put(105 ,-3){$t$}   
 \put(84 ,40){$\alpha(\mu\vert t)$}
 \end{picture}
\vspace*{-2mm}
 \caption{(Color online) Evolution of the log-likelihood, $\hat{F}_N(t)\equiv\hat{F}_N(\cmatrix(t),\xmatrix)$,  and the fraction of data in cluster $\mu$, $\alpha(\mu\vert t)\equiv\alpha(\mu\vert\cmatrix(t))$, where $\mu=\{1,2,\ldots,7\}$, shown as functions of time (normalised number of   `moves') in the gradient descent algorithm evolving from a random unbiased initial partition. The  assumed number of clusters is $K=7$. Blue horizontal lines correspond to the  levels $3/8$, $4/8$ and $5/8$.  } 
 \label{figure:dynK7} 
 \end{figure}

\begin{figure}[t]
 \setlength{\unitlength}{0.67mm}
 \hspace*{-4mm}
 \begin{picture}(151,53)
 \put(0,0){\includegraphics[height=50\unitlength]{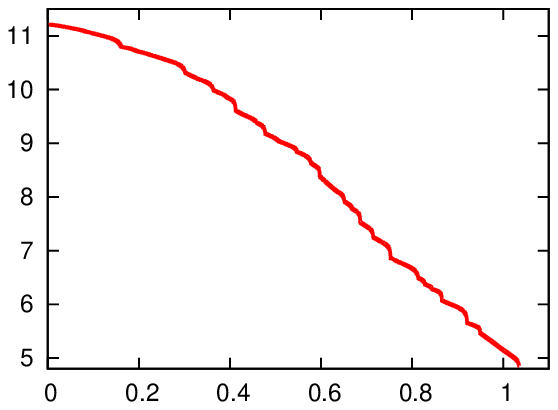}}
 \put(67,0){\includegraphics[height=50\unitlength]{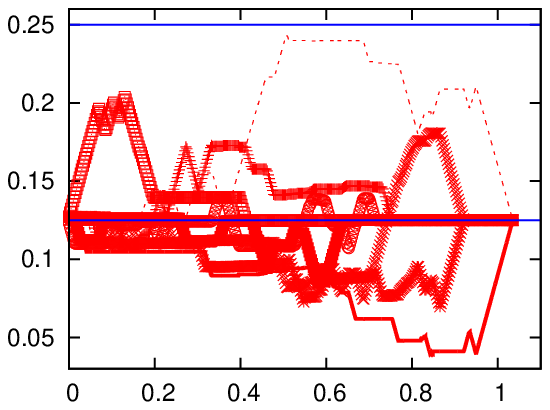}}
 \put(49,40){$\hat{F}_N(t)$}  
 \put(37 ,-3){$t$}   \put(105 ,-3){$t$}   
 \put(84 ,40){$\alpha(\mu\vert t)$}
 \end{picture}
\vspace*{-2mm}
 \caption{(Color online) Evolution of the log-likelihood, $\hat{F}_N(t)\equiv\hat{F}_N(\cmatrix(t),\xmatrix)$,  and the fraction of data in cluster $\mu$, $\alpha(\mu\vert t)\equiv\alpha(\mu\vert\cmatrix(t))$, where $\mu=\{1,2,\ldots,8\}$, shown as functions of time (normalised number of   `moves') in the gradient descent algorithm evolving from a random unbiased initial partition. The  assumed number of clusters is $K=8$. Blue horizontal lines correspond to the  levels $3/8$, $4/8$ and $5/8$.  } 
 \label{figure:dynK8} 
 \end{figure}

 Those turning  points of $F(\alpham)$ that are  of the form $\left[\alpham\right]_{\nu\mu}=\Ind[\nu\in S_\mu]\gamma(\nu)$  also act as  dynamic  `attractors'.   This can be seen by comparing Figure  \ref{figure:FhistK2} to Figure  \ref{figure:dynK2}, and Figure  \ref{figure:FhistK3} to Figure   \ref{figure:dynK3}, etc.  Here $\hat{F}_N(t)\equiv\hat{F}_N\left(\cmatrix\,(t),\,\xmatrix\right)$, as computed during the simulated process,   is seen to evolve  from plateau to plateau by a succession of  rapid relaxations, and the value of $\hat{F}_N(t)$ at the beginning of each plateau can be (approximately) mapped  to the value of $F(\alpham)$  via  the  fractions $\alpha(\mu)=\sum_{\nu=1}^L\Ind[\nu\in S_\mu]\gamma(\nu)$ of data in clusters $\mu$.  However as $K$  is increased, more and more attractors are not of the  form  $\Ind[\nu\in S_\mu]\gamma(\nu)$  (see  Figures   \ref{figure:dynK3},   \ref{figure:dynK7} and \ref{figure:dynK8}). 

\begin{figure}[t]
 \setlength{\unitlength}{0.67mm}
 \begin{picture}(100,71)
 \put(0,0){\includegraphics[height=73\unitlength]{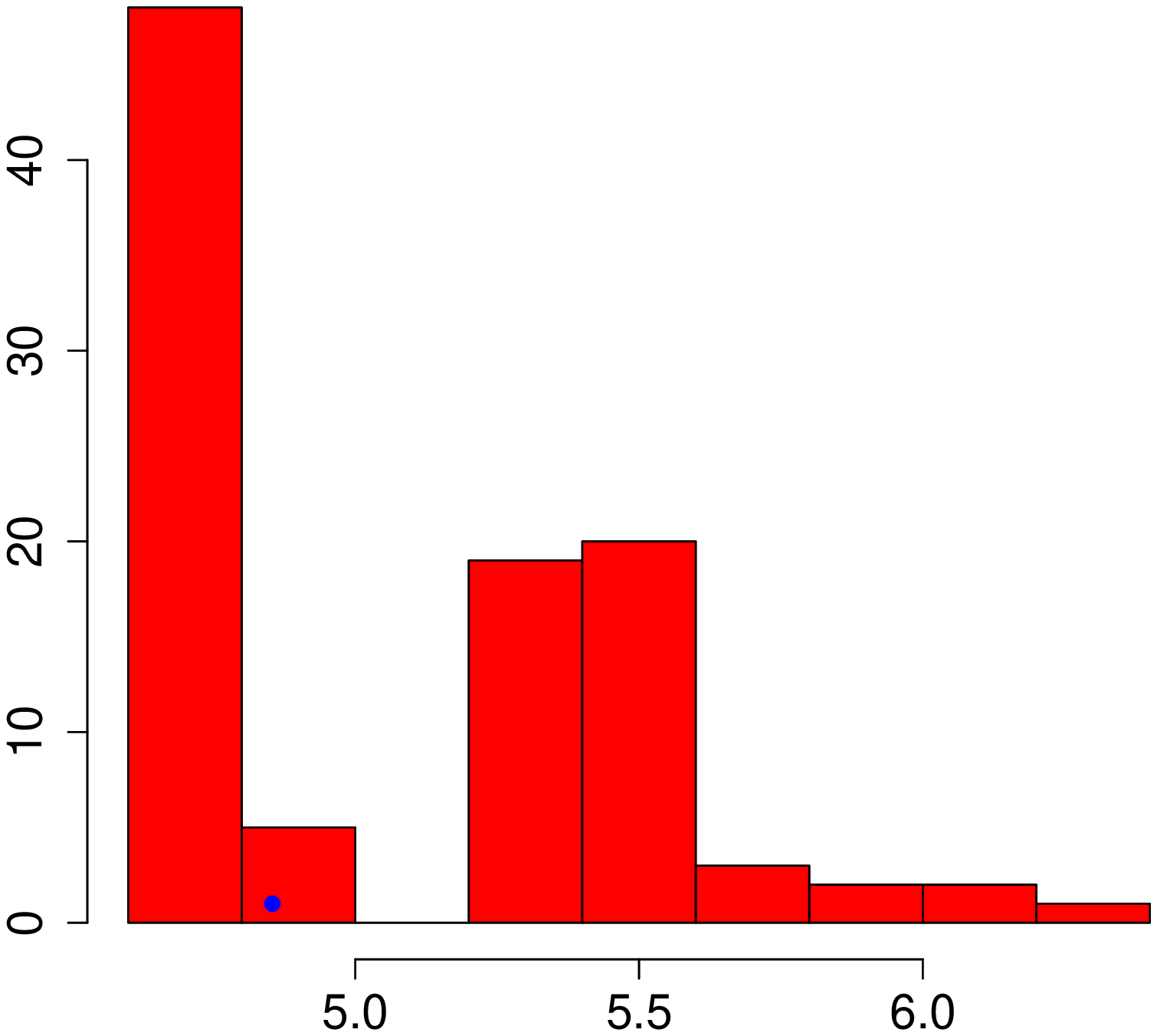}}
  \put(22 ,1){$\hat{F}_N(\cmatrix\,(\infty), \xmatrix)$} 
\end{picture}
 \caption{(Color online) 
 Histogram of  the log-likelihood values obtained  by running the gradient descent algorithm  from a $100$ different random unbiased partitions,   with  the assumed number $K=9$ of clusters.  The blue filled circle corresponds to the MF lower bound $\sum_{\nu=1}^L\gamma(\nu)H(q_\nu)=4.853905$.
 }
 \label{figure:FhistK9} 
 \end{figure}

The predictions of the mean-field log-likelihood $F(\alpham)$ for  $\min_{\cmatrix}\hat{F}_N(\cmatrix,\xmatrix)$ are incorrect when $K>L$.  The log-likelihood $F(\alpham)$ is bounded from below by the average entropy $\sum_{\nu=1}^L\gamma(\nu)H(q_\nu)$,  but in this regime the gap between this lower bound and $\min_{\cmatrix}\hat{F}_N(\cmatrix,\xmatrix)$ is widening as  we increase the number of assumed clusters $K$. This effect can be  clearly seen in Figure \ref{figure:FhistK9}.
 We also see in this Figure that $\sum_{\nu=1}^L\gamma(\nu)H(q_\nu)$ separates the low entropy states obtained by gradient descent into two sets. The first set, which includes $\argmin_{\cmatrix}\hat{F}_N(\cmatrix,\xmatrix)$, is given by\footnote{The equality in this definition can only be true when $K=L$ (see Figure \ref{figure:FhistK8}).} $\{\cmatrix:~\hat{F}_N(\cmatrix,\xmatrix)\leq \sum_{\nu=1}^L\gamma(\nu)H(q_\nu)\}$, and the second set is given by $\{\cmatrix:~\hat{F}_N(\cmatrix,\xmatrix)>\sum_{\nu=1}^L\gamma(\nu)H(q_\nu)\}$. Since for $K>L$ we have $F(\alpham)>\sum_{\nu=1}^L\gamma(\nu)H(q_\nu)$,  we expect that $\min_{\alpham}F(\alpham)$ gives correct predictions for at least some of the low entropy states in the second set.  

%
%
%
%

\clearpage

 \begin{figure}[t]
 \setlength{\unitlength}{0.67mm}
 \hspace*{-4mm}
 \begin{picture}(151,50)
 \put(0,0){\includegraphics[height=50\unitlength]{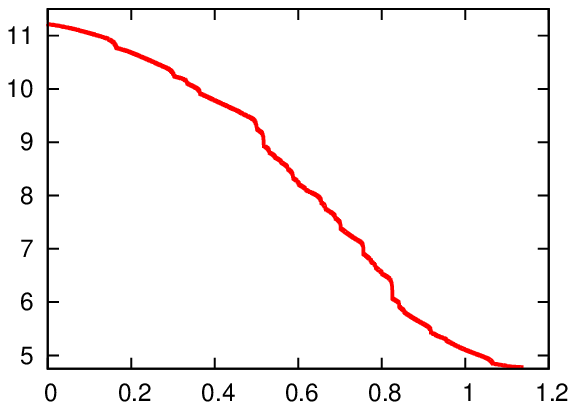}}
 \put(67,0){\includegraphics[height=50\unitlength]{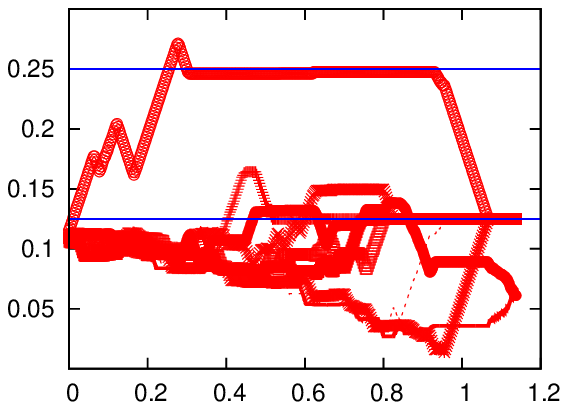}}
 \put(49,40){$\hat{F}_N(t)$}  
 \put(37 ,-3){$t$}   \put(105 ,-3){$t$}   
 \put(84 ,41){$\alpha(\mu\vert t)$}
 \end{picture}
\vspace*{-2mm}
 \caption{(Color online) Evolution of the log-likelihood, $\hat{F}_N(t)\equiv\hat{F}_N(\cmatrix(t),\xmatrix)$,  and the fraction of data in cluster $\mu$, $\alpha(\mu\vert t)\equiv\alpha(\mu\vert\cmatrix(t))$, where $\mu=\{1,2,\ldots,9\}$, shown as functions of time (normalised number of   `moves') in the gradient descent algorithm evolving from a random unbiased initial partition. The  assumed number of clusters is $K=9$. Blue horizontal lines correspond to the  levels $3/8$, $4/8$ and $5/8$.  } 
 \label{figure:dynK9} 
 \end{figure}
 \vspace*{-12mm}

\section{Estimation of differential entropy\label{section:log-det}}
In this section we compute  the finite sample-size corrections to the MF entropy   (\ref{eq:F-Norm}).  In order to do this we first note that for a sample  $\{\x_1,\ldots, \x_N\}$, where  each $\x_i \in \mathbb{R}^d$ is drawn from  the multivariate Gaussian distribution $\mathcal{N}(\x\vert\m,\Lmatrix)$, the empirical  covariance matrix  $\hat{\Lmatrix}=N^{-1}\sum_{i=1}^N(\x_i\!-\!\hat{\m})(\x_i\!-\!\hat{\m})^T$, where $\hat{\m}=\frac{1}{N}\sum_{i=1}^N\x_i$ is the empirical mean, obeys the following asymptotic law:  $[\log\vert\hat{\Lmatrix}\vert-\log\left\vert \Lmatrix\right\vert-d(d\!+\!1)/2N ]/\sqrt{2d/N}\rightarrow \mathcal{N}(0,1)$ as $N\rightarrow\infty$  (see~\cite{Cai2015}  and references therein).  This is equivalent to stating $\log|\hat{\Lmatrix}|\rightarrow \log|\Lmatrix|+d(d\!+\!1)/2N+z\sqrt{2d/N}$, where $z\sim\mathcal{N}(0,1)$. 

Let us assume that the above is true for the empirical covariance matrices that feature in the log-likelihood  (\ref{eq:F-hat-Norm}) and evaluate $\hat{F}_N(\cmatrix)$ for large $N$:
\begin{eqnarray}
\hat{F}_N(\cmatrix)&=&\sum_{\mu=1}^K\frac{ M_\mu\left(\cmatrix\right)}{N}  \frac{1}{2}  \log \left((2\pi\rme)^{d}\left\vert \Lmatrix_{\mu}^{-1}\left(\cmatrix\right)\right\vert  \right)  \nonumber \\
&=&\sum_{\mu=1}^K \frac{ M_\mu\left(\cmatrix\right)}{2N}  \Bigg\{ \log \left((2\pi\rme)^{d} \left\vert \Lmatrix_{\mu}^{-1}(\alpham)\right\vert \right)\nonumber
\\&&\hspace*{10mm} +~\frac{d(d\!+\!1)}{2M_\mu(\cmatrix)}+z_\mu\sqrt{\frac{2d}{M_\mu(\cmatrix)}}  \Bigg\} \nonumber\\
&=&F(\alpham)+\sum_{\mu=1}^K \frac{ M_\mu(\cmatrix)}{2N}  \Bigg\{\frac{d(d\!+\!1)}{2M_\mu\left(\cmatrix\right)}
+z_\mu\sqrt{\frac{2d}{M_\mu(\cmatrix)}}  \Bigg  \} \nonumber\\
&=&F(\alpham)+\frac{Kd(d\!+\!1)}{4N} + \sum_{\mu=1}^K z_\mu\sqrt{\frac{d\,\alpha(\mu)}{2N }}
 \label{eq:F-corr-2}
\end{eqnarray}
The average and variance  of the above random variable are given by $F(\alpham)+Kd(d\!+\!1)/4N$ and $d/2N$, respectively.  We expect the above result to be exact when  $F(\alpham)=\sum_{\nu=1}^L\gamma(\nu)H \left(q_\nu \right)$, which can only happen when $K=L$, and all  $q_\nu(\x)$ are Gaussian distributions.



%

\end{document}